\documentclass[aps,twocolumn,letterpaper,superscriptaddress]{revtex4-1}
\usepackage[T1]{fontenc}
\usepackage[latin9]{inputenc}
\usepackage{amsmath}
\usepackage{amssymb}
\usepackage{stmaryrd}

\usepackage[usenames,dvipsnames]{xcolor}
\usepackage[hyperindex,pdftex, breaklinks,colorlinks = true,linkcolor = blue,urlcolor=blue,citecolor=blue]{hyperref}

\begin{document}
	
	\title{Microscopic polarization and magnetization fields in extended systems}

	\author{ Perry T. Mahon }
	\email{pmahon@physics.utoronto.ca}
	\affiliation{ Department of Physics, University of Toronto, Toronto, Ontario M5S 1A7, Canada }

	\author{ Rodrigo A. Muniz }
	\email{rodrigo.a.muniz@gmail.com}
	
	\affiliation{ Department of Physics, University of Toronto, Toronto, Ontario M5S 1A7, Canada }
	\affiliation{ Department of Electrical Engineering and Computer Science, University of Michigan, Ann Arbor, MI 48109, USA}

	\author{ J. E. Sipe }
	\email{sipe@physics.utoronto.ca}
	
	\affiliation{ Department of Physics, University of Toronto, Toronto, Ontario M5S 1A7, Canada }

	\date{\today }
	
	\begin{abstract}
		We introduce microscopic polarization and magnetization fields at
		each site of an extended system, as well as free charge and current
		density fields associated with charge movement from site to site,
		by employing a lattice gauge approach based on a set of orthogonal
		orbitals associated with each site. These microscopic fields are defined using
		a single-particle electron Green function, and the equations
		governing its evolution under excitation by an electromagnetic field at
		arbitrary frequency involve the electric and magnetic fields rather than the scalar and
		vector potentials. If the sites are taken to be far from each other,
		we recover the limit of isolated atoms. For an infinite crystal we
		choose the orbitals to be maximally-localized Wannier functions, and
		in the long-wavelength limit we recover the expected linear response
		of an insulator, including the zero frequency transverse conductivity of a topologically nontrivial insulator. For a topologically trivial insulator we recover the expected expressions
		for the macroscopic polarization and magnetization in the ground state,
		and find that the linear response to excitation at arbitrary frequency
		is described solely by the microscopic polarization and magnetization
		fields. For very general optical response calculations the microscopic
		fields necessarily satisfy charge conservation, even under basis
		truncation, and do not suffer from the false divergences at zero frequency
		that can plague response calculations using other approaches.
	\end{abstract}
	
	\maketitle

\section{Introduction}

The optical properties of materials were first related to the microscopic
structure and properties of matter near the start of the twentieth
century \cite{Lorentz}. Macroscopic polarization and magnetization
fields, $\boldsymbol{P}(\boldsymbol{x},t)$ and $\boldsymbol{M}(\boldsymbol{x},t)$,
which appear in the description of the macroscopic charge and current
densities, 
\begin{align}
\varrho(\boldsymbol{x},t) & =-\boldsymbol{\nabla}\boldsymbol{\cdot}\boldsymbol{P}(\boldsymbol{x},t)+\varrho_{F}(\boldsymbol{x},t),\label{eq:macro}\\
\boldsymbol{J}(\boldsymbol{x},t) & =\frac{\partial\boldsymbol{P}(\boldsymbol{x},t)}{\partial t}+c\boldsymbol{\nabla\times M}(\boldsymbol{x},t)+\boldsymbol{J}_{F}(\boldsymbol{x},t),\nonumber 
\end{align}
were associated with charges ``bound'' in molecules and attributed
to the electric and magnetic dipole moments of those molecules. The
``free'' charge and current densities, $\varrho_{F}(\boldsymbol{x},t)$
and $\boldsymbol{J}_{F}(\boldsymbol{x},t)$, were associated with
charged carriers free to move through the medium if driven by applied
fields, as in a conductor. Later developments extended the definitions
of $\boldsymbol{P}(\boldsymbol{x},t)$ and $\boldsymbol{M}(\boldsymbol{x},t)$
to include contributions from higher multipole moments of constituent
molecules \cite{RosenfeldBook,Fokker}. The work of Power, Zienau,
Wooley \footnote{For a review and references to original work see C. Cohen-Tannoudji,
J. Dupont-Roc, and G. Grynberg, \textit{Photons and Atoms. Introduction
to Quantum Electrodynamics}, John Wiley and Sons, Inc., 1989.}, and Healy \cite{Healy1982} established a framework in which the
interaction of a molecule with the electromagnetic field, fundamentally
described by the minimal coupling Hamiltonian, could be written involving
microscopic polarization and magnetization fields associated with
each molecule and the microscopic electromagnetic field in the neighborhood
of that molecule. An expansion of the electromagnetic field about
a point in the molecule then leads to a Hamiltonian involving the
multipole moments of the molecule. For fluid systems, this was combined
with the definition of the macroscopic polarization and magnetization
as ensemble averages of the densities of multipoles and their derivatives
\cite{deGroot}, establishing the basis of the modern theory of the
optical properties of fluids.

The approach is based on the identification of molecules as stable
units, with charges moving within units but not moving from unit to
unit. Sometimes this can be a good approximation even for solids,
as in the treatment of molecular crystals. Yet at a fundamental level
it appears problematic: Valence electrons in a periodic solid are
typically associated with Bloch waves extending throughout the crystal,
and as the electron motion is perturbed by an electromagnetic field
it is not clear how a multipole expansion about any point would make
sense. Of course, one can always work directly with the minimal coupling
Hamiltonian and simply calculate the charge and current densities
induced by any applied field, bypassing a description in terms of
polarization and magnetization fields. But the physical insight that
such a description would provide is lacking, and when the basis states
used in calculations are truncated, as they inevitably must be, the
use of the electromagnetic potentials in a perturbation calculation
can lead to unphysical divergences due to the violation of certain
sum rules \cite{Cardona,Ghahramani}. A strategy developed by Adams,
Blount, and others \cite{Blount} in the 1960s relied on the introduction
of a macroscopic polarization associated with the position operator,
but that operator is poorly defined in a periodic solid and the calculations
must be treated with care \cite{Aversa,RestaX}.

A different strategy is that of the ``modern theory of polarization''
\cite{Resta1,Resta}, which has focussed on gapped systems where physical
intuition suggests that to lowest order none of the response to slowly
varying fields should be identified with ``free'' currents. Then,
under the application of a uniform applied electric field, a change
in polarization is associated with the induced current density $\boldsymbol{J}$
via the equation $d\boldsymbol{P}/dt\equiv\boldsymbol{J}.$ Since
$\boldsymbol{J}$ can be calculated for an adiabatically applied uniform
electric field, at least the change in $\boldsymbol{P}$ can be identified.
Magnetizations, $\boldsymbol{M}$, associated with unperturbed systems
can also be identified \cite{Mdef1,Mdef2}, and for topologically
trivial insulators the ground state polarization and magnetization
are associated with the electric and magnetic dipole moments of the
filled Wannier functions, respectively. For the polarization, however,
there is a ``quantum of ambiguity'' that arises and is related to
the way one associates Wannier functions with lattice sites; this
is related to the gauge choice made in the definition of the Bloch
eigenstates. There are also subtleties arising in the definition of
the magnetization that are associated with the itinerant motion of
charges between lattice sites \cite{Mdef1}. Recently the effects
of electron-electron interactions on these quantities have been addressed
\cite{Shi,Nouraflkan1,Nouraflkan2,Aryasetiawan}. Yet treatments following
this strategy have generally been aimed only at describing the ground
state, or the response to uniform, adiabatically varying applied fields
\footnote{For a treatment of the magnetic response, see, e.g., Andrei Malashevich,
Ivo Souza, Sinisa Coh, and David Vanderbilt, New Journal of Physics
\textbf{12}, 053032 (2010).}.

A more general approach would be to seek \emph{microscopic} polarization
and magnetization fields $\boldsymbol{p}(\boldsymbol{x},t)$ and $\boldsymbol{m}(\boldsymbol{x},t)$,
and \emph{microscopic} ``free'' charge and current densities $\rho_{F}(\boldsymbol{x},t)$
and $\boldsymbol{j}_{F}(\boldsymbol{x},t)$, such that the expectation
value of the microscopic charge and current density operators would
be given by 
\begin{align}
\left\langle \hat{\rho}(\boldsymbol{x},t)\right\rangle  & =-\boldsymbol{\nabla}\boldsymbol{\cdot}\boldsymbol{p}(\boldsymbol{x},t)+\rho_{F}(\boldsymbol{x},t),\label{eq:micro}\\
\left\langle \boldsymbol{\hat{j}}(\boldsymbol{x},t)\right\rangle  & =\frac{\partial\boldsymbol{p}(\boldsymbol{x},t)}{\partial t}+c\boldsymbol{\nabla\times m}(\boldsymbol{x},t)+\boldsymbol{j}_{F}(\boldsymbol{x},t).\nonumber 
\end{align}
For a specified $\left\langle \hat{\rho}(\boldsymbol{x},t)\right\rangle $
and $\langle\boldsymbol{\hat{j}}(\boldsymbol{x},t)\rangle$, a set
of such quantities $\big(\boldsymbol{p}(\boldsymbol{x},t),\boldsymbol{m}(\boldsymbol{x},t),\rho_{F}(\boldsymbol{x},t),\boldsymbol{j}_{F}(\boldsymbol{x},t)\big)$
required only to satisfy (\ref{eq:micro}) is far from unique \footnote{As a trivial example, consider $(\boldsymbol{p}(\boldsymbol{x},t),\boldsymbol{m}(\boldsymbol{x},t),$
$\rho_{F}(\boldsymbol{x},t),\boldsymbol{j}_{F}(\boldsymbol{x},t))=(-\boldsymbol{e}(\boldsymbol{x},t)/4\pi,\boldsymbol{b}(\boldsymbol{x},t)/4\pi,0,0)$,
where $\boldsymbol{e}(\boldsymbol{x},t)$ and $\boldsymbol{b}(\boldsymbol{x},t)$
are the total microscopic electric and magnetic fields, including
contributions from the charge and current densities themselves. Then Maxwell's equations
guarantee that (\ref{eq:micro}) are satisfied. Indeed, one could
even just choose $\boldsymbol{e}(\boldsymbol{x},t)$ and $\boldsymbol{b}(\boldsymbol{x},t)$
to be just the electrostatic and magnetostatic fields generated by
the charge current distribution. See \cite{Sipe83}.}. Yet to have clear physical significance we could seek polarization
and magnetization fields that consist of contributions associated
with different sites, labeled by $\boldsymbol{R}$, 
\begin{align}
\boldsymbol{p}(\boldsymbol{x},t) & =\sum_{\boldsymbol{R}}\boldsymbol{p}_{\boldsymbol{R}}(\boldsymbol{x},t),\label{eq:p_and_m}\\
\boldsymbol{m}(\boldsymbol{x},t) & =\sum_{\boldsymbol{R}}\boldsymbol{m}_{\boldsymbol{R}}(\boldsymbol{x},t),\nonumber 
\end{align}
with $\boldsymbol{p}_{\boldsymbol{R}}(\boldsymbol{x},t)$ and $\boldsymbol{m}_{\boldsymbol{R}}(\boldsymbol{x},t)$ explicitly related to the microscopic charge and current densities in the neighborhood of the site $\boldsymbol{R}$. 

Although there is no unique choice for the site(s), there may be a natural one. For an isolated atom, if the nucleus is taken to be fixed in space, the position of the nucleus can be taken to identify the single site $\boldsymbol{R}$. For a crystal with a single atom per unit cell, if we neglect the motion of the ion cores, we can take the sites to be the positions of the ions, such that the $\{\boldsymbol{R}\}$ form a Bravais lattice. For a crystal with multiple atoms per unit cell, there may not be such a natural choice for the sites. But we can always choose there to be one site in each unit cell such that the $\{\boldsymbol{R}\}$ again form a Bravais lattice. The positions of the ions composing the crystal lattice are then described by a Bravais lattice, consisting of the sites $\{\boldsymbol{R}\}$, with a basis; in this case, there will necessarily be positive charge located away from each site. We refer to a collection of sites that form a Bravais lattice as the ``lattice sites;'' dealing with the periodic structures that host such lattice sites is the primary focus of this paper. The total charge and current densities will be unaffected by the choice of the lattice sites, although the unit cell quantities $\boldsymbol{p}_{\boldsymbol{R}}(\boldsymbol{x},t)$ and $\boldsymbol{m}_{\boldsymbol{R}}(\boldsymbol{x},t)$ will depend on that choice. In most of the paper we focus only on the charge and current densities associated with the valence electrons. In the Conclusion we indicate how the charge density of the ions in the unit cell can be included in our unit cell quantities, in the limit that the ions are considered fixed; in a future publication we plan to generalize to include the motion of the ions.

The macroscopic fields corresponding to $\boldsymbol{p}(\boldsymbol{x},t)$ and $\boldsymbol{m}(\boldsymbol{x},t)$ would be obtained by taking spatial
averages of these microscopic fields, and a moment expansion of the
$\boldsymbol{p}_{\boldsymbol{R}}(\boldsymbol{x},t)$ and $\boldsymbol{m}_{\boldsymbol{R}}(\boldsymbol{x},t)$
about their lattice sites would generate a moment expansion in the
resulting macroscopic fields $\boldsymbol{P}(\boldsymbol{x},t)$ and
$\boldsymbol{M}(\boldsymbol{x},t)$. Such an approach could describe
the effects of variations of the electromagnetic field over the unit
cell on the optical response, and could treat the optical response
at arbitrary frequency; it would not be restricted to treating excitation
by adiabatically applied uniform optical fields. The microscopic polarization
and magnetization fields, which could vary significantly over the
unit cell, would give a physical picture of the effects of optical
excitation.

This is the approach we develop in this paper. Overall, the framework
of the dynamical equations that arise provides the microscopic underpinning
of a lattice gauge theory, involving matrices that include labels
for basis functions associated with each lattice site. The site charges
and link currents are identified with the free charge and current
densities, respectively, and the polarization and magnetization fields
(\ref{eq:p_and_m}) arise from the matrices that are associated with
the lattice sites. And since we find that the free charge and current
densities themselves satisfy continuity by construction, regardless
of the approximations made in the calculation of $\boldsymbol{p}(\boldsymbol{x},t)$
and\textbf{ $\boldsymbol{m}(\boldsymbol{x},t)$}, the values of $\left\langle \hat{\rho}(\boldsymbol{x},t)\right\rangle $
and $\langle\boldsymbol{\hat{j}}(\boldsymbol{x},t)\rangle$ determined
from (\ref{eq:micro}) necessarily satisfy continuity. Therefore,
in this approach charge conservation is completely robust against
any approximations.

In constructing the description of the dynamics of the fields $\boldsymbol{p}(\boldsymbol{x},t),$
$\boldsymbol{m}(\boldsymbol{x},t)$, $\rho_{F}(\boldsymbol{x},t)$,
and $\boldsymbol{j}_{F}(\boldsymbol{x},t)$ we are naturally led to
describe the effects of the electromagnetic field by the electric
and magnetic fields themselves, rather than the scalar and vector
potentials. Thus, calculations following this strategy should be free
of the kind of unphysical divergences that can plague the use of the
minimal coupling Hamiltonian \cite{Cardona,Ghahramani}. The transformation
from an interaction involving the scalar and vector potentials to
one involving the electric and magnetic fields arises because we borrow
some of our strategy from the theory of the optical response of molecules
\cite{Healybook}. Yet there are important differences between that
approach and ours. Because charges can move from site to site, we
do not attempt to construct a Hamiltonian involving operators for
$\boldsymbol{p}_{\boldsymbol{R}}(\boldsymbol{x},t)$, and $\boldsymbol{m}_{\boldsymbol{R}}(\boldsymbol{x},t)$,
as is usually done for atoms and molecules, since there is no reasonable
protocol by which charges that would be associated with each site
could be identified \cite{Sipe83}. Rather, we construct expressions
for $\boldsymbol{p}_{\boldsymbol{R}}(\boldsymbol{x},t)$ and $\boldsymbol{m}_{\boldsymbol{R}}(\boldsymbol{x},t)$
in terms of electron Green functions and their expansions in terms
of localized basis functions associated with each site $\boldsymbol{R}$.
In treating a large molecule these localized orbitals could be convenient
orthonormal molecular orbitals \cite{Mayer}, but for the problem
of a periodic solid that is the focus of this paper we choose the
maximally-localized Wannier functions that can be constructed from
each set of bands that is topologically trivial \cite{Brouder,Marzari,Panati,Troyer}; 
in general we make no assumption about the initial
occupation of any of these bands. In the process of constructing
$\boldsymbol{p}_{\boldsymbol{R}}(\boldsymbol{x},t)$ and $\boldsymbol{m}_{\boldsymbol{R}}(\boldsymbol{x},t)$
we can confirm that they are indeed related to the microscopic charge
and current densities near $\boldsymbol{R}$. Thus, for lattice constants
much less than the wavelength of light, a multipole expansion about
each lattice site is justified.

In this first paper we consider an incident classical electromagnetic
field and neglect interactions between electrons except as could be
captured in a simple mean field treatment. While the
formalism does not require it, the restriction to a single-particle
Hamiltonian eases our initial formulation, for it is sufficient to
consider only the lesser, equal time single-particle 
Green function. We omit the spin contribution to the magnetization;
it could be easily included, and does not affect the kind of issues
that arise here. We also assume the ions are fixed, but do allow for
a general dependence of the unperturbed Hamiltonian on position and
momentum. Our goal here is to present the basic formalism, and so
we will only be able to allude to some of the physical points made
above, and some made below. We plan to return to many of the issues
raised here in future publications, and to present a treatment of
electron-electron interactions that will rely on a more general Green
function framework.

After deriving the basic equations for an arbitrary applied electromagnetic
field here, however, we consider four important limits that can be
reached from the general results; these limits serve as a form of validation for the formalism we develop, in addition to being helpful examples of its use. We first confirm that for isolated atoms on a lattice our result reduces to what would be expected from
the usual treatment of atoms, to all orders in the multipole moments.
We then consider the limit that is often of interest in optical response,
where the applied electric field is approximated as uniform and the
applied magnetic field as vanishing; this is the so-called ``long-wavelength
limit''. We show how the current density in that limit consists of
the time derivative of the polarization and a free current density.
For a topologically nontrivial insulator, the linear response of the
free current density to an applied field includes the current perpendicular
to the applied field described by a transverse conductivity, as expected
\cite{TIs,3DQAH}.

For the other two limits under investigation we restrict ourselves
to topologically trivial insulators. In the first we show that the
usual expressions for the bulk polarization and magnetization from
the ``modern theory'' are reproduced in our treatment of the ground
state. We then show that in the linear response to an electromagnetic
field of arbitrary wavelength there is no free current or change in
the free charge density induced; the first order response is completely
described by the polarization and magnetization. This is what one
would physically expect of the kind of approach we develop here; it
is only to higher order, when injected electrons and holes can be
driven by the electromagnetic field, that one would expect the appearance
of free charges and currents. We note that this would hold even for
excitation by x-rays. In future publications we will extend these
investigations to systems that are topologically nontrivial, as well
as consider the description of linear and nonlinear response of a
range of materials.

In Section \ref{sectionII} we present the derivation, in Section
\ref{sectionIII} we present the limits mentioned above, and in Section
\ref{sectionIV} we conclude. Many details of the derivation have
been relegated to the Appendices.

\section{Microscopic polarization and magnetization fields}

\label{sectionII} We work in the Heisenberg picture with the lesser,
equal time single-particle Green function 
\begin{align}
 & G_{mc}(\boldsymbol{x},\boldsymbol{y};t)=i\left\langle \psi^{\dagger}(\boldsymbol{y},t)\psi(\boldsymbol{x},t)\right\rangle ,\label{eq:Gmc_def}
\end{align}
where $\psi(\boldsymbol{x},t)$ is the fermionic electron
field operator, and the subscript $mc$ denotes the usual minimal
coupling procedure to include the effect of an external, classical
electromagnetic field specified by a scalar potential $\phi(\boldsymbol{x},t)$
and a vector potential $\boldsymbol{A}(\boldsymbol{x},t)$; interactions
between the electrons, except as they might be included within the
use of a self-consistent electromagnetic field, are neglected. The
Green function satisfies the dynamical equation 
\begin{align}
 & i\hbar\frac{\partial G_{mc}(\boldsymbol{x},\boldsymbol{y};t)}{\partial t}=\mathcal{K}_{mc}(\boldsymbol{x},\boldsymbol{y};t)G_{mc}(\boldsymbol{x},\boldsymbol{y};t),\label{eq:mc_dynamical}
\end{align}
where 
\begin{align}
 & \mathcal{K}_{mc}(\boldsymbol{x},\boldsymbol{y};t)=\mathcal{H}_{mc}(\boldsymbol{x},t)-\mathcal{H}_{mc}^{*}(\boldsymbol{y},t),\label{eq:K_script}
\end{align}
with the ``script'' fonts (such as $\mathcal{K}_{mc}$ and $\mathcal{H}_{mc}$)
denoting differential operators acting on the functions that follow,
at and in the neighborhood of the spatial variable(s) indicated. We
have 
\begin{align}
\mathcal{H}_{mc}(\boldsymbol{x},t)=H_{0}\big(\boldsymbol{x},\mathfrak{p}_{mc}(\boldsymbol{x},t)\big)+e\phi(\boldsymbol{x},t),
\end{align}
where
\begin{align}
\mathfrak{p}_{mc}(\boldsymbol{x},t)=\mathfrak{p}(\boldsymbol{x})-\frac{e}{c}\boldsymbol{A}(\boldsymbol{x},t),\label{eq:pmc_def}
\end{align}
with 
\begin{align}
\mathfrak{p}(\boldsymbol{x})=\frac{\hbar}{i}\frac{\partial}{\partial\boldsymbol{x}}-\frac{e}{c}\boldsymbol{A}_{static}(\boldsymbol{x}).\label{eq:pS}
\end{align}
In (\ref{eq:pS}) we have included the possibility of the presence
of a static, periodic magnetic field described by a vector potential
$\boldsymbol{A}_{static}(\boldsymbol{x})$, where $\boldsymbol{A}_{static}(\boldsymbol{x})=\boldsymbol{A}_{static}(\boldsymbol{x}+\boldsymbol{R})$
for any lattice site position $\boldsymbol{R}$; note that we distinguish
such a static, periodic vector potential from the ``external'' electromagnetic
field associated with $\phi(\boldsymbol{x},t)$ and $\boldsymbol{A}(\boldsymbol{x},t)$.
The use of ``normal'' fonts (such as $H_{0}$) denotes a function
of the quantities indicated as variables, some of which may be differential
operators. For the usual Schrödinger Hamiltonian, for example, in
the absence of an external electromagnetic field we have 
\begin{align}
 & H_{0}^{Sch}\big(\boldsymbol{x},\mathfrak{p}(\boldsymbol{x})\big)=\frac{\big(\mathfrak{p}(\boldsymbol{x})\big)^{2}}{2m}+V(\boldsymbol{x}),\label{eq:Hsch}
\end{align}
but more general forms can be considered. In a system that is periodic
before the application of an external electromagnetic field, we have
$H_{0}\big(\boldsymbol{x}+\boldsymbol{R},\mathfrak{p}(\boldsymbol{x})\big)=H_{0}\big(\boldsymbol{x},\mathfrak{p}(\boldsymbol{x})\big)$,
but for much of the following this assumption is not necessary; a
large part of the formalism developed here could be used to introduce
microscopic polarization and magnetization fields to describe the
optical response of large molecules. The dynamical equation (\ref{eq:mc_dynamical})
was obtained from the simpler equation governing the dynamics of the
Green function in the absence of an external electromagnetic field
by the usual minimal coupling prescription, which in our notation
is written 
\begin{align*}
 & H_{0}\big(\boldsymbol{x},\mathfrak{p}(\boldsymbol{x})\big)\rightarrow H_{0}\big(\boldsymbol{x},\mathfrak{p}_{mc}(\boldsymbol{x},t)\big)+e\phi(\boldsymbol{x},t).
\end{align*}
The expectation value of the electronic charge density operator, $\hat{\rho}(\boldsymbol{x},t)$,
and the electronic current density operator, $\boldsymbol{\hat{j}}(\boldsymbol{x},t)$,
are given by 
\begin{align}
\left\langle \hat{\rho}(\boldsymbol{x},t)\right\rangle  & =-ie\Big[G_{mc}(\boldsymbol{x},\boldsymbol{y};t)\Big]_{\boldsymbol{y}\rightarrow\boldsymbol{x}},\label{eq:charge_current_mc}\\
\left\langle \boldsymbol{\hat{j}}(\boldsymbol{x},t)\right\rangle  & =\Big[\mathcal{J}_{mc}(\boldsymbol{x},\boldsymbol{y};t)G_{mc}(\boldsymbol{x},\boldsymbol{y};t)\Big]_{\boldsymbol{y}\rightarrow\boldsymbol{x}},\nonumber 
\end{align}
where 
\begin{align*}
 & \mathcal{J}_{mc}(\boldsymbol{x},\boldsymbol{y};t)=-\frac{i}{2}\Big[\boldsymbol{J}\big(\boldsymbol{x},\mathfrak{p}_{mc}(\boldsymbol{x},t)\big)+\boldsymbol{J}\big(\boldsymbol{y},\mathfrak{p}_{mc}^{*}(\boldsymbol{y},t)\big)\Big],
\end{align*}
and the function $\boldsymbol{J}\big(\boldsymbol{x},\mathfrak{p}_{mc}(\boldsymbol{x},t)\big)$
follows from $H_{0}\big(\boldsymbol{x},\mathfrak{p}_{mc}(\boldsymbol{x},t)\big)$
in the usual fashion \footnote{See, e.g., Michael E. Peskin and Daniel V. Schroeder, \textit{An Introduction
to quantum field theory}, Addison-Wesley, Reading, USA, 1995}. For the Schrödinger Hamiltonian, for example, we have 
\begin{align*}
 & \boldsymbol{J}\big(\boldsymbol{x},\mathfrak{p}_{mc}(\boldsymbol{x},t)\big)=\frac{e}{m}\mathfrak{p}_{mc}(\boldsymbol{x},t).
\end{align*}

Many of the expression above involve the scalar and vector potentials
explicitly. It is possible to replace that set of expressions by a
corresponding set of expressions that involve only the electric and
magnetic fields associated with these potentials, 
\begin{align*}
\boldsymbol{E}(\boldsymbol{x},t) & =-\boldsymbol{\nabla}\phi(\boldsymbol{x},t)-\frac{1}{c}\frac{\partial\boldsymbol{A}(\boldsymbol{x},t)}{\partial t},\\
\boldsymbol{B}(\boldsymbol{x},t) & =\boldsymbol{\nabla\times A}(\boldsymbol{x},t).
\end{align*}
The strategy for doing this was introduced many years ago for atoms
and molecules where a ``special point'', such as the centre-of-mass
of the atom or molecule, or the position of a nucleus assumed fixed,
is identified \cite{Healybook}. We implement an analogous strategy
in the paragraph below without identifying such a special point, but
still employing quantities used in problems in atomic and molecular
physics. Those quantities, or ``relators'', are defined \cite{Sylvia1,Sylvia2}
as 
\begin{align}
s^{i}(\boldsymbol{w};\boldsymbol{x},\boldsymbol{y}) & =\int_{C(\boldsymbol{x},\boldsymbol{y})}dz^{i}\delta(\boldsymbol{w}-\boldsymbol{z}),\label{eq:relator_definitions}\\
\alpha^{jk}(\boldsymbol{w};\boldsymbol{x},\boldsymbol{y}) & =\epsilon^{jmn}\int_{C(\boldsymbol{x},\boldsymbol{y})}dz^{m}\frac{\partial z^{n}}{\partial x^{k}}\delta(\boldsymbol{w}-\boldsymbol{z}),\nonumber \\
\beta^{jk}(\boldsymbol{w};\boldsymbol{x},\boldsymbol{y}) & =\epsilon^{jmn}\int_{C(\boldsymbol{x},\boldsymbol{y})}dz^{m}\frac{\partial z^{n}}{\partial y^{k}}\delta(\boldsymbol{w}-\boldsymbol{z}),\nonumber 
\end{align}
where for each two points $\boldsymbol{x}$ and $\boldsymbol{y}$
we use $C(\boldsymbol{x},\boldsymbol{y})$ to indicate some path from
\textbf{$\boldsymbol{y}$} to \textbf{$\boldsymbol{x}$}. Here and
below superscripts indicate Cartesian components, $\epsilon^{jmn}$
is the Levi-Civita symbol, and repeated Cartesian components are summed
over \footnote{As the metric for spatial components is identity, covariant and contravariant
objects transform trivially into one another.}. Regardless of the path chosen, the quantities (\ref{eq:relator_definitions})
satisfy 
\begin{align}
 & \frac{\partial s^{i}(\boldsymbol{w};\boldsymbol{x},\boldsymbol{y})}{\partial w^{i}}=-\delta(\boldsymbol{w}-\boldsymbol{x})+\delta(\boldsymbol{w}-\boldsymbol{y}),\label{eq:relator_equations}\\
 & \frac{\partial s^{i}(\boldsymbol{w};\boldsymbol{x},\boldsymbol{y})}{\partial x^{k}}=\delta^{ik}\delta(\boldsymbol{w}-\boldsymbol{x})-\epsilon^{ipj}\frac{\partial\alpha^{jk}(\boldsymbol{w};\boldsymbol{x},\boldsymbol{y})}{\partial w^{p}},\nonumber \\
 & \frac{\partial s^{i}(\boldsymbol{w};\boldsymbol{x},\boldsymbol{y})}{\partial y^{k}}=-\delta^{ik}\delta(\boldsymbol{w}-\boldsymbol{y})-\epsilon^{ipj}\frac{\partial\beta^{jk}(\boldsymbol{w};\boldsymbol{x},\boldsymbol{y})}{\partial w^{p}},\nonumber 
\end{align}
(See Appendix A). It is useful to choose a ``symmetric'' set of
paths, by which we mean that for each and every $\boldsymbol{x}$
and $\boldsymbol{y}$ the path $C(\boldsymbol{x},\boldsymbol{y})$
is the ``reverse'' of the path $C(\boldsymbol{y},\boldsymbol{x}).$
For a symmetric set of paths, we call the resulting sets of relators
``symmetric'', and they satisfy 
\begin{align}
s^{i}(\boldsymbol{w};\boldsymbol{x},\boldsymbol{y})+s^{i}(\boldsymbol{w};\boldsymbol{y},\boldsymbol{x}) & =0,\label{eq:symmetric}\\
\alpha^{jk}(\boldsymbol{w};\boldsymbol{x},\boldsymbol{y})+\beta^{jk}(\boldsymbol{w};\boldsymbol{y},\boldsymbol{x}) & =0.\nonumber 
\end{align}
(See Appendix A). The second of these means that $\beta^{jk}(\boldsymbol{w};\boldsymbol{y},\boldsymbol{x})$
can be eliminated in favor of $\alpha^{jk}(\boldsymbol{w};\boldsymbol{x},\boldsymbol{y})$,
which we generally do in the formulas below.

\subsection{The global Green function}

We now use these relators to introduce a new Green function, $G_{gl}(\boldsymbol{x},\boldsymbol{y};t)$
\footnote{This extends earlier work \cite{Levanda, Sylvia1, Sylvia2}}.
The subscript $gl$ denotes ``global'', in that no special point
is introduced, and yet the dynamical equation for $G_{gl}(\boldsymbol{x},\boldsymbol{y};t)$
and the expressions for the electronic charge and current densities in terms
of it are gauge invariant in that they depend directly on the electric
and magnetic fields, and not explicitly on the scalar and vector potentials.
The gauge freedom in choosing the scalar and vector potentials for
a given electromagnetic field has in a sense been replaced by a similar
freedom in choosing the paths $C(\boldsymbol{x},\boldsymbol{y}),$
or in choosing sets of relators satisfying (\ref{eq:relator_equations})
that are even more general than (\ref{eq:relator_definitions}). While
straight lines between $\boldsymbol{y}$ and $\boldsymbol{x}$ are
almost always chosen for the paths $C(\boldsymbol{x},\boldsymbol{y})$
in applications (see Appendix A), the freedom to choose different
sets of relators is still there. To avoid confusion, we use the term
``gauge invariant'' here to refer specifically to quantities that
do not depend on the scalar and vector potentials explicitly, but
only on the electric and magnetic fields.

To obtain the global Green function, $G_{gl}(\boldsymbol{x},\boldsymbol{y};t)$,
we introduce 
\begin{align}
 & \Phi(\boldsymbol{x},\boldsymbol{y};t)\equiv\frac{e}{\hbar c}\int s^{i}(\boldsymbol{w};\boldsymbol{x},\boldsymbol{y})A^{i}(\boldsymbol{w},t)d\boldsymbol{w},\label{eq:PHI_def}
\end{align}
for a general path -- note that if a straight-line path is chosen
this is just the standard Peierls phase -- and put 
\begin{align}
 & G_{gl}(\boldsymbol{x},\boldsymbol{y};t)=e^{-i\Phi(\boldsymbol{x},\boldsymbol{y};t)}G_{mc}(\boldsymbol{x},\boldsymbol{y};t).\label{eq:Ggl_def}
\end{align}
As we always use symmetric relators we obtain 
\begin{align*}
 & G_{gl}^{*}(\boldsymbol{x},\boldsymbol{y};t)=-G_{gl}(\boldsymbol{y},\boldsymbol{x};t),
\end{align*}
following from the fact that $G_{mc}(\boldsymbol{x},\boldsymbol{y};t)$
trivially satisfies this relation. While each factor on the right-hand-side
of (\ref{eq:Ggl_def}) is gauge dependent, their product is not. For
the dynamical equation for $G_{gl}(\boldsymbol{x},\boldsymbol{y};t)$
is found to be 
\begin{align}
 & i\hbar\frac{\partial G_{gl}(\boldsymbol{x},\boldsymbol{y};t)}{\partial t}=\mathcal{K}_{gl}(\boldsymbol{x},\boldsymbol{y};t)G_{gl}(\boldsymbol{x},\boldsymbol{y};t),
\end{align}
where 
\begin{align}
\mathcal{K}_{gl}(\boldsymbol{x},\boldsymbol{y};t) & =H_{0}\big(\boldsymbol{x},\mathfrak{p}(\boldsymbol{x},\boldsymbol{y};t)\big)-H_{0}^{*}\big(\boldsymbol{y},\mathfrak{p}(\boldsymbol{y},\boldsymbol{x};t)\big)\nonumber \\
 & -e\Omega_{\boldsymbol{y}}^{0}(\boldsymbol{x},t),
\end{align}
with 
\begin{align}
 & \mathfrak{p}^{k}(\boldsymbol{x},\boldsymbol{y};t)\equiv\mathfrak{p}^{k}(\boldsymbol{x})-\frac{e}{c}\Omega_{\boldsymbol{y}}^{k}(\boldsymbol{x},t),\label{eq:pref_def}
\end{align}
and 
\begin{align}
\Omega_{\boldsymbol{y}}^{0}(\boldsymbol{x},t) & \equiv\int s^{i}(\boldsymbol{w};\boldsymbol{x},\boldsymbol{y})E^{i}(\boldsymbol{w},t)d\boldsymbol{w},\label{eq:Omega_defs}\\
\Omega_{\boldsymbol{y}}^{k}(\boldsymbol{x},t) & \equiv\int\alpha^{lk}(\boldsymbol{w};\boldsymbol{x},\boldsymbol{y})B^{l}(\boldsymbol{w},t)d\boldsymbol{w}.\nonumber 
\end{align}
Now using (\ref{eq:charge_current_mc},\ref{eq:Ggl_def}), we can
write the expectation value of the electronic charge and current densities as 
\begin{align}
\left\langle \hat{\rho}(\boldsymbol{x},t)\right\rangle  & =-ie\Big[G_{gl}(\boldsymbol{x},\boldsymbol{y};t)\Big]_{\boldsymbol{y}\rightarrow\boldsymbol{x}},\label{eq:charge_current_gl}\\
\left\langle \boldsymbol{\hat{j}}(\boldsymbol{x},t)\right\rangle  & =\Big[\mathcal{J}_{gl}(\boldsymbol{x},\boldsymbol{y};t)G_{gl}(\boldsymbol{x},\boldsymbol{y};t)\Big]_{\boldsymbol{y}\rightarrow\boldsymbol{x}},\nonumber 
\end{align}
where 
\begin{align*}
\mathcal{J}_{gl}(\boldsymbol{x},\boldsymbol{y};t)=-\frac{i}{2}\Big[ \boldsymbol{J}\big(\boldsymbol{x},\mathfrak{p}(\boldsymbol{x},\boldsymbol{y};t)\big)+\boldsymbol{J}^{*}\big(\boldsymbol{y},\mathfrak{p}(\boldsymbol{y},\boldsymbol{x};t)\big)\Big].%\label{eq:current_{t}erm_{g}lobal}
\end{align*}
Each these quantities is clearly gauge invariant.

\subsection{Wannier functions and adjusted Wannier functions}

While the global gauge invariant Green function $G_{gl}(\boldsymbol{x},\boldsymbol{y};t)$
is interesting in its own right, our goal here is to use it to associate
charge and current densities with each lattice site, and then associate
microscopic polarizations and magnetizations with those sites in a
way similar, as much as possible, with what one would do to treat
a model consisting of ``isolated atoms'' where charges could not
move from atom to atom. To do this, we introduce a set of localized
basis functions $\{W_{\alpha\boldsymbol{R}}(\boldsymbol{x})\},$ or
``orbitals,'' labeled by the lattice site $\boldsymbol{R}$ and
a ``type index'' $\alpha$. We take the functions in the set $\left\{ W_{\alpha\boldsymbol{R}}(\boldsymbol{x})\right\} $
to be orthogonal, 
\begin{align}
 & \int W_{\beta\boldsymbol{R'}}^{*}(\boldsymbol{x})W_{\alpha\boldsymbol{R}}(\boldsymbol{x})d\boldsymbol{x}=\delta_{\beta\alpha}\delta_{\boldsymbol{R'}\boldsymbol{R}}.\label{eq:Worthogonality}
\end{align}
In the special case of a periodic system, which is our main focus
here, we have 
\begin{align}
 & W_{\alpha\boldsymbol{R}}(\boldsymbol{x})=W_{\alpha}(\boldsymbol{x}-\boldsymbol{R}),\label{eq:Tinv}
\end{align}
where we identify $W_{\alpha}(\boldsymbol{x})$ as a function localized
near the origin -- in general the functions will be centered at different points for different $\alpha$ -- and $\left\{ W_{\alpha\boldsymbol{R}}(\boldsymbol{x})\right\}$ is a set of Wannier functions. We make a particular choice for these
Wannier functions by requiring that they are maximally-localized functions;
then each $N$ element subset of the Wannier functions considered
will be constructed from a subset of $N$ bands that do not intersect
in energy with elements of other subsets of bands, and where the subset
of bands used is, as a whole, topologically trivial \footnote{In 2D this condition implies the net Chern number associated with
the bands in the subset used in constructing Wannier functions is
zero.} \cite{Brouder,Marzari,Panati,Troyer}; at the moment we make no assumption
about the initial occupation of any of these bands.

In the presence of a \textit{uniform} vector potential, $\boldsymbol{A}(\boldsymbol{x},t)\rightarrow\boldsymbol{A}(t)$,
the set $\left\{ W'_{\alpha\boldsymbol{R}}(\boldsymbol{x},t)\right\} $
of modified orbitals 
\begin{align}
 & W'_{\alpha\boldsymbol{R}}(\boldsymbol{x},t)=e^{i\Phi(\boldsymbol{x},\boldsymbol{R};t)}W_{\alpha\boldsymbol{R}}(\boldsymbol{x})\label{eq:Wprime_def}
\end{align}
are orthonormal and often useful in calculations. Yet for \textit{nonuniform}
vector potentials the functions in the set $\left\{ W'_{\alpha\boldsymbol{R}}(\boldsymbol{x})\right\} $
are not orthonormal, nor even are their overlap integrals gauge invariant.
However, Löwdin's method of symmetric orthogonalization \cite{Mayer}
can be used to construct an orthonormal set of functions $\left\{ \bar{W}{}_{\alpha\boldsymbol{R}}(\boldsymbol{x},t)\right\} $,
\begin{align}
 & \int\bar{W}_{\beta\boldsymbol{R'}}^{*}(\boldsymbol{x},t)\bar{W}_{\alpha\boldsymbol{R}}(\boldsymbol{x},t)d\boldsymbol{x}=\delta_{\beta\alpha}\delta_{\boldsymbol{R'}\boldsymbol{R}},\label{eq:Wbar_orthonormality}
\end{align}
which are as close as possible to the functions $\left\{ W'_{\alpha\boldsymbol{R}}(\boldsymbol{x},t)\right\} $,
in the sense that 
\begin{align}
 & \sum_{\alpha,\boldsymbol{R}}\int\Big|\bar{W}_{\alpha\boldsymbol{R}}(\boldsymbol{x},t)-W'_{\alpha\boldsymbol{R}}(\boldsymbol{x},t)\Big|^{2}d\boldsymbol{x}\label{eq:TBminimized}
\end{align}
is a minimum at each time $t$. We refer to these new functions as
``adjusted Wannier functions.'' For a finite system the set $\left\{ \bar{W}{}_{\alpha\boldsymbol{R}}(\boldsymbol{x},t)\right\} $
can be found numerically if the sum over $\alpha$ is truncated. In
any case, it always follows that the $\left\{ \bar{W}_{\alpha\boldsymbol{R}}(\boldsymbol{x},t)\right\} $
are of the form 
\begin{align}
 & \bar{W}_{\alpha\boldsymbol{R}}(\boldsymbol{x},t)=e^{i\Phi(\boldsymbol{x},\boldsymbol{R};t)}\chi_{\alpha\boldsymbol{R}}(\boldsymbol{x},t),\label{eq:chi_introduce}
\end{align}
where the functions in the set $\left\{ \chi_{\alpha\boldsymbol{R}}(\boldsymbol{x},t)\right\} $
are generally \textit{not} orthonormal, but can be written in a gauge
invariant way, and in fact depend only on the magnetic field and not
on the electric field (see Appendix B). In the limit of a weak applied
magnetic field one can construct a perturbative expansion for $\chi_{\alpha\boldsymbol{R}}(\boldsymbol{x},t)$;
the first two terms are 
\begin{align}
\chi_{\alpha\boldsymbol{R}}(\boldsymbol{x},t)&=W_{\alpha\boldsymbol{R}}(\boldsymbol{x})-\frac{1}{2}i\sum_{\beta,\boldsymbol{R'}}W_{\beta\boldsymbol{R'}}(\boldsymbol{x})\label{eq:chi_expand}\\
 &\times\left[\int W_{\beta\boldsymbol{R'}}^{*}(\boldsymbol{y})\Delta(\boldsymbol{R'},\boldsymbol{y},\boldsymbol{R};t)W_{\alpha\boldsymbol{R}}(\boldsymbol{y})d\boldsymbol{y}\right]+...\nonumber 
\end{align}
(see Appendix B), where very generally 
\begin{align}
 & \Delta(\boldsymbol{x},\boldsymbol{y},\boldsymbol{z};t)\equiv\Phi(\boldsymbol{z},\boldsymbol{x};t)+\Phi(\boldsymbol{x},\boldsymbol{y};t)+\Phi(\boldsymbol{y},\boldsymbol{z};t)\label{eq:Delta_def}
\end{align}
is a gauge invariant quantity, since it involves only the flux of
the magnetic field through the surface identified by $\boldsymbol{x},$
$\boldsymbol{y}$, $\boldsymbol{z}$ and the paths connecting them.

Neglecting the spin degree of freedom, which could be included in
a straight-forward way, we expand the fermionic electron field operator,
\begin{align}
 & \psi(\boldsymbol{x},t)=\sum_{\alpha,\boldsymbol{R}}a_{\alpha\boldsymbol{R}}(t)\bar{W}_{\alpha\boldsymbol{R}}(\boldsymbol{x},t),\label{eq:field_expand}
\end{align}
where 
\begin{align}
\left\{ a_{\alpha\boldsymbol{R}}(t),a_{\beta\boldsymbol{R'}}(t)\right\}  & =0,\\
\Big\{ a_{\alpha\boldsymbol{R}}(t),a_{\beta\boldsymbol{R'}}^{\dagger}(t)\Big\}  & =\delta_{\alpha\beta}\delta_{\boldsymbol{R}\boldsymbol{R'}}.\nonumber 
\end{align}
Formally, of course we can take the number of type indices to be infinite,
so we have a complete set of basis functions in the expansion (\ref{eq:field_expand}).
However, we will derive expressions for the microscopic charge and
current density involving intermediate quantities such that charge
is explicitly conserved, and so in evaluating those intermediate quantities
it will be possible to truncate the basis without violating continuity.
In this basis the Green function $G_{mc}(\boldsymbol{x},\boldsymbol{y};t)$
takes the form 
\begin{align}
 & G_{mc}(\boldsymbol{x},\boldsymbol{y};t)=i\sum_{\alpha,\beta,\boldsymbol{R},\boldsymbol{R'}}\breve{\eta}_{\alpha\boldsymbol{R};\beta\boldsymbol{R'}}(t)\bar{W}_{\beta\boldsymbol{R'}}^{*}(\boldsymbol{y},t)\bar{W}_{\alpha\boldsymbol{R}}(\boldsymbol{x},t),\label{eq:Gmc_decompose}
\end{align}
where 
\begin{align}
 & \breve{\eta}_{\alpha\boldsymbol{R};\beta\boldsymbol{R'}}(t)=\Big\langle a_{\beta\boldsymbol{R'}}^{\dagger}(t)a_{\alpha\boldsymbol{R}}(t)\Big\rangle .\label{eq:sigma_up}
\end{align}
From the dynamical equation (\ref{eq:mc_dynamical}) for $G_{mc}(\boldsymbol{x},\boldsymbol{y};t)$
we immediately find the equations of motion for the $\breve{\eta}_{\alpha\boldsymbol{R};\beta\boldsymbol{R'}}(t)$,
\begin{align}
 & i\hbar\frac{\partial\breve{\eta}_{\alpha\boldsymbol{R};\beta\boldsymbol{R'}}(t)}{\partial t}=\label{eq:gen_lattice_gauge}\\
 &\qquad\sum_{\lambda,\boldsymbol{R''}}\left(e^{i\Phi(\boldsymbol{R},\boldsymbol{R}'';t)}{\displaystyle \bar{H}_{\alpha\boldsymbol{R};\lambda\boldsymbol{R''}}(t)}\breve{\eta}_{\lambda\boldsymbol{R''};\beta\boldsymbol{R'}}(t)\right.\nonumber \\
 & \quad\qquad\qquad\left.-\breve{\eta}_{\alpha\boldsymbol{R};\lambda\boldsymbol{R''}}(t)\bar{H}_{\lambda\boldsymbol{R''};\beta\boldsymbol{R'}}(t)e^{i\Phi(\boldsymbol{R}'',\boldsymbol{R}';t)}\right)\nonumber \\
 & \qquad-e\Omega_{\boldsymbol{R'}}^{\phi}(\boldsymbol{R},t)\breve{\eta}_{\alpha\boldsymbol{R};\beta\boldsymbol{R'}}(t),\nonumber 
\end{align}
where 
\begin{align*}
 & \Omega_{\boldsymbol{y}}^{\phi}(\boldsymbol{x},t)\equiv-\int s^{i}(\boldsymbol{z};\boldsymbol{x},\boldsymbol{y})\frac{\partial\phi(\boldsymbol{z},t)}{\partial z^{i}}d\boldsymbol{z},
\end{align*}
and the matrix $\bar{H}$ is Hermitian, $\bar{H}_{\alpha\boldsymbol{R};\lambda\boldsymbol{R''}}^{*}(t)=\bar{H}_{\lambda\boldsymbol{R''};\alpha\boldsymbol{R}}(t)$,
with elements that are gauge invariant, 
\begin{widetext}
\begin{align}
\bar{H}_{\alpha\boldsymbol{R};\lambda\boldsymbol{R''}}(t) & =\frac{1}{2}\int\chi_{\alpha\boldsymbol{R}}^{*}(\boldsymbol{x},t)e^{i\Delta(\boldsymbol{R},\boldsymbol{x},\boldsymbol{R}'';t)}H_{0}\big(\boldsymbol{x},\mathfrak{p}(\boldsymbol{x},\boldsymbol{R''};t)\big)\chi_{\lambda\boldsymbol{R''}}(\boldsymbol{x},t)d\boldsymbol{x}\label{eq:Hmatrix_def}\\
 & +\frac{1}{2}\int\Big(H_{0}\big(\boldsymbol{x},\mathfrak{p}(\boldsymbol{x},\boldsymbol{R};t)\big)\chi_{\alpha\boldsymbol{R}}(\boldsymbol{x},t)\Big)^{*}\chi_{\lambda\boldsymbol{R''}}(\boldsymbol{x},t)e^{i\Delta(\boldsymbol{R},\boldsymbol{x},\boldsymbol{R''};t)}d\boldsymbol{x}\nonumber \\
 & -\frac{e}{2}\int e^{i\Delta(\boldsymbol{R},\boldsymbol{x},\boldsymbol{R''};t)}\chi_{\alpha\boldsymbol{R}}^{*}(\boldsymbol{x},t)\Big(\Omega_{\boldsymbol{R''}}^{0}(\boldsymbol{x},t)+\Omega_{\boldsymbol{R}}^{0}(\boldsymbol{x},t)\Big)\chi_{\lambda\boldsymbol{R''}}(\boldsymbol{x},t)d\boldsymbol{x}\nonumber \\
 & -\frac{1}{2}i\hbar\int e^{i\Delta(\boldsymbol{R},\boldsymbol{x},\boldsymbol{R''};t)}\left(\chi_{\alpha\boldsymbol{R}}^{*}(\boldsymbol{x},t)\frac{\partial\chi_{\lambda\boldsymbol{R''}}(\boldsymbol{x},t)}{\partial t}-\chi_{\lambda\boldsymbol{R''}}(\boldsymbol{x},t)\frac{\partial\chi_{\alpha\boldsymbol{R}}^{*}(\boldsymbol{x},t)}{\partial t}\right)d\boldsymbol{x}.\nonumber 
\end{align}
\end{widetext}

The equations (\ref{eq:gen_lattice_gauge},\ref{eq:Hmatrix_def})
can be understood as the microscopic underpinnings of the equations
of a lattice gauge theory. These expressions display five main features: (a) the functions $\Phi(\boldsymbol{R},\boldsymbol{R''};t)$ and $\Omega_{\boldsymbol{R''}}^{\phi}(\boldsymbol{R},t)$
allow for an arbitrary variation of the scalar and vector potential
as one moves between neighboring lattice sites; (b) the ``hopping'' matrix
elements $\bar{H}_{\alpha\boldsymbol{R};\lambda\boldsymbol{R''}}(t)$
involve a set of states (labeled by Greek letters) at each site, and
thus are matrices even for fixed $\boldsymbol{R}$ and $\boldsymbol{R}''$;
(c) in principle the hopping matrix elements connect each site to
every other site, although in practice any site will only be coupled
to sites close to it; (d) the matrix elements are also more complicated
than in tight-binding models, containing the electric and magnetic
fields, and are therefore time dependent; note, however, that they
are gauge invariant; (e) since the complete electronic charge and current densities
can be constructed once the $\breve{\eta}_{\alpha\boldsymbol{R};\beta\boldsymbol{R'}}(t)$
are found and $G_{mc}(\boldsymbol{x},\boldsymbol{y};t)$ identified
from (\ref{eq:Gmc_decompose}), the solution of (\ref{eq:gen_lattice_gauge})
in fact allows for a determination of the full densities within the
model identified by our original orbitals, and not just site charges
and link currents. Indeed, with the use of the matrices $\bar{H}_{\alpha\boldsymbol{R};\lambda\boldsymbol{R''}}(t)$
we will construct the microscopic polarization and magnetization fields
associated with each lattice site, and the site charges and link currents
identified from the lattice gauge perspective will be associated with
the microscopic version of the ``free charges'' and ``free currents''
of elementary electrodynamics.

\subsection{Density operator dynamics}

It is convenient to move to a fully gauge-invariant description by
defining a single-particle density matrix 
\begin{align}
 & \eta_{\alpha\boldsymbol{R};\beta\boldsymbol{R'}}(t)\equiv\breve{\eta}_{\alpha\boldsymbol{R};\beta\boldsymbol{R'}}(t)e^{i\Phi(\boldsymbol{R}',\boldsymbol{R};t)},\label{eq:toGI}
\end{align}
from which it follows that the dynamics of $\eta_{\alpha\boldsymbol{R};\beta\boldsymbol{R'}}(t)$
are specified by 
\begin{align}
 & i\hbar\frac{\partial\eta_{\alpha\boldsymbol{R};\beta\boldsymbol{R'}}(t)}{\partial t}=\label{eq:density_operator_dynamical_result}\\
 & \quad\sum_{\lambda,\boldsymbol{R''}}e^{i\Delta(\boldsymbol{R},\boldsymbol{R''},\boldsymbol{R'};t)}\Big(\bar{H}_{\alpha\boldsymbol{R};\lambda\boldsymbol{R''}}(t)\eta_{\lambda\boldsymbol{R''};\beta\boldsymbol{R'}}(t)\nonumber \\
 & \quad-\eta_{\alpha\boldsymbol{R};\lambda\boldsymbol{R''}}(t)\bar{H}_{\lambda\boldsymbol{R''};\beta\boldsymbol{R'}}(t)\Big)-e\Omega_{\boldsymbol{R'}}^{0}(\boldsymbol{R};t)\eta_{\alpha\boldsymbol{R};\beta\boldsymbol{R'}}(t).\nonumber 
\end{align}
Although the form of (\ref{eq:density_operator_dynamical_result})
is a bit unusual because of the factor $\text{exp}(i\Delta(\boldsymbol{R},\boldsymbol{R''},\boldsymbol{R'};t)),$
note that all terms appearing in this equation for $\eta_{\alpha\boldsymbol{R};\beta\boldsymbol{R'}}(t)$
are gauge invariant. Also, since any initial $\eta_{\alpha\boldsymbol{R};\beta\boldsymbol{R'}}(t)$
before any fields are applied is trivially gauge invariant, then $\eta_{\alpha\boldsymbol{R};\beta\boldsymbol{R'}}(t)$
itself will be gauge-invariant even as the fields are introduced.
We can write $G_{gl}(\boldsymbol{x},\boldsymbol{y};t)$ in terms of
the $\eta_{\alpha\boldsymbol{R};\beta\boldsymbol{R'}}(t)$ via (\ref{eq:Ggl_def},\ref{eq:Gmc_decompose},\ref{eq:toGI}),
and from the orthonormality relations (\ref{eq:Wbar_orthonormality})
we can write the $\eta_{\alpha\boldsymbol{R};\beta\boldsymbol{R'}}(t)$
in terms of $G_{gl}(\boldsymbol{x},\boldsymbol{y};t);$ writing these
results in terms of the $\chi_{\alpha\boldsymbol{R}}(\boldsymbol{x},t)$
of (\ref{eq:chi_introduce}) we have 
\begin{align}
G_{gl}(\boldsymbol{x},\boldsymbol{y};t) & =i\sum_{\alpha,\beta,\boldsymbol{R},\boldsymbol{R'}}\eta_{\alpha\boldsymbol{R};\beta\boldsymbol{R'}}(t)e^{i\Delta(\boldsymbol{x},\boldsymbol{R},\boldsymbol{R'},\boldsymbol{y};t)}\label{eq:Ggl_and_sigma}\\
 & \qquad\qquad\quad\times\chi_{\beta\boldsymbol{R'}}^{*}(\boldsymbol{y},t)\chi_{\alpha\boldsymbol{R}}(\boldsymbol{x},t),\nonumber \\
\eta_{\alpha\boldsymbol{R};\beta\boldsymbol{R'}}(t) & =-i\int\chi_{\beta\boldsymbol{R'}}(\boldsymbol{y},t)\chi_{\alpha\boldsymbol{R}}^{*}(\boldsymbol{x},t)e^{-i\Delta(\boldsymbol{x},\boldsymbol{R},\boldsymbol{R'},\boldsymbol{y};t)}\nonumber \\
 & \qquad\qquad\quad\times G_{gl}(\boldsymbol{x},\boldsymbol{y};t)d\boldsymbol{x}d\boldsymbol{y},\nonumber 
\end{align}
where in general 
\begin{align*}
\Delta(\boldsymbol{x},\boldsymbol{u},\boldsymbol{v},\boldsymbol{y};t) & =\Phi(\boldsymbol{y},\boldsymbol{x};t)+\Phi(\boldsymbol{x},\boldsymbol{u};t)\\
 & +\Phi(\boldsymbol{u},\boldsymbol{v};t)+\Phi(\boldsymbol{v},\boldsymbol{y};t)
\end{align*}
is gauge invariant. While the first equation of (\ref{eq:Ggl_and_sigma})
involves a double sum over lattice sites, we now introduce a Green
function $G_{\boldsymbol{R}}(\boldsymbol{x},\boldsymbol{y};t)$ associated
with each lattice site $\boldsymbol{R},$ with the goal of writing
$G_{gl}(\boldsymbol{x},\boldsymbol{y};t)$ involving a sum over these.
To guarantee that each $G_{\boldsymbol{R}}(\boldsymbol{x},\boldsymbol{y};t)$
satisfies the property 
\begin{align*}
 & G_{\boldsymbol{R}}^{*}(\boldsymbol{x},\boldsymbol{y};t)=-G_{\boldsymbol{R}}(\boldsymbol{y},\boldsymbol{x};t),
\end{align*}
as do $G_{mc}(\boldsymbol{x},\boldsymbol{y};t)$ and $G_{gl}(\boldsymbol{x},\boldsymbol{y};t)$
(see (\ref{eq:Gmc_def},\ref{eq:Ggl_def})), we take 
\begin{align}
 & G_{\boldsymbol{R}}(\boldsymbol{x},\boldsymbol{y};t)=\label{eq:GR_def}\\
 & \qquad\frac{i}{2}\sum_{\alpha,\beta,\boldsymbol{R'}}\eta_{\alpha\boldsymbol{R};\beta\boldsymbol{R'}}(t)e^{i\Delta(\boldsymbol{R'},\boldsymbol{y},\boldsymbol{R};t)}\chi_{\beta\boldsymbol{R'}}^{*}(\boldsymbol{y},t)\chi_{\alpha\boldsymbol{R}}(\boldsymbol{x},t)\nonumber \\
 & \qquad+\frac{i}{2}\sum_{\alpha,\beta,\boldsymbol{R'}}\eta_{\beta\boldsymbol{R'};\alpha\boldsymbol{R}}(t)e^{i\Delta(\boldsymbol{R},\boldsymbol{x},\boldsymbol{R'};t)}\chi_{\alpha\boldsymbol{R}}^{*}(\boldsymbol{y},t)\chi_{\beta\boldsymbol{R'}}(\boldsymbol{x},t)\nonumber 
\end{align}
and then have 
\begin{align}
 & G_{gl}(\boldsymbol{x},\boldsymbol{y};t)=\sum_{\boldsymbol{R}}e^{-i\Delta(\boldsymbol{x},\boldsymbol{y},\boldsymbol{R};t)}G_{\boldsymbol{R}}(\boldsymbol{x},\boldsymbol{y};t).\label{eq:Ggl_decomp}
\end{align}
The quantities (\ref{eq:GR_def}) are chosen so that the gauge invariant
phase factors $\text{exp}(-i\Delta(\boldsymbol{x},\boldsymbol{y},\boldsymbol{R};t))$
appear in the decomposition (\ref{eq:Ggl_decomp}) for reasons that
will later become clear \footnote{This is a further development of a strategy introduced earlier \cite{Sylvia2}.}.

We use the $G_{\boldsymbol{R}}(\boldsymbol{x},\boldsymbol{y};t)$
to identify components of the full electronic charge and current densities that
we associate with each lattice site, $\rho_{\boldsymbol{R}}(\boldsymbol{x},t)$
and $\boldsymbol{j}_{\boldsymbol{R}}(\boldsymbol{x},t).$ From (\ref{eq:charge_current_gl},\ref{eq:Ggl_decomp})
we have 
\begin{align}
\left\langle \hat{\rho}(\boldsymbol{x},t)\right\rangle  & =\sum_{\boldsymbol{R}}\rho_{\boldsymbol{R}}(\boldsymbol{x},t),\label{eq:site_decompose}\\
\left\langle \boldsymbol{\hat{j}}(\boldsymbol{x},t)\right\rangle  & =\sum_{\boldsymbol{R}}\boldsymbol{j}_{\boldsymbol{R}}(\boldsymbol{x},t),\nonumber 
\end{align}
where 
\begin{align}
\rho_{\boldsymbol{R}}(\boldsymbol{x},t) & =-ie\Big[G_{\boldsymbol{R}}(\boldsymbol{x},\boldsymbol{y};t)\Big]_{\boldsymbol{y\rightarrow}\boldsymbol{x}}\label{eq:charge_current_site}\\
\boldsymbol{j}_{\boldsymbol{R}}(\boldsymbol{x},t) & =\Big[\mathcal{J}_{\boldsymbol{R}}(\boldsymbol{x},\boldsymbol{y};t)G_{\boldsymbol{R}}(\boldsymbol{x},\boldsymbol{y};t)\Big]_{\boldsymbol{y}\rightarrow\boldsymbol{x}},\nonumber 
\end{align}
with 
\begin{align}
\mathcal{J}_{\boldsymbol{R}}(\boldsymbol{x},\boldsymbol{y};t)=-\frac{i}{2}\Big( & \boldsymbol{J}\big(\boldsymbol{x},\mathfrak{p}(\boldsymbol{x},\boldsymbol{R};t)\big)+\boldsymbol{J}^{*}\big(\boldsymbol{y},\mathfrak{p}(\boldsymbol{y},\boldsymbol{R};t)\big)\Big).\label{eq:current_term_site}
\end{align}
In particular, we obtain 
\begin{align}
\rho_{\boldsymbol{R}}(\boldsymbol{x},t) & =\sum_{\alpha,\beta,\boldsymbol{R'},\boldsymbol{R}''}\rho_{\beta\boldsymbol{R}';\alpha\boldsymbol{R}''}(\boldsymbol{x},\boldsymbol{R};t)\eta_{\alpha\boldsymbol{R}'';\beta\boldsymbol{R'}}(t),\label{eq:rho_and_j_decomp}\\
\boldsymbol{j}_{\boldsymbol{R}}(\boldsymbol{x},t) & =\sum_{\alpha,\beta,\boldsymbol{R}',\boldsymbol{R}''}\boldsymbol{j}_{\beta\boldsymbol{R}';\alpha\boldsymbol{R}''}(\boldsymbol{x},\boldsymbol{R};t)\eta_{\alpha\boldsymbol{R''};\beta\boldsymbol{R'}}(t),\nonumber 
\end{align}
where 
\begin{align}
\rho_{\beta\boldsymbol{R'};\alpha\boldsymbol{R''}}(\boldsymbol{x},\boldsymbol{R};t) & =\frac{e}{2}\Big(\delta_{\boldsymbol{R}\boldsymbol{R'}}+\delta_{\boldsymbol{R}\boldsymbol{R''}}\Big)e^{i\Delta(\boldsymbol{R'},\boldsymbol{x},\boldsymbol{R''};t)}\nonumber \\
 & \qquad\times\chi_{\beta\boldsymbol{R'}}^{*}(\boldsymbol{x},t)\chi_{\alpha\boldsymbol{R''}}(\boldsymbol{x},t)\label{eq:rho_decomp_term}
\end{align}
and 
\begin{widetext}
\begin{align}
\boldsymbol{j}_{\beta\boldsymbol{R'};\alpha\boldsymbol{R''}}(\boldsymbol{x},\boldsymbol{R};t) & =\frac{1}{4}\delta_{\boldsymbol{R}\boldsymbol{R''}}e^{i\Delta(\boldsymbol{R'},\boldsymbol{x},\boldsymbol{R}'';t)}\chi_{\beta\boldsymbol{R'}}^{*}(\boldsymbol{x},t)\Big(\boldsymbol{J}\big(\boldsymbol{x},\mathfrak{p}(\boldsymbol{x},\boldsymbol{R};t)\big)\chi_{\alpha\boldsymbol{R}''}(\boldsymbol{x},t)\Big)\label{eq:j_decomp_term}\\
 & +\frac{1}{4}\delta_{\boldsymbol{R}\boldsymbol{R'}}\Big(\boldsymbol{J}^{*}\big(\boldsymbol{x},\mathfrak{p}(\boldsymbol{x},\boldsymbol{R};t)\big)\chi_{\beta\boldsymbol{R}'}^{*}(\boldsymbol{x},t)\Big)e^{i\Delta(\boldsymbol{R}',\boldsymbol{x},\boldsymbol{R''};t)}\chi_{\alpha\boldsymbol{R''}}(\boldsymbol{x},t)\nonumber \\
 & +\frac{1}{4}\delta_{\boldsymbol{R}\boldsymbol{R''}}\left(\boldsymbol{J}^{*}\big(\boldsymbol{x},\mathfrak{p}(\boldsymbol{x},\boldsymbol{R};t)\big)e^{i\Delta(\boldsymbol{R'},\boldsymbol{x},\boldsymbol{R}'';t)}\chi_{\beta\boldsymbol{R'}}^{*}(\boldsymbol{x},t)\right)\chi_{\alpha\boldsymbol{R}''}(\boldsymbol{x},t)\nonumber \\
 & +\frac{1}{4}\delta_{\boldsymbol{R}\boldsymbol{R'}}\chi_{\beta\boldsymbol{R}'}^{*}(\boldsymbol{x},t)\left(\boldsymbol{J}\big(\boldsymbol{x},\mathfrak{p}(\boldsymbol{x},\boldsymbol{R};t)\big)e^{i\Delta(\boldsymbol{R}',\boldsymbol{x},\boldsymbol{R''};t)}\chi_{\alpha\boldsymbol{R''}}(\boldsymbol{x},t)\right).\nonumber 
\end{align}
\end{widetext}
Note that in the expression (\ref{eq:GR_def}) for $G_{\boldsymbol{R}}(\boldsymbol{x},\boldsymbol{y};t)$
the lattice site $\boldsymbol{R}$ always appears as one of the indices
of the single-particle density matrix. Then, since $\chi_{\alpha\boldsymbol{R}}(\boldsymbol{x},t)$
can be reasonably expected to be nonzero only for $\boldsymbol{x}$
close to $\boldsymbol{R}$ (see the expansion (\ref{eq:chi_expand})),
we can expect the $\rho_{\boldsymbol{R}}(\boldsymbol{x},t)$ and $\boldsymbol{j}_{\boldsymbol{R}}(\boldsymbol{x},t)$
of (\ref{eq:charge_current_site}) to be nonzero only for $\boldsymbol{x}$
close to $\boldsymbol{R}$ as well.

\subsection{Site polarizations and magnetizations}

The equations (\ref{eq:charge_current_site},\ref{eq:current_term_site})
are precisely the ones that would be written down in a model for isolated
atoms, where there $G_{\boldsymbol{R}}(\boldsymbol{x},\boldsymbol{y};t)$
would be the Green function for the atom at $\boldsymbol{R}$. The
``special point'' $\boldsymbol{R}$ has been identified for the
charge-current distribution about lattice site $\boldsymbol{R}$,
and appears in the expression for the current density in precisely
the way that a ``special point'' is identified in the treatment
of an atom; it appears in (\ref{eq:charge_current_site}) in that
way because the phase factors $\text{exp}(-i\Delta(\boldsymbol{x},\boldsymbol{y},\boldsymbol{R};t))$
were introduced in (\ref{eq:Ggl_decomp}). A difference between our
problem and that of isolated atoms, of course, is that in general
\begin{align}
 & K_{\boldsymbol{R}}(\boldsymbol{x},t)\equiv\boldsymbol{\nabla}\boldsymbol{\cdot}\boldsymbol{j}_{\boldsymbol{R}}(\boldsymbol{x},t)+\frac{\partial\rho_{\boldsymbol{R}}(\boldsymbol{x},t)}{\partial t}\neq0,\label{eq:KRterm}
\end{align}
since electrons can move from the region nearest one lattice site
to regions nearest others. Thus generally the site charges 
\begin{align}
 & Q_{\boldsymbol{R}}(t)\equiv\int\rho_{\boldsymbol{R}}(\boldsymbol{x},t)d\boldsymbol{x}=e\sum_{\alpha}\eta_{\alpha\boldsymbol{R};\alpha\boldsymbol{R}}(t)\label{eq:QR_def}
\end{align}
are time dependent. From the dynamical equation (\ref{eq:density_operator_dynamical_result})
we find that we can write 
\begin{align}
 & \frac{dQ_{\boldsymbol{R}}(t)}{dt}=\sum_{\boldsymbol{R'}}I(\boldsymbol{R},\boldsymbol{R}';t),\label{eq:site_charge_dynamical}
\end{align}
where 
\begin{align}
I(\boldsymbol{R},\boldsymbol{R'};t)=\frac{e}{i\hbar}\sum_{\alpha,\lambda}\Big( & \bar{H}_{\alpha\boldsymbol{R};\lambda\boldsymbol{R'}}(t)\eta_{\lambda\boldsymbol{R'};\alpha\boldsymbol{R}}(t)\nonumber \\
 & -\eta_{\alpha\boldsymbol{R};\lambda\boldsymbol{R'}}(t)\bar{H}_{\lambda\boldsymbol{R'};\alpha\boldsymbol{R}}(t)\Big).\label{eq:I_identify}
\end{align}
Since 
\begin{align}
 & I(\boldsymbol{R}',\boldsymbol{R};t)=-I(\boldsymbol{R},\boldsymbol{R}';t),\label{eq:back_flow}
\end{align}
$I(\boldsymbol{R},\boldsymbol{R}';t)$ can be interpreted as the net
current flowing from site $\boldsymbol{R}'$ to site $\boldsymbol{R}$,
and thus as a link current.

With the site charges and link currents identified, we can now define
microscopic ``free'' charge and current densities associated with
them, 
\begin{align}
\rho_{F}(\boldsymbol{x},t) & \equiv\sum_{\boldsymbol{R}}Q_{\boldsymbol{R}}(t)\delta(\boldsymbol{x}-\boldsymbol{R}),\label{eq:rhoF_def}\\
\boldsymbol{j}_{F}(\boldsymbol{x},t) & \equiv\frac{1}{2}\sum_{\boldsymbol{R},\boldsymbol{R'}}\boldsymbol{s}(\boldsymbol{x};\boldsymbol{R},\boldsymbol{R}')I(\boldsymbol{R},\boldsymbol{R}';t).\label{eq:jF_def}
\end{align}
The first simply takes the free charge density to be the sum of the
charges associated with each lattice site placed at that lattice site.
The second introduces a microscopic current density by distributing
the net current from site $\boldsymbol{R}'$ to $\boldsymbol{R}$
along the path from $\boldsymbol{R}'$ to $\boldsymbol{R}$ defined
by $C(\boldsymbol{R},\boldsymbol{R}').$ From (\ref{eq:site_charge_dynamical},\ref{eq:back_flow})
and the second of the properties (\ref{eq:relator_equations}) we
immediately find 
\begin{align}
 & \boldsymbol{\nabla}\boldsymbol{\cdot}\boldsymbol{j}_{F}(\boldsymbol{x},t)+\frac{\partial\rho_{F}(\boldsymbol{x},t)}{\partial t}=0.\label{eq:free_continuity}
\end{align}
That is, the microscopic free charge and current densities themselves
satisfy continuity. We shall write the remaining contributions to
the total charge and current densities in terms of microscopic polarization
and magnetization fields associated with each site. We begin by introducing
these terms as we would were we dealing with isolated atoms. We define
site polarization fields $\boldsymbol{p}_{\boldsymbol{R}}(\boldsymbol{x},t)$
as 
\begin{align}
 & \boldsymbol{p}_{\boldsymbol{R}}(\boldsymbol{x},t)\equiv\int\boldsymbol{s}(\boldsymbol{x};\boldsymbol{y},\boldsymbol{R})\rho_{\boldsymbol{R}}(\boldsymbol{y},t)d\boldsymbol{y},\label{eq:site_polarization}
\end{align}
and define preliminary site magnetization fields as 
\begin{align}
 & \bar{m}_{\boldsymbol{R}}^{j}(\boldsymbol{x},t)\equiv\frac{1}{c}\int\alpha^{jk}(\boldsymbol{x};\boldsymbol{y},\boldsymbol{R})j_{\boldsymbol{R}}^{k}(\boldsymbol{y},t)d\boldsymbol{y}.\label{eq:site_magntization1}
\end{align}
Since $\rho_{\boldsymbol{R}}(\boldsymbol{x},t)$ and $\boldsymbol{j}_{\boldsymbol{R}}(\boldsymbol{x},t)$
are nonzero only for $\boldsymbol{x}$ close to $\boldsymbol{R},$
we can expect $\boldsymbol{p}_{\boldsymbol{R}}(\boldsymbol{x},t)$
and $\boldsymbol{\bar{m}}_{\boldsymbol{R}}(\boldsymbol{x},t)$ to
share that property as well. Introducing associated total microscopic
polarization and magnetization fields, 
\begin{align}
\boldsymbol{p}(\boldsymbol{x},t) & =\sum_{\boldsymbol{R}}\boldsymbol{p}_{\boldsymbol{R}}(\boldsymbol{x},t),\label{eq:first_pm_def}\\
\boldsymbol{\bar{m}}(\boldsymbol{x},t) & =\sum_{\boldsymbol{R}}\boldsymbol{\bar{m}}_{\boldsymbol{R}}(\boldsymbol{x},t),\nonumber 
\end{align}
with the use of (\ref{eq:relator_equations}) we obtain 
\begin{align}
\left\langle \hat{\rho}(\boldsymbol{x},t)\right\rangle  & =-\boldsymbol{\nabla}\boldsymbol{\cdot}\boldsymbol{p}(\boldsymbol{x},t)+\rho_{F}(\boldsymbol{x},t),\label{eq:expectation_write1}\\
\left\langle \boldsymbol{\hat{j}}(\boldsymbol{x},t)\right\rangle  & =\frac{\partial\boldsymbol{p}(\boldsymbol{x},t)}{\partial t}+c\boldsymbol{\nabla\times\bar{m}}(\boldsymbol{x},t)+\boldsymbol{\tilde{j}}(\boldsymbol{x},t)+\boldsymbol{j}_{F}(\boldsymbol{x},t),\nonumber 
\end{align}
where 
\begin{align}
 & \boldsymbol{\tilde{j}}(\boldsymbol{x},t)=-\sum_{\boldsymbol{R}}\int\boldsymbol{s}(\boldsymbol{x};\boldsymbol{y},\boldsymbol{R})K_{\boldsymbol{R}}(\boldsymbol{y},t)d\boldsymbol{y}-\boldsymbol{j}_{F}(\boldsymbol{x},t).\label{eq:jbar_def}
\end{align}
In contrast with the problem of isolated atoms, here in general we
have a time dependent $\rho_{F}(\boldsymbol{x},t)$, a nonzero $\boldsymbol{j}_{F}(\boldsymbol{x},t)$,
and a nonzero $\boldsymbol{\tilde{j}}(\boldsymbol{x},t),$ all arising
because the functions in $\{K_{\boldsymbol{R}}(\boldsymbol{x},t)\}$
are generally nonzero as charge moves from site to site. Nonetheless,
since total charge is conserved we have 
\begin{align*}
 & \sum_{\boldsymbol{R}}K_{\boldsymbol{R}}(\boldsymbol{x},t)=0,
\end{align*}
and it is easy to confirm that 
\begin{align*}
 & \boldsymbol{\nabla}\boldsymbol{\cdot}\boldsymbol{\tilde{j}}(\boldsymbol{x},t)=0.
\end{align*}

To complete our treatment of the site polarizations and magnetizations
we express the divergenceless $\boldsymbol{\tilde{j}}(\boldsymbol{x},t)$
in terms of the curls of magnetizations associated with the different
lattice sites. We begin by writing 
\begin{align*}
 & \boldsymbol{\tilde{j}}(\boldsymbol{x},t)=\sum_{\alpha,\beta,\boldsymbol{R''},\boldsymbol{R'''}}\boldsymbol{\tilde{j}}_{\beta\boldsymbol{R'''};\alpha\boldsymbol{R''}}(\boldsymbol{x},t)\eta_{\alpha\boldsymbol{R''};\beta\boldsymbol{R'''}}(t),
\end{align*}
where the expressions (\ref{eq:KRterm}) and (\ref{eq:jF_def}) for
$K_{\boldsymbol{R}}(\boldsymbol{y},t)$ and $\boldsymbol{j}_{F}(\boldsymbol{x},t)$
respectively are used to identify $\boldsymbol{\tilde{j}}_{\beta\boldsymbol{R'''};\alpha\boldsymbol{R''}}(\boldsymbol{x},t)$,
\begin{align}
\boldsymbol{\tilde{j}}_{\beta\boldsymbol{R'''};\alpha\boldsymbol{R''}}(\boldsymbol{x},t) & =-\sum_{\boldsymbol{R}}\int\boldsymbol{s}(\boldsymbol{x};\boldsymbol{y},\boldsymbol{R})\Gamma_{\boldsymbol{R}}^{\alpha\boldsymbol{R''};\beta\boldsymbol{R'''}}(\boldsymbol{y},t)d\boldsymbol{y}\nonumber \\
 & -\frac{1}{2}\sum_{\boldsymbol{R},\boldsymbol{R'}}\boldsymbol{s}(\boldsymbol{x};\boldsymbol{R},\boldsymbol{R'})\varsigma_{\boldsymbol{R}\boldsymbol{R'}}^{\alpha\boldsymbol{R''};\beta\boldsymbol{R'''}}(t),\label{eq:jtilde_work2}
\end{align}
where 
\begin{align}
\varsigma_{\boldsymbol{R}\boldsymbol{R'}}^{\alpha\boldsymbol{R''};\beta\boldsymbol{R'''}}(t) & =\frac{e}{i\hbar}\Big(\delta_{\boldsymbol{R}'''\boldsymbol{R}}\delta_{\boldsymbol{R''}\boldsymbol{R'}}\bar{H}_{\beta\boldsymbol{R};\alpha\boldsymbol{R'}}(t)\label{eq:eta_def} \\
 & \qquad\qquad-\delta_{\boldsymbol{R''}\boldsymbol{R}}\delta_{\boldsymbol{R'''}\boldsymbol{R'}}\bar{H}_{\beta\boldsymbol{R'};\alpha\boldsymbol{R}}(t)\Big),\nonumber
\end{align}
and 
\begin{align}
 & \Gamma_{\boldsymbol{R}}^{\alpha\boldsymbol{R''};\beta\boldsymbol{R'}}(\boldsymbol{x},t)=\label{eq:Gamma_def}\\
 & \qquad\boldsymbol{\nabla}\boldsymbol{\cdot}\boldsymbol{j}_{\beta\boldsymbol{R'};\alpha\boldsymbol{R''}}(\boldsymbol{x},\boldsymbol{R};t)+\frac{\partial\rho_{\beta\boldsymbol{R'};\alpha\boldsymbol{R''}}(\boldsymbol{x},\boldsymbol{R};t)}{\partial t}\nonumber \\
 & \qquad+\frac{1}{i\hbar}\sum_{\mu,\nu,\boldsymbol{R}_{1},\boldsymbol{R}_{2}}\rho_{\nu\boldsymbol{R}_{2};\mu\boldsymbol{R}_{1}}(\boldsymbol{x},\boldsymbol{R};t)\mathfrak{F}_{\mu\boldsymbol{R}_{1};\nu\boldsymbol{R}_{2}}^{\alpha\boldsymbol{R''};\beta\boldsymbol{R'}}(t),\nonumber 
\end{align}
with 
\begin{align}
\mathfrak{F}_{\mu\boldsymbol{R}_{1};\nu\boldsymbol{R}_{2}}^{\alpha\boldsymbol{R''};\beta\boldsymbol{R'}}(t) & =\delta_{\beta\nu}\delta_{\boldsymbol{R}_{2}\boldsymbol{R}'}e^{i\Delta(\boldsymbol{R}_{1},\boldsymbol{R}'',\boldsymbol{R}_{2};t)}\bar{H}_{\mu\boldsymbol{R}_{1};\alpha\boldsymbol{R}''}(t)\nonumber \\
 & -\delta_{\alpha\mu}\delta_{\boldsymbol{R}''\boldsymbol{R}_{1}}e^{i\Delta(\boldsymbol{R}_{1},\boldsymbol{R}',\boldsymbol{R}_{2};t)}\bar{H}_{\beta\boldsymbol{R}';\nu\boldsymbol{R}_{2}}(t)\nonumber \\
 & -e\delta_{\beta\nu}\delta_{\alpha\mu}\delta_{\boldsymbol{R}_{2}\boldsymbol{R}'}\delta_{\boldsymbol{R}_{1}\boldsymbol{R}''}\Omega^0_{\boldsymbol{R}_{2}}(\boldsymbol{R}_{1};t).\label{eq:Fscript}
\end{align}
We associate portions of $\boldsymbol{\tilde{j}}(\boldsymbol{x},t)$
with each lattice site in an obvious way, 
\begin{align*}
 & \boldsymbol{\tilde{j}}(\boldsymbol{x},t)=\sum_{\boldsymbol{R}}\boldsymbol{\tilde{j}}_{\boldsymbol{R}}(\boldsymbol{x},t),
\end{align*}
where 
\begin{align}
 & \boldsymbol{\tilde{j}}_{\boldsymbol{R}}(\boldsymbol{x},t)=\sum_{\alpha,\beta,\boldsymbol{R'},\boldsymbol{R''}}\boldsymbol{\tilde{j}}_{\beta\boldsymbol{R'};\alpha\boldsymbol{R''}}(\boldsymbol{x},\boldsymbol{R};t)\eta_{\alpha\boldsymbol{R''};\beta\boldsymbol{R'}}(t),\label{eq:jtilde_site}
\end{align}
with 
\begin{align}
 & \boldsymbol{\tilde{j}}_{\beta\boldsymbol{R'};\alpha\boldsymbol{R''}}(\boldsymbol{x},\boldsymbol{R};t)=\frac{1}{2}\big(\delta_{\boldsymbol{R}\boldsymbol{R''}}+\delta_{\boldsymbol{R}\boldsymbol{R'}}\big)\boldsymbol{\tilde{j}}_{\beta\boldsymbol{R'};\alpha\boldsymbol{R''}}(\boldsymbol{x},t).\label{eq:jtilde_site_decompose}
\end{align}
We then define 
\begin{align}
 & \tilde{m}_{\boldsymbol{R}}^{j}(\boldsymbol{x},t)\equiv\frac{1}{c}\int\alpha^{jk}(\boldsymbol{x};\boldsymbol{y},\boldsymbol{R})\tilde{j}_{\boldsymbol{R}}^{k}(\boldsymbol{y},t)d\boldsymbol{y},\label{eq:site_magnetization2}
\end{align}
which, following the arguments used above for $\boldsymbol{p}_{\boldsymbol{R}}(\boldsymbol{x},t)$
and $\boldsymbol{m}_{\boldsymbol{R}}(\boldsymbol{x},t)$, can be expected
to be nonzero only for $\boldsymbol{x}$ close to $\boldsymbol{R}$.
Using the relator properties (\ref{eq:relator_equations}), we obtain
\begin{align*}
 & \boldsymbol{\tilde{j}}_{\boldsymbol{R}}(\boldsymbol{x},t)=c\boldsymbol{\nabla\times\tilde{m}}_{\boldsymbol{R}}(\boldsymbol{x},t),
\end{align*}
as desired. We can now write 
\begin{align*}
 & \boldsymbol{m}_{\boldsymbol{R}}(\boldsymbol{x},t)\equiv\boldsymbol{\bar{m}}_{\boldsymbol{R}}(\boldsymbol{x},t)+\boldsymbol{\tilde{m}}_{\boldsymbol{R}}(\boldsymbol{x},t)
\end{align*}
and, with 
\begin{align}
 & \boldsymbol{m}(\boldsymbol{x},t)=\sum_{\boldsymbol{R}}\boldsymbol{m}_{\boldsymbol{R}}(\boldsymbol{x},t),\label{eq:second_m_def}
\end{align}
we can write the second of (\ref{eq:expectation_write1}) as 
\begin{align}
 & \left\langle \boldsymbol{\hat{j}}(\boldsymbol{x},t)\right\rangle =\frac{\partial\boldsymbol{p}(\boldsymbol{x},t)}{\partial t}+c\boldsymbol{\nabla\times m}(\boldsymbol{x},t)+\boldsymbol{j}_{F}(\boldsymbol{x},t).\label{eq:full_j}
\end{align}

\subsection{Summary}
\label{Summary}

We can now separate the ``free'' and ``bound'' charge and current
densities, writing the total expectation values of the microscopic
electronic charge and current density operators as 
\begin{align}
\left\langle \hat{\rho}(\boldsymbol{x},t)\right\rangle  & =\rho_{B}(\boldsymbol{x},t)+\rho_{F}(\boldsymbol{x},t),\label{eq:summary_equations}\\
\left\langle \boldsymbol{\hat{j}}(\boldsymbol{x},t)\right\rangle  & =\boldsymbol{j}_{B}(\boldsymbol{x},t)+\boldsymbol{j}_{F}(\boldsymbol{x},t).\nonumber 
\end{align}
The free charge density is given in terms of site charges $Q_{\boldsymbol{R}}(t)$
by (\ref{eq:rhoF_def}) and the free current density in terms of link
currents $I(\boldsymbol{R},\boldsymbol{R'};t)$ by (\ref{eq:jF_def});
the site charges evolve via the link currents according to (\ref{eq:site_charge_dynamical}),
and the link currents evolve according to (\ref{eq:I_identify}).
This guarantees that the free charge and current densities satisfy
continuity (\ref{eq:free_continuity}). The bound charge and current
densities are given by
\begin{align}
\rho_{B}(\boldsymbol{x},t) & =-\boldsymbol{\nabla}\boldsymbol{\cdot}\boldsymbol{p}(\boldsymbol{x},t),\label{eq:bound_charge_current}\\
\boldsymbol{j}_{B}(\boldsymbol{x},t) & =\frac{\partial\boldsymbol{p}(\boldsymbol{x},t)}{\partial t}+c\boldsymbol{\nabla\times m}(\boldsymbol{x},t),\nonumber 
\end{align}
which guarantee that the bound charge and current densities satisfy
continuity as well, 
\begin{align}
 & \boldsymbol{\nabla}\boldsymbol{\cdot}\boldsymbol{j}_{B}(\boldsymbol{x},t)+\frac{\partial\rho_{B}(\boldsymbol{x},t)}{\partial t}=0.\label{eq:bound_continuity}
\end{align}
The microscopic polarization and magnetization fields can be broken
up into site contributions, 
\begin{align}
\boldsymbol{p}(\boldsymbol{x},t) & =\sum_{\boldsymbol{R}}\boldsymbol{p}_{\boldsymbol{R}}(\boldsymbol{x},t),\label{eq:pm_decomp}\\
\boldsymbol{m}(\boldsymbol{x},t) & =\sum_{\boldsymbol{R}}\boldsymbol{m}_{\boldsymbol{R}}(\boldsymbol{x},t).\nonumber 
\end{align}
The site polarizations are given by (\ref{eq:site_polarization}),
or 
\begin{align}
\boldsymbol{p}_{\boldsymbol{R}}(\boldsymbol{x},t) & =\sum_{\alpha,\beta,\boldsymbol{R'},\boldsymbol{R}''}\left[\int\boldsymbol{s}(\boldsymbol{x};\boldsymbol{y},\boldsymbol{R})\rho_{\beta\boldsymbol{R}';\alpha\boldsymbol{R}''}(\boldsymbol{y},\boldsymbol{R};t)d\boldsymbol{y}\right]\nonumber \\
 & \quad\qquad\qquad\times\eta_{\alpha\boldsymbol{R}'';\beta\boldsymbol{R'}}(t),\label{eq:pRgeneral}
\end{align}
where $\rho_{\beta\boldsymbol{R}';\alpha\boldsymbol{R}''}(\boldsymbol{y},\boldsymbol{R};t)$
is given by (\ref{eq:rho_decomp_term}), and we have used the first
of (\ref{eq:rho_and_j_decomp}). The site magnetizations are given
by the sums of (\ref{eq:site_magntization1}) and (\ref{eq:site_magnetization2}),
or 
\begin{widetext}
\begin{align}
 & m_{\boldsymbol{R}}^{j}(\boldsymbol{x},t)=\frac{1}{c}\sum_{\alpha,\beta,\boldsymbol{R'},\boldsymbol{R''}}\left[\int\alpha^{jk}(\boldsymbol{x};\boldsymbol{y},\boldsymbol{R})\Big(j_{\beta\boldsymbol{R}';\alpha\boldsymbol{R}''}^{k}(\boldsymbol{y},\boldsymbol{R};t)+\tilde{j}{}_{\beta\boldsymbol{R'};\alpha\boldsymbol{R''}}^{k}(\boldsymbol{y},\boldsymbol{R};t)\Big)d\boldsymbol{y}\right]\eta_{\alpha\boldsymbol{R''};\beta\boldsymbol{R'}}(t),\label{eq:mRgeneral}
\end{align}
\end{widetext}
where $j_{\beta\boldsymbol{R}';\alpha\boldsymbol{R}''}^{k}(\boldsymbol{y},\boldsymbol{R};t)$
and $\tilde{j}{}_{\beta\boldsymbol{R'};\alpha\boldsymbol{R''}}^{k}(\boldsymbol{y},\boldsymbol{R};t)$
are given by (\ref{eq:j_decomp_term}) and (\ref{eq:jtilde_site_decompose})
respectively, and we have used the second of (\ref{eq:rho_and_j_decomp})
and (\ref{eq:jtilde_site}). Each of these quantities involve the
single particle density matrix $\eta_{\alpha\boldsymbol{R''};\beta\boldsymbol{R'}}(t),$
the dynamics of which is governed by (\ref{eq:density_operator_dynamical_result}).
We have omitted the spin degree of freedom and its contribution to
the magnetization, but that could be easily included.

We emphasize that as long as in any approximations made the quantities
$I(\boldsymbol{R},\boldsymbol{R'};t)$ that result still satisfy (\ref{eq:back_flow}),
and the evolution of the site charges is governed by (\ref{eq:site_charge_dynamical}),
the approximate free charge and current densities will satisfy continuity.
Moreover, for $any$ approximations made in evaluating $\left\{ \boldsymbol{p}_{\boldsymbol{R}}(\boldsymbol{x},t)\right\} $
and $\left\{ \boldsymbol{m}_{\boldsymbol{R}}(\boldsymbol{x},t)\right\} $
the bound charge and current densities (\ref{eq:bound_charge_current})
will satisfy continuity (\ref{eq:bound_continuity}). Thus in this
description full charge conservation at the microscopic level is extremely
robust against approximations.

\section{Some Limits of Interest}

\label{sectionIII} 

\subsection{The isolated atom limit}

We first consider the limit where the Wannier functions $W_{\alpha\boldsymbol{R}}(\boldsymbol{x})$
and $W_{\beta\boldsymbol{R'}}(\boldsymbol{x})$ are assumed to have
no common support if $\boldsymbol{R}\neq\boldsymbol{R'}$. Then we
would expect our treatment of the solid to reduce to that of isolated
atoms positioned at the lattice sites. That is, we would expect to
find 
\begin{align}
\left\langle \hat{\rho}(\boldsymbol{x},t)\right\rangle  & =\sum_{\boldsymbol{R}}\Big(-\boldsymbol{\nabla}\boldsymbol{\cdot}\boldsymbol{p}_{\boldsymbol{R}}(\boldsymbol{x},t)+Q_{\boldsymbol{R}}\delta(\boldsymbol{x}-\boldsymbol{R})\Big),\label{eq:isolated_limit0}\\
\left\langle \boldsymbol{\hat{j}}(\boldsymbol{x},t)\right\rangle  & =\sum_{\boldsymbol{R}}\left(\frac{\partial\boldsymbol{p}_{\boldsymbol{R}}(\boldsymbol{x},t)}{\partial t}+\boldsymbol{\nabla\times m}_{\boldsymbol{R}}(\boldsymbol{x},t)\right),\nonumber 
\end{align}
where $Q_{\boldsymbol{R}}$ is the fixed electronic charge associated
with lattice site $\boldsymbol{R}$, and $\boldsymbol{p}_{\boldsymbol{R}}(\boldsymbol{x},t)$
and $\boldsymbol{m}_{\boldsymbol{R}}(\boldsymbol{x},t)$ are the polarization
and magnetization expressions we would expect from isolated atoms
placed at the indicated lattice sites. In Appendix C we review the
usual results for an isolated atom, and in Appendix D we show that
indeed in the limit of no common support of Wannier functions at different
sites our expressions do reduce to (\ref{eq:isolated_limit0}). Note
that this holds even if the electromagnetic field varies strongly
over the extension of the Wannier functions.

\subsection{The long-wavelength limit}

We next consider our equations in the long-wavelength limit, taking
$\boldsymbol{E}(\boldsymbol{x},t)\rightarrow\boldsymbol{E}(t),$ and
restrict ourselves to an infinite, periodic crystal. The magnetic
field must then be time-independent, and we take any applied constant
magnetic field to vanish; microscopic magnetic fields that have the
periodicity of the lattice can be taken into account in the unperturbed
Hamiltonian, $H_{0}\big(\boldsymbol{x},\mathfrak{p}(\boldsymbol{x})\big)$,
using the expression (\ref{eq:pS}). This is a standard model often
used to calculate the optical response of crystals \cite{Aversa}.
For the paths $C(\boldsymbol{x},\boldsymbol{y})$ in (\ref{eq:relator_definitions})
we take straight lines.

With $\boldsymbol{B}(\boldsymbol{x},t)=0$ and $\boldsymbol{E}(t)$
uniform we have $\Delta(\boldsymbol{x},\boldsymbol{y},\boldsymbol{z};t)\rightarrow0$
from (\ref{eq:Delta_def}), $\chi_{\alpha\boldsymbol{R}}(\boldsymbol{x},t)\rightarrow W_{\alpha\boldsymbol{R}}(\boldsymbol{x})$
from (\ref{eq:chi_expand}) and the discussion preceding it, $\Omega_{\boldsymbol{x}}^{0}(\boldsymbol{y};t)\rightarrow(\boldsymbol{y}-\boldsymbol{x})\boldsymbol{\cdot}\boldsymbol{E}(t)$
from (\ref{eq:Omega_defs}) and the choice of a straight-line path,
and $\mathfrak{p}^{k}(\boldsymbol{x},\boldsymbol{y};t)\rightarrow\mathfrak{p}^{k}(\boldsymbol{x})$
from (\ref{eq:pref_def}). The dynamical equation (\ref{eq:density_operator_dynamical_result})
then simplifies to 
\begin{align}
 & i\hbar\frac{\partial\eta_{\alpha\boldsymbol{R};\beta\boldsymbol{R'}}(t)}{\partial t}=\label{eq:density_operator_dynamical_result_simplified}\\
 & \sum_{\lambda,\boldsymbol{R''}}\Big(H_{\alpha\boldsymbol{R};\lambda\boldsymbol{R''}}(t)\eta_{\lambda\boldsymbol{R''};\beta\boldsymbol{R'}}(t)-\eta_{\alpha\boldsymbol{R};\lambda\boldsymbol{R''}}(t)H_{\lambda\boldsymbol{R''};\beta\boldsymbol{R'}}(t)\Big)\nonumber \\
 & -\big(e(\boldsymbol{R}-\boldsymbol{R'})\boldsymbol{\cdot}\boldsymbol{E}(t)\big)\eta_{\alpha\boldsymbol{R};\beta\boldsymbol{R'}}(t),\nonumber 
\end{align}
where from (\ref{eq:Hmatrix_def}) we have used 
\begin{align}
\bar{H}_{\alpha\boldsymbol{R};\lambda\boldsymbol{R''}}(t) & \rightarrow H_{\alpha\boldsymbol{R};\lambda\boldsymbol{R''}}(t)\label{eq:Hsimple}\\
 & =\int W_{\alpha\boldsymbol{R}}^{*}(\boldsymbol{x})H_{0}\big(\boldsymbol{x},\mathfrak{p}(\boldsymbol{x})\big)W_{\lambda\boldsymbol{R''}}(\boldsymbol{x})d\boldsymbol{x}\nonumber \\
 & -\frac{e}{2}\boldsymbol{E}(t)\boldsymbol{\cdot}\int\Big(W_{\alpha\boldsymbol{R}}^{*}(\boldsymbol{x})\big(\boldsymbol{x}-\boldsymbol{R''}\big)W_{\lambda\boldsymbol{R''}}(\boldsymbol{x})\nonumber \\
 & \qquad\qquad+W_{\alpha\boldsymbol{R}}^{*}(\boldsymbol{x})\big(\boldsymbol{x}-\boldsymbol{R}\big)W_{\lambda\boldsymbol{R''}}(\boldsymbol{x})\Big)d\boldsymbol{x},\nonumber 
\end{align}
as well as the Hermiticity of $H_{0}\big(\boldsymbol{x},\mathfrak{p}(\boldsymbol{x})\big)$.
Using the orthogonality of the Wannier functions (\ref{eq:Worthogonality})
and translational invariance (\ref{eq:Tinv}) we can manipulate this
into the form 
\begin{align*}
 & H_{\alpha\boldsymbol{R};\lambda\boldsymbol{R''}}(t)=H_{\alpha\lambda}(\boldsymbol{R}-\boldsymbol{R}'';t),
\end{align*}
where 
\begin{align}
H_{\alpha\lambda}(\boldsymbol{R};t) & \equiv\int W_{\alpha}^{*}(\boldsymbol{x-}\boldsymbol{R})H_{0}\big(\boldsymbol{x},\mathfrak{p}(\boldsymbol{x})\big)W_{\lambda}(\boldsymbol{x})d\boldsymbol{x}\label{eq:HmatrixR}\\
 & -e\boldsymbol{E}(t)\boldsymbol{\cdot}\int W_{\alpha}^{*}(\boldsymbol{x-}\boldsymbol{R})\boldsymbol{x}W_{\lambda}(\boldsymbol{x})d\boldsymbol{x}.\nonumber 
\end{align}
We now look at the transform of $\eta_{\alpha\boldsymbol{R};\beta\boldsymbol{R'}}(t)$
into crystal momentum space. Using the identities 
\begin{align}
 & \frac{\Omega_{uc}}{(2\pi)^{3}}\int_{BZ}e^{i\boldsymbol{k}\boldsymbol{\cdot}(\boldsymbol{R}-\boldsymbol{R}')}d\boldsymbol{k}=\delta_{\boldsymbol{R}\boldsymbol{R'}},\label{eq:BZidentities}\\
 & \frac{\Omega_{uc}}{(2\pi)^{3}}\sum_{\boldsymbol{R}}e^{i(\boldsymbol{k}-\boldsymbol{k'})\boldsymbol{\cdot}\boldsymbol{R}}=\delta(\boldsymbol{k-}\boldsymbol{k'}),\nonumber 
\end{align}
where we restrict $\boldsymbol{k}$ and $\boldsymbol{k'}$ to the
first Brillouin zone, and $\boldsymbol{R}$ and $\boldsymbol{R'}$
are lattice sites, we introduce the Fourier transform 
\begin{align}
 & \eta_{\alpha\boldsymbol{k};\beta\boldsymbol{k'}}(t)\equiv\frac{\Omega_{uc}}{(2\pi)^{3}}\sum_{\boldsymbol{R},\boldsymbol{R'}}e^{i(\boldsymbol{k'}\boldsymbol{\cdot}\boldsymbol{R'}-\boldsymbol{k}\boldsymbol{\cdot}\boldsymbol{R})}\eta_{\alpha\boldsymbol{R};\beta\boldsymbol{R'}}(t),\label{eq:FT1}
\end{align}
and from the dynamical equation (\ref{eq:density_operator_dynamical_result_simplified})
for $\eta_{\alpha\boldsymbol{R};\beta\boldsymbol{R'}}(t)$ we obtain
the corresponding equations for $\eta_{\alpha\boldsymbol{k};\beta\boldsymbol{k'}}(t)$,
\begin{align}
 & i\hbar\frac{\partial\eta_{\alpha\boldsymbol{k};\beta\boldsymbol{k'}}(t)}{\partial t}=\label{eq:density_operator_dynamical_simplified_k}\\
 & \qquad\sum_{\gamma}\Big(H_{\alpha\gamma}(\boldsymbol{k};t)\eta_{\gamma\boldsymbol{k};\beta\boldsymbol{k'}}(t)-\eta_{\alpha\boldsymbol{k};\gamma\boldsymbol{k'}}(t)H_{\gamma\beta}(\boldsymbol{k'};t)\Big)\nonumber \\
 & \qquad-ie\boldsymbol{E}(t)\boldsymbol{\cdot}\left(\frac{\partial}{\partial\boldsymbol{k}}+\frac{\partial}{\partial\boldsymbol{k'}}\right)\eta_{\alpha\boldsymbol{k};\beta\boldsymbol{k'}}(t),\nonumber 
\end{align}
where 
\begin{align}
 & H_{\alpha\gamma}(\boldsymbol{k};t)\equiv\sum_{\boldsymbol{R}}e^{-i\boldsymbol{k}\boldsymbol{\cdot}\boldsymbol{R}}H_{\alpha\gamma}(\boldsymbol{R};t).\label{eq:HmatrixK}
\end{align}
From the inverse Fourier transform of (\ref{eq:FT1}), 
\begin{align}
 & \eta_{\alpha\boldsymbol{R};\beta\boldsymbol{R'}}(t)=\frac{\Omega_{uc}}{(2\pi)^{3}}\int\int e^{-i(\boldsymbol{k'}\boldsymbol{\cdot}\boldsymbol{R'}-\boldsymbol{k}\boldsymbol{\cdot}\boldsymbol{R})}\eta_{\alpha\boldsymbol{k};\beta\boldsymbol{k'}}(t)d\boldsymbol{k}d\boldsymbol{k'},\label{eq:FT2}
\end{align}
where we have used (\ref{eq:BZidentities}), we see that if we have
a state that shares the translational symmetry of the lattice, for
which $\eta_{\alpha\boldsymbol{R};\beta\boldsymbol{R'}}(t)$ depends
only on $(\boldsymbol{R}-\boldsymbol{R'})$, we have $\eta_{\alpha\boldsymbol{k};\beta\boldsymbol{k'}}(t)$
of the form 
\begin{align}
 & \eta_{\alpha\boldsymbol{k};\beta\boldsymbol{k'}}(t)=\eta_{\alpha\beta}(\boldsymbol{k};t)\delta(\boldsymbol{k}-\boldsymbol{k'}).\label{eq:FTform}
\end{align}
If this holds initially then the dynamical equation (\ref{eq:density_operator_dynamical_simplified_k})
guarantees that it will hold at all later times, with 
\begin{align}
 & i\hbar\frac{\partial\eta_{\alpha\beta}(\boldsymbol{k};t)}{\partial t}=\label{eq:density_operator_dynamical_simplified_k-1}\\
 & \qquad\sum_{\gamma}\Big(H_{\alpha\gamma}(\boldsymbol{k};t)\eta_{\gamma\beta}(\boldsymbol{k};t)-\eta_{\alpha\gamma}(\boldsymbol{k};t)H_{\gamma\beta}(\boldsymbol{k};t)\Big)\nonumber \\
 & \qquad-ie\boldsymbol{E}(t)\boldsymbol{\cdot}\frac{\partial\eta_{\alpha\beta}(\boldsymbol{k};t)}{\partial\boldsymbol{k}}.\nonumber 
\end{align}
We can write (\ref{eq:HmatrixK}) in a more familiar form by introduce
a Bloch function associated with each Wannier function, 
\begin{align}
\phi_{\alpha\boldsymbol{k}}(\boldsymbol{x}) & =\sqrt{\frac{\Omega_{uc}}{(2\pi)^{3}}}\sum_{\boldsymbol{R}}e^{i\boldsymbol{k}\boldsymbol{\cdot}\boldsymbol{R}}W_{\alpha\boldsymbol{R}}(\boldsymbol{x})\label{eq:Bloch_introduce}\\
 & =\frac{1}{\sqrt{(2\pi)^{3}}}e^{i\boldsymbol{k}\boldsymbol{\cdot}\boldsymbol{x}}u_{\alpha\boldsymbol{k}}(\boldsymbol{x}).\nonumber 
\end{align}
They are orthonormal according to 
\begin{align*}
 & \int\phi_{\alpha\boldsymbol{k}}^{*}(\boldsymbol{x})\phi_{\beta\boldsymbol{k'}}(\boldsymbol{x})d\boldsymbol{x}=\delta_{\alpha\beta}\delta(\boldsymbol{k-}\boldsymbol{k'}),
\end{align*}
and in the second line of (\ref{eq:Bloch_introduce}) we have introduced
the periodic function $u_{\alpha\boldsymbol{k}}(\boldsymbol{x})$,
where $u_{\alpha\boldsymbol{k}}(\boldsymbol{x})=u_{\alpha\boldsymbol{k}}(\boldsymbol{x}+\boldsymbol{R})$
for any lattice constant $\boldsymbol{R}$. Using the inverse relation
of (\ref{eq:Bloch_introduce}), 
\begin{align*}
 & W_{\alpha\boldsymbol{R}}(\boldsymbol{x})=\sqrt{\frac{\Omega_{uc}}{(2\pi)^{3}}}\int e^{-i\boldsymbol{k}\boldsymbol{\cdot}\boldsymbol{R}}\phi_{\alpha\boldsymbol{k}}(\boldsymbol{x})d\boldsymbol{k}
\end{align*}
in (\ref{eq:HmatrixR}), we find we can write (\ref{eq:HmatrixK})
as 
\begin{align}
 & H_{\alpha\gamma}(\boldsymbol{k};t)=H_{\alpha\gamma}^{0}(\boldsymbol{k})-e\boldsymbol{\xi}_{\alpha\gamma}(\boldsymbol{k})\boldsymbol{\cdot}\boldsymbol{E}(t),\label{eq:H_pieces}
\end{align}
where 
\begin{align*}
H_{\alpha\gamma}^{0}(\boldsymbol{k}) & =\sum_{\boldsymbol{R}}e^{-i\boldsymbol{k}\boldsymbol{\cdot}\boldsymbol{R}}\int W_{\alpha}^{*}(\boldsymbol{x}-\boldsymbol{R})H_{0}\big(\boldsymbol{x},\mathfrak{p}(\boldsymbol{x})\big)W_{\gamma}(\boldsymbol{x})d\boldsymbol{x}\\
 & =\frac{1}{\Omega_{uc}}\int_{uc}u_{\alpha\boldsymbol{k}}^{*}(\boldsymbol{x})H_{0}\big(\boldsymbol{x},\mathfrak{p}(\boldsymbol{x})+\hbar\boldsymbol{k}\big)u_{\gamma\boldsymbol{k}}(\boldsymbol{x})d\boldsymbol{x},
\end{align*}
with the second integral ranging over any unit cell, and 
\begin{align}
\boldsymbol{\xi}_{\alpha\gamma}(\boldsymbol{k}) & =\sum_{\boldsymbol{R}}e^{-i\boldsymbol{k}\boldsymbol{\cdot}\boldsymbol{R}}\int W_{\alpha}^{*}(\boldsymbol{x}-\boldsymbol{R})\boldsymbol{x}W_{\gamma}(\boldsymbol{x})d\boldsymbol{x}\label{eq:xi_Wannier}\\
 & =\frac{i}{\Omega_{uc}}\int_{uc}u_{\alpha\boldsymbol{k}}^{*}(\boldsymbol{x})\frac{\partial u_{\gamma\boldsymbol{k}}(\boldsymbol{x})}{\partial\boldsymbol{k}}d\boldsymbol{x},\nonumber 
\end{align}
where $\boldsymbol{\xi}_{\alpha\gamma}(\boldsymbol{k})$ is a non-Abelian
connection associated with the polarization; this object is discussed
at length in earlier work, including for example Aversa \cite{Aversa}
and Resta \cite{Resta1}. Dropping the matrix indices and writing,
for example, $\eta(\boldsymbol{k};t)$ for the matrix with elements
$\eta_{\alpha\beta}(\boldsymbol{k};t)$, we can then write (\ref{eq:density_operator_dynamical_simplified_k-1})
in the standard matrix form, 
\begin{align}
 & i\hbar\frac{\partial\eta(\boldsymbol{k};t)}{\partial t}=\label{eq:sigma_dynamical}\\
 & \qquad\Big[H^{0}(\boldsymbol{k})-e\boldsymbol{\xi}(\boldsymbol{k})\boldsymbol{\cdot}\boldsymbol{E}(t),\eta(\boldsymbol{k};t)\Big]-ie\boldsymbol{E}(t)\boldsymbol{\cdot}\frac{\partial\eta(\boldsymbol{k},t)}{\partial\boldsymbol{k}}.\nonumber 
\end{align}

When considering optical response, with the electric field treated
as uniform, one is usually interested in the spatially averaged current
density $\boldsymbol{J}(t)$. Returning to the expression (\ref{eq:full_j})
for $\langle\boldsymbol{\hat{j}}(\boldsymbol{x},t)\rangle$, since
for excitation by a uniform electric field we expect this quantity
to share the translational symmetry of the lattice, the spatial average
of the magnetization term will vanish, and we will only recover contributions
from the polarization $\boldsymbol{p}(\boldsymbol{x},t)$ and the
free current density $\boldsymbol{j}_{F}(\boldsymbol{x},t)$. Using
the decomposition $(\ref{eq:pm_decomp})$ of $\boldsymbol{p}(\boldsymbol{x},t)$
into contributions from different lattice sites, and noting that the
$\boldsymbol{p}_{\boldsymbol{R}}(\boldsymbol{x},t)$ for different
$\boldsymbol{R}$ will be the same except for a translation associated
with the difference in the lattice sites, we can introduce a spatially
averaged polarization 
\begin{align}
 & \boldsymbol{P}(t)=\frac{1}{\Omega_{uc}}\int\boldsymbol{p}_{\boldsymbol{R}}(\boldsymbol{x},t)d\boldsymbol{x},
\end{align}
which will be the same regardless of the $\boldsymbol{R}$ chosen
to evaluate it. Similarly decomposing the free current density (\ref{eq:jF_def})
into contributions from different lattice sites, 
\begin{align}
 & \boldsymbol{j}_{F}(\boldsymbol{x},t)=\sum_{\boldsymbol{R}}\boldsymbol{j}_{F;\boldsymbol{R}}(\boldsymbol{x},t)
\end{align}
where 
\begin{align}
 & \boldsymbol{j}_{F;\boldsymbol{R}}(\boldsymbol{x},t)=\frac{1}{2}\sum_{\boldsymbol{R'}}\boldsymbol{s}(\boldsymbol{x};\boldsymbol{R},\boldsymbol{R}')I(\boldsymbol{R},\boldsymbol{R}';t),\label{eq:jFR}
\end{align}
we can introduce a free current density, 
\begin{align}
 & \boldsymbol{J}_{F}(t)=\frac{1}{\Omega_{uc}}\int\boldsymbol{j}_{F;\boldsymbol{R}}(\boldsymbol{x},t)d\boldsymbol{x},\label{eq:JFdef}
\end{align}
which will be independent of the $\boldsymbol{R}$ chosen to evaluate
it. From (\ref{eq:full_j}) we then have 
\begin{align}
 & \boldsymbol{J}(t)=\frac{d\boldsymbol{P}(t)}{dt}+\boldsymbol{J}_{F}(t).\label{eq:Jresult}
\end{align}
Using the expression (\ref{eq:site_polarization},\ref{eq:rho_and_j_decomp})
for $\boldsymbol{p}_{\boldsymbol{R}}(\boldsymbol{x},t)$ and $\rho_{\boldsymbol{R}}(\boldsymbol{y},t)$,
together with the form (\ref{eq:relator_definitions}) for $\boldsymbol{s}(\boldsymbol{x};\boldsymbol{y},\boldsymbol{R})$,
the use of the Fourier decomposition (\ref{eq:FT2}) yields 
\begin{align}
 & \boldsymbol{P}(t)=e\sum_{\alpha,\beta}\int\frac{d\boldsymbol{k}}{(2\pi)^{3}}\boldsymbol{\xi}_{\beta\alpha}(\boldsymbol{k})\eta_{\alpha\beta}(\boldsymbol{k};t).\label{eq:Presult}
\end{align}
Similarly, performing the integral (\ref{eq:jF_def}), and in the
expression (\ref{eq:I_identify}) for $I(\boldsymbol{R},\boldsymbol{R}';t)$
replacing the general $\bar{H}_{\alpha\boldsymbol{R};\lambda\boldsymbol{R''}}(t)$
by $H_{\alpha\lambda}(\boldsymbol{R}-\boldsymbol{R}'';t)$ (see (\ref{eq:HmatrixR})
and preceding), we obtain 
\begin{align}
\boldsymbol{J}_{F}(t) & =\frac{e}{\hbar}\sum_{\alpha,\lambda}\int\frac{d\boldsymbol{k}}{(2\pi)^{3}}\Bigg(\frac{\partial}{\partial\boldsymbol{k}}\Big(H_{\alpha\lambda}^{0}(\boldsymbol{k})\label{eq:JFresult}\\
 & \qquad\qquad\qquad\qquad-e\boldsymbol{\xi}_{\alpha\lambda}(\boldsymbol{k})\boldsymbol{\cdot}\boldsymbol{E}(t)\Big)\Bigg)\eta_{\lambda\alpha}(\boldsymbol{k};t),\nonumber 
\end{align}
where we have used (\ref{eq:H_pieces}) for the transform (\ref{eq:HmatrixK})
$H_{\alpha\lambda}(\boldsymbol{k};t).$ In matrix form we write (\ref{eq:Presult},\ref{eq:JFresult})
as 
\begin{align}
\boldsymbol{P}(t) & =e\int\frac{d\boldsymbol{k}}{(2\pi)^{3}}\text{Tr}\left[\boldsymbol{\xi}\eta(t)\right],\label{eq:PJFsimple}\\
\boldsymbol{J}_{F}(t) & =\frac{e}{\hbar}\int\frac{d\boldsymbol{k}}{(2\pi)^{3}}\text{Tr}\left[\Big(\boldsymbol{\partial}\big(H^{0}-e\boldsymbol{\xi}\boldsymbol{\cdot}\boldsymbol{E}(t)\big)\Big)\eta(t)\right],\nonumber 
\end{align}
where $\partial^{n}=\partial/\partial k^{n}$. Using these results
in the expression (\ref{eq:Jresult}) for $\boldsymbol{J}(t)$, together
with the dynamical equation (\ref{eq:sigma_dynamical}) we obtain
\begin{align}
 & \boldsymbol{J}(t)=e\int\frac{d\boldsymbol{k}}{(2\pi)^{3}}\text{Tr}\big[\boldsymbol{v}\eta(t)\big],\label{eq:Jresult1}
\end{align}
where the matrix 
\begin{align}
 & \boldsymbol{v}=\boldsymbol{\partial}H^{0}-\frac{i}{\hbar}\left[\boldsymbol{\xi},H^{0}\right].\label{eq:vresult1}
\end{align}
We note that these results can be derived via an entirely different
strategy that begins directly in the long-wavelength limit and calculates
the response of the system to an applied electric field $\boldsymbol{E}(t)$
\footnote{Rodrigo A. Muniz, J. L. Cheng, and J. E. Sipe, in preparation};
such an approach is of course more direct and much easier if only
the long-wavelength limit is desired. 

\subsubsection{Basis transformations}

Rather than work with the basis functions $\left\{ \phi_{\alpha\boldsymbol{k}}(\boldsymbol{x})\right\} $
of (\ref{eq:Bloch_introduce}) at each $\boldsymbol{k}$ it is often
convenient to work with a new set of basis functions $\left\{ \bar{\phi}_{n\boldsymbol{k}}(\boldsymbol{x})\right\} ,$
related to the original set at each $\boldsymbol{k}$ by a unitary
transformation, 
\begin{align}
 & \bar{\phi}_{n\boldsymbol{k}}(\boldsymbol{x})=\sum_{\alpha}\phi_{\alpha\boldsymbol{k}}(\boldsymbol{x})U_{\alpha n}(\boldsymbol{k}),\label{eq:unitary_transformation}
\end{align}
where at each $\boldsymbol{k}$ the $U_{\alpha n}(\boldsymbol{k})$
are elements of a unitary matrix $U$. With 
\begin{align*}
\bar{H}^{0} & \equiv U^{\dagger}H^{0}U,\\
\bar{\eta} & \equiv U^{\dagger}\eta U,
\end{align*}
in our short-hand notation, we find that the equation for $\bar{\eta}(t)$
that follows from (\ref{eq:sigma_dynamical}) is 
\begin{align}
 & i\hbar\frac{\partial\bar{\eta}(t)}{\partial t}=\left[\bar{H}^{0}-e\boldsymbol{\bar{\xi}}\boldsymbol{\cdot}\boldsymbol{E}(t),\bar{\eta}(t)\right]-ie\boldsymbol{E}(t)\boldsymbol{\cdot}\boldsymbol{\partial}\bar{\eta}(t),\label{eq:sigmabar_dynamical}
\end{align}
where 
\begin{align}
 & \boldsymbol{\bar{\xi}}\equiv U^{\dagger}\boldsymbol{\xi}U+iU^{\dagger}\boldsymbol{\partial}U,
\end{align}
and the matrix $\boldsymbol{\bar{\xi}}$ at $\boldsymbol{k}$ has
components 
\begin{align}
 & \boldsymbol{\bar{\xi}}_{nm}(\boldsymbol{k})=\frac{i}{\Omega_{uc}}\int_{uc}\bar{u}_{n\boldsymbol{k}}^{*}(\boldsymbol{x})\frac{\partial\bar{u}_{m\boldsymbol{k}}(\boldsymbol{x})}{\partial\boldsymbol{k}}d\boldsymbol{x},
\end{align}
(\textit{cf. }(\ref{eq:xi_Wannier})) where $\bar{u}_{n\boldsymbol{k}}(\boldsymbol{x})$
is the periodic component, $\bar{u}_{n\boldsymbol{k}}(\boldsymbol{x}+\boldsymbol{R})=\bar{u}_{n\boldsymbol{k}}(\boldsymbol{x})$
for any lattice vector $\boldsymbol{R}$, associated with the basis
function $\bar{\phi}_{n\boldsymbol{k}}(\boldsymbol{x})$, 
\begin{align}
 & \bar{\phi}_{n\boldsymbol{k}}(\boldsymbol{x})=\frac{1}{\sqrt{(2\pi)^{3}}}e^{i\boldsymbol{k}\boldsymbol{\cdot}\boldsymbol{x}}\bar{u}_{n\boldsymbol{k}}(\boldsymbol{x}).
\end{align}
Such a transformation is often done as a prelude to a perturbation
calculation, and chosen so that $\bar{H}^{0}$ is diagonal, but more
general transformations can be considered. In terms of the new matrices
we find that the current density (\ref{eq:Jresult1}) can be written
as 
\begin{align}
 & \boldsymbol{J}(t)=e\int\frac{d\boldsymbol{k}}{(2\pi)^{3}}\text{Tr}\left[\boldsymbol{\bar{v}}\bar{\eta}(t)\right],\label{eq:Jresult2}
\end{align}
where the matrix 
\begin{align}
 & \boldsymbol{\bar{v}}=\boldsymbol{\partial}\bar{H}^{0}-\frac{i}{\hbar}\left[\boldsymbol{\bar{\xi}},\bar{H}^{0}\right]\label{eq:vresult2}
\end{align}
(\textit{cf. }(\ref{eq:vresult1})). Thus the form of both the dynamical
equations (\ref{eq:sigmabar_dynamical}) and the expression (\ref{eq:Jresult2})
for the current density are invariant under such a set of unitary
transformations $\left\{ U(\boldsymbol{k})\right\} $. The same does
\textit{not} hold for the polarization and free current density that
lead to the current density via (\ref{eq:Jresult}). In place of (\ref{eq:PJFsimple})
we obtain 
\begin{align}
\boldsymbol{P}(t) & =e\int\frac{d\boldsymbol{k}}{(2\pi)^{3}}\text{Tr}\left[\boldsymbol{\bar{\xi}}\bar{\eta}(t)\right]\nonumber \\
 & -ie\int\frac{d\boldsymbol{k}}{(2\pi)^{3}}\text{Tr}\left[\left(U^{\dagger}\boldsymbol{\partial}U\right)\bar{\eta}(t)\right],\label{eq:P_general}\\
\boldsymbol{J}_{F}(t) & =\frac{e}{\hbar}\int\frac{d\boldsymbol{k}}{(2\pi)^{3}}\text{Tr}\left[\left(\boldsymbol{\partial}\left(\bar{H}^{0}-e\boldsymbol{\bar{\xi}}\boldsymbol{\cdot}\boldsymbol{E}(t)\right)\right)\bar{\eta}(t)\right]\nonumber \\
 & +\frac{e}{\hbar}\int\frac{d\boldsymbol{k}}{(2\pi)^{3}}\text{Tr}\Big[\left[\left(U^{\dagger}\boldsymbol{\partial}U\right),\left(\bar{H}^{0}-e\boldsymbol{\bar{\xi}}\boldsymbol{\cdot}\boldsymbol{E}(t)\right)\right]\bar{\eta}(t)\Big]\nonumber \\
 & +i\frac{e^{2}}{\hbar}\int\frac{d\boldsymbol{k}}{(2\pi)^{3}}\text{Tr}\Big[\left(\boldsymbol{E}(t)\boldsymbol{\cdot}\boldsymbol{\partial}\right)\left(U^{\dagger}\boldsymbol{\partial}U\right)\bar{\eta}(t)\Big].\label{eq:JF_general}
\end{align}
It is only in the Bloch basis (\ref{eq:Bloch_introduce}) associated
with the maximally localized Wannier functions that our expressions
for $\boldsymbol{P}(t)$ and $\boldsymbol{J}_{F}(t)$ take a simple
form (\ref{eq:PJFsimple}), since it is those Wannier functions that
were used for the introduction of our quantities $\boldsymbol{p}(\boldsymbol{x},t)$,
$\boldsymbol{m}(\boldsymbol{x},t)$, $\rho_{F}(\boldsymbol{x},t)$
and $\boldsymbol{j}_{F}(\boldsymbol{x},t)$. This kind of dependence
on the details of a unitary transformation changing the Bloch basis
is not unique to our approach; indeed, the appearance of the second
term on the right-hand side of (\ref{eq:P_general}) arises as well
in the ``modern theory of polarization'', even in the limit of a
topologically trivial insulator, and is associated with the ``quantum
of ambiguity'' \cite{Mdef2}. In the following section we show that
for a topologically trivial insulator we find the same expressions
for the ground state polarization and magnetization as in the ``modern
theory of polarization and magnetization'', and it is easy to show
that our expression for $\boldsymbol{J}_{F}$ in such a ground state
vanishes even more generally, as expected. We will turn to our expressions
for the ground state polarization and magnetization for metals and
topologically nontrivial insulators in a future publication.

\subsubsection{Perturbative calculation}

We close this section by using our approach to calculate the linear
response of an insulator in the long-wavelength limit. We choose the
transformation (\ref{eq:unitary_transformation}) to be that which
diagonalizes the Hamiltonian, so 
\begin{align*}
 & \bar{H}_{nm}^{0}(\boldsymbol{k})=\delta_{nm}\hbar\omega_{n}(\boldsymbol{k}),
\end{align*}
and introducing a perturbation expansion of $\eta_{nm}(\boldsymbol{k};t)$
in orders of the electric field, 
\begin{align*}
 & \eta_{nm}(\boldsymbol{k};t)=\eta_{nm}^{(0)}(\boldsymbol{k})+\eta_{nm}^{(1)}(\boldsymbol{k};t)+...
\end{align*}
we take 
\begin{align*}
 & \eta_{nm}^{(0)}(\boldsymbol{k})=f_{n}\delta_{nm},
\end{align*}
where $f_{n}=0$ or $1$ for each $n$, independent of $\boldsymbol{k}.$
For an electric field 
\begin{align*}
 & \boldsymbol{E}(t)=\boldsymbol{E}(\omega)e^{-i\omega t}+\boldsymbol{E}(-\omega)e^{i\omega t},
\end{align*}
with $\boldsymbol{E}(-\omega)=\boldsymbol{E}^{*}(\omega)$, the usual
perturbation treatment of (\ref{eq:sigmabar_dynamical}) gives 
\begin{align*}
 & \eta_{nm}^{(1)}(\boldsymbol{k};t)=\eta_{nm}^{(1)}(\boldsymbol{k};\omega)e^{-i\omega t}+\eta_{nm}^{(1)}(\boldsymbol{k};-\omega)e^{i\omega t},
\end{align*}
where 
\begin{align*}
 & \eta_{nm}^{(1)}(\boldsymbol{k};\omega)=\frac{e\boldsymbol{\bar{\xi}}_{nm}(\boldsymbol{k})\boldsymbol{\cdot}\boldsymbol{E}(\omega)f_{mn}}{\hbar(\omega_{nm}(\boldsymbol{k})-\omega)},
\end{align*}
where $f_{mn}=f_{m}-f_{n}$ and likewise for $\omega_{nm}(\boldsymbol{k})$.
Then either calculating the first order result for 
\begin{align*}
 & \boldsymbol{J}^{(1)}(t)=\boldsymbol{J}(\omega)e^{-i\omega t}+\boldsymbol{J}(-\omega)e^{i\omega t}
\end{align*}
directly from (\ref{eq:Jresult2}), or calculating the first order
contributions to $\boldsymbol{P}(t)$ and $\boldsymbol{J}_{F}(t)$
from (\ref{eq:Jresult2},\ref{eq:vresult2}) and using the expressions
in (\ref{eq:Jresult}), we obtain 
\begin{align*}
 & J^{a}(\omega)=\sigma^{ab}(\omega)E^{b}(\omega),
\end{align*}
where the conductivity $\sigma^{ab}(\omega)$ is given by 
\begin{align*}
\sigma^{ab}(\omega) & =-i\omega\frac{e^{2}}{\hbar}\sum_{n,m}\int\frac{d\boldsymbol{k}}{(2\pi)^{3}}\frac{f_{m}\omega_{nm}(\bar{\xi}_{mn}^{a}\bar{\xi}_{nm}^{b}+\bar{\xi}_{mn}^{b}\bar{\xi}_{nm}^{a})}{(\omega_{nm}^{2}-\omega^{2})}\\
 & -i\frac{e^{2}}{\hbar}\sum_{n,m}\int\frac{d\boldsymbol{k}}{(2\pi)^{3}}\frac{f_{m}\omega_{nm}^{2}(\bar{\xi}_{mn}^{a}\bar{\xi}_{nm}^{b}-\bar{\xi}_{mn}^{b}\bar{\xi}_{nm}^{a})}{(\omega_{nm}^{2}-\omega^{2})}.
\end{align*}
Note that in general the conductivity can be finite as $\omega\rightarrow0$,
\begin{align}
 & \lim_{\omega\rightarrow0}\sigma^{ab}(\omega)=\frac{e^{2}}{\hbar}\sum_{m}\int\frac{d\boldsymbol{k}}{(2\pi)^{3}}f_{m}\left(\frac{\partial\bar{\xi}_{mm}^{a}}{\partial k^{b}}-\frac{\partial\bar{\xi}_{mm}^{b}}{\partial k^{a}}\right),\label{eq:trans_cond}
\end{align}
which is the well-known expression for the zero frequency transverse conductivity
of a topologically nontrivial insulator arising from the Kubo formula \footnote{In 2D this is proportional to the net Chern number of the filled bands.}\cite{TIs,DiXiao,3DQAH};
here we have used 
\begin{align*}
 & \boldsymbol{\partial\times\xi}-i\boldsymbol{\xi\times\xi}=0,
\end{align*}
and the fact that the corresponding equation holds for $\boldsymbol{\bar{\xi}}$
as well. Note that the zero frequency transverse conductivity is identified
in our treatment with the ``free'' current density $\boldsymbol{J}_{F}$,
as follows immediately from the expression (\ref{eq:Jresult}) for
$\boldsymbol{J}(t)$. This zero frequency response vanishes for a
topologically trivial insulator; in fact, we show in Section \ref{Summary}
that in linear response the full microscopic free current density
$\boldsymbol{j}_{F}(\boldsymbol{x},t)$ of a topologically trivial
insulator vanishes. In the long-wavelength limit being considered
here, the only linear response of a topologically trivial insulator
is from the polarization, $\boldsymbol{J}^{(1)}(t)=d\boldsymbol{P}^{(1)}(t)/dt$,
where 
\begin{align*}
 & \boldsymbol{P}^{(1)}(t)=\boldsymbol{P}(\omega)e^{-i\omega t}+\boldsymbol{P}(-\omega)e^{i\omega t},
\end{align*}
and 
\begin{align*}
 & \boldsymbol{P}(\omega)=\frac{e^{2}}{\hbar}\sum_{n,m}\int\frac{d\boldsymbol{k}}{(2\pi)^{3}}\frac{f_{mn}\boldsymbol{\bar{\xi}}_{mn}(\boldsymbol{k})\big(\boldsymbol{\bar{\xi}}_{nm}(\boldsymbol{k})\boldsymbol{\cdot}\boldsymbol{E}(\omega)\big)}{(\omega_{nm}(\boldsymbol{k})-\omega)},
\end{align*}
the usual result from perturbation theory \cite{Aversa}.

\subsection{Ground state moments in a topologically trivial insulator}

Next we consider both the electric and magnetic dipole moments associated
with the lattice sites in the ground state, restricting ourselves
to a topologically trivial insulator at zero temperature. In such a system all bands are
either completely empty or completely filled, and each set of energy
overlapping bands is topologically trivial as a whole. Then a maximally
localized set of Wannier functions are associated with the filled
bands, and another set with the empty bands \cite{Brouder,Marzari,Panati,Troyer},
so we have 
\begin{align*}
 & \eta_{\alpha\boldsymbol{R};\beta\boldsymbol{R'}}=f_{\alpha}\delta_{\alpha\beta}\delta_{\boldsymbol{R}\boldsymbol{R'}},
\end{align*}
where $f_{\alpha}=0$ or $1$, in the expressions (\ref{eq:pRgeneral},\ref{eq:mRgeneral})
for $\boldsymbol{p}_{\boldsymbol{R}}(\boldsymbol{x},t)$ and $\boldsymbol{m}_{\boldsymbol{R}}(\boldsymbol{x},t)$.
Those quantities are then of course independent of time, and the expressions
simplify considerably. Taking straight-line paths for $C(\boldsymbol{x},\boldsymbol{y})$
(see Appendix A) we have 
\begin{align*}
\int s^{i}(\boldsymbol{w};\boldsymbol{x},\boldsymbol{y})d\boldsymbol{w} & =(x^{i}-y^{i}),\\
\int\alpha^{jk}(\boldsymbol{w};\boldsymbol{x},\boldsymbol{y})d\boldsymbol{w} & =\frac{1}{2}\epsilon^{jmk}(x^{m}-y^{m}),
\end{align*}
and we find that the electric dipole moment $\boldsymbol{\mu}$ and
magnetic dipole moment $\boldsymbol{\nu}$ of the charge-current distribution
associated with each site are just 
\begin{align*}
\boldsymbol{\mu} & =\int\boldsymbol{p}_{\boldsymbol{R}}(\boldsymbol{x})d\boldsymbol{x},\\
\boldsymbol{\nu} & =\int\boldsymbol{m}_{\boldsymbol{R}}(\boldsymbol{x})d\boldsymbol{x},
\end{align*}
and are independent of $\boldsymbol{R}$, as expected. We obtain 
\begin{align*}
\boldsymbol{\mu} & =\sum_{\alpha}f_{\alpha}\boldsymbol{\mu}_{\alpha},\\
\boldsymbol{\nu} & =\sum_{\alpha}f_{\alpha}\boldsymbol{\nu}_{\alpha},
\end{align*}
where 
\begin{align*}
 & \boldsymbol{\mu}_{\alpha}=e\int W_{\alpha}^{*}(\boldsymbol{x})\boldsymbol{x}W_{\alpha}(\boldsymbol{x})d\boldsymbol{x},
\end{align*}
with $W_{\alpha}(\boldsymbol{x})\equiv W_{\alpha\boldsymbol{0}}(\boldsymbol{x})$,
while $\boldsymbol{\nu}_{\alpha}$ is the sum of two contributions,
\begin{align*}
 & \boldsymbol{\nu}_{\alpha}=\boldsymbol{\bar{\nu}}_{\alpha}+\boldsymbol{\tilde{\nu}}_{\alpha},
\end{align*}
arising from the two contributions to $\boldsymbol{m}_{\boldsymbol{R}}(\boldsymbol{x})$
in (\ref{eq:mRgeneral}). The first term is an atomic-like contribution,
\begin{align}
\bar{\nu}_{\alpha}^{j} & =\frac{1}{4c}\epsilon^{jmk}\int x^{m}W_{\alpha}^{*}(\boldsymbol{x})\Big(J^{k}\big(\boldsymbol{x},\mathfrak{p}(\boldsymbol{x})\big)W_{\alpha}(\boldsymbol{x})\Big)d\boldsymbol{x}\nonumber \\
 & +\frac{1}{4c}\epsilon^{jmk}\int x^{m}\Big(J^{k}\big(\boldsymbol{x},\mathfrak{p}(\boldsymbol{x})\big)W_{\alpha}(\boldsymbol{x})\Big)^{*}W_{\alpha}(\boldsymbol{x})d\boldsymbol{x},
\end{align}
which, for a Hamiltonian of Schrödinger form (\ref{eq:Hsch}), gives
\begin{align*}
\boldsymbol{\bar{\nu}}_{\alpha} & =\frac{e}{2mc}\int W_{\alpha}^{*}(\boldsymbol{x})\Big(\boldsymbol{x}\boldsymbol{\times}\Big(\frac{\hbar}{i}\frac{\partial W_{\alpha}(\boldsymbol{x})}{\partial\boldsymbol{x}}\\
 & \quad\qquad\qquad\qquad\qquad\qquad-\frac{e}{c}\boldsymbol{A}_{static}(\boldsymbol{x})W_{\alpha}(\boldsymbol{x})\Big)\Big)d\boldsymbol{x},
\end{align*}
which is familiar from the corresponding expression in atomic physics,
and 
\begin{align}
 & \tilde{\nu}_{\alpha}^{j}=\frac{1}{2\hbar c}\epsilon^{jmk}\sum_{\lambda,\boldsymbol{R}_{1}}R_{1}^{m}\text{Im}\left[H_{\alpha\boldsymbol{0};\lambda\boldsymbol{R}_{1}}^{(0)}\mu_{\lambda\boldsymbol{R}_{1};\alpha\boldsymbol{0}}^{k}(\boldsymbol{R}_{1})\right],
\end{align}
where $H_{\alpha\boldsymbol{0};\lambda\boldsymbol{R}_{1}}^{(0)}$
is given by (\ref{eq:Hsimple}) in the limit of no applied field,
or more generally by 
\begin{align}
 & H_{\alpha\boldsymbol{R}_{1};\lambda\boldsymbol{R}_{2}}^{(0)}\equiv\int W_{\alpha\boldsymbol{R}_{1}}^{*}(\boldsymbol{x})H_{0}\big(\boldsymbol{x},\mathfrak{p}(\boldsymbol{x})\big)W_{\lambda\boldsymbol{R}_{2}}(\boldsymbol{x})d\boldsymbol{x},\label{eq:Hnought}
\end{align}
where we have defined 
\begin{align*}
 & \boldsymbol{\mu}_{\beta\boldsymbol{R}_{1};\alpha\boldsymbol{R}_{2}}(\boldsymbol{R})=e\int W_{\beta\boldsymbol{R}_{1}}^{*}(\boldsymbol{x})\big(\boldsymbol{x}-\boldsymbol{R}\big)W_{\alpha\boldsymbol{R}_{2}}(\boldsymbol{x})d\boldsymbol{x}.
\end{align*}
Here $\boldsymbol{\tilde{\nu}}_{\alpha}$ is the itinerant contribution
to magnetic moment defined earlier \cite{Mdef1}; taking the macroscopic
polarization and magnetization to be given by 
\begin{align*}
\boldsymbol{P} & =\frac{\boldsymbol{\mu}}{\Omega_{uc}},\\
\boldsymbol{M} & =\frac{\boldsymbol{\nu}}{\Omega_{uc}}.
\end{align*}
We are in agreement with earlier results from the modern theory of
polarization and magnetization \cite{DiXiao}. In a calculation such
as this, using the basis of Wannier functions, the quantum of ambiguity
arises because one must choose the lattice site with which a representative
Wannier function of given type $\alpha$ is associated; the lattice
sites with which the rest of the Wannier functions of that type are
associated follow from translational symmetry.

\subsection{Perturbative result for a topologically trivial insulator}

We finally consider the general nature of the linear response of a
topologically trivial insulator to an applied electromagnetic field.
We look at the expression (\ref{eq:I_identify}) for the link current,
and expand it into terms involving powers of the electromagnetic field,
\begin{align*}
 & I(\boldsymbol{R},\boldsymbol{R'};t)=I^{(0)}(\boldsymbol{R},\boldsymbol{R'})+I^{(1)}(\boldsymbol{R},\boldsymbol{R'};t)+...,
\end{align*}
which is achieved by expanding the terms of which it is composed in
a similar way, 
\begin{align*}
\bar{H}_{\alpha\boldsymbol{R};\lambda\boldsymbol{R'}}(t) & =\bar{H}_{\alpha\boldsymbol{R};\lambda\boldsymbol{R'}}^{(0)}+\bar{H}_{\alpha\boldsymbol{R};\lambda\boldsymbol{R'}}^{(1)}(t)+...,\\
\eta_{\alpha\boldsymbol{R};\lambda\boldsymbol{R'}}(t) & =\eta_{\alpha\boldsymbol{R};\lambda\boldsymbol{R'}}^{(0)}+\eta_{\alpha\boldsymbol{R};\lambda\boldsymbol{R'}}^{(1)}(t)+...
\end{align*}
For a topologically trivial insulator we have 
\begin{align}
 & \eta_{\alpha\boldsymbol{R};\lambda\boldsymbol{R'}}^{(0)}=f_{\alpha}\delta_{\alpha\lambda}\delta_{\boldsymbol{R}\boldsymbol{R'}}\label{eq:TTI}
\end{align}
as above, and we identify $\bar{H}_{\alpha\boldsymbol{R};\lambda\boldsymbol{R'}}^{(0)}=H_{\alpha\boldsymbol{R};\lambda\boldsymbol{R'}}^{(0)},$
with $H_{\alpha\boldsymbol{R};\lambda\boldsymbol{R'}}^{(0)}$ given
by (\ref{eq:Hnought}). Then it immediately follows that $I^{(0)}(\boldsymbol{R},\boldsymbol{R'})=0,$
while 
\begin{align}
 & I^{(1)}(\boldsymbol{R},\boldsymbol{R'};t)=\label{eq:I1work}\\
 & \frac{e}{i\hbar}\sum_{\alpha,\lambda}\left(\bar{H}_{\alpha\boldsymbol{R};\lambda\boldsymbol{R'}}^{(1)}(t)\eta_{\lambda\boldsymbol{R'};\alpha\boldsymbol{R}}^{(0)}(t)-\eta_{\alpha\boldsymbol{R};\lambda\boldsymbol{R'}}^{(0)}(t)\bar{H}_{\lambda\boldsymbol{R'};\alpha\boldsymbol{R}}^{(1)}(t)\right)\nonumber \\
 & +\frac{e}{i\hbar}\sum_{\alpha,\lambda}\left(H_{\alpha\boldsymbol{R};\lambda\boldsymbol{R'}}^{(0)}\eta_{\lambda\boldsymbol{R'};\alpha\boldsymbol{R}}^{(1)}(t)-\eta_{\alpha\boldsymbol{R};\lambda\boldsymbol{R'}}^{(1)}(t)H_{\lambda\boldsymbol{R'};\alpha\boldsymbol{R}}^{(0)}\right).\nonumber 
\end{align}
The first of these two terms gives 
\begin{align*}
 & \frac{e}{i\hbar}\sum_{\alpha,\lambda}\left(\bar{H}_{\alpha\boldsymbol{R};\lambda\boldsymbol{R'}}^{(1)}(t)-\bar{H}_{\lambda\boldsymbol{R'};\alpha\boldsymbol{R}}^{(1)}(t)\right)f_{\alpha}\delta_{\alpha\lambda}\delta_{\boldsymbol{R}\boldsymbol{R'}},
\end{align*}
which vanishes regardless of the form of $\bar{H}_{\alpha\boldsymbol{R};\lambda\boldsymbol{R'}}^{(1)}(t)$.
To investigate the second term in (\ref{eq:I1work}), we look at the
equation for $\eta_{\alpha\boldsymbol{R};\lambda\boldsymbol{R'}}^{(1)}(t)$
that follows from a perturbative analysis of the general dynamical
equation (\ref{eq:density_operator_dynamical_result}). Expanding
terms in the usual way we obtain 
\begin{align*}
 & i\hbar\frac{\partial\eta_{\alpha\boldsymbol{R};\beta\boldsymbol{R'}}^{(1)}(t)}{\partial t}=\\
 & \qquad\sum_{\lambda,\boldsymbol{R''}}\Big(H_{\alpha\boldsymbol{R};\lambda\boldsymbol{R''}}^{(0)}\eta_{\lambda\boldsymbol{R''};\beta\boldsymbol{R'}}^{(1)}(t)-\eta_{\alpha\boldsymbol{R};\lambda\boldsymbol{R''}}^{(1)}H_{\lambda\boldsymbol{R''};\beta\boldsymbol{R'}}^{(0)}\Big)\\
 & \qquad+f_{\beta\alpha}H_{\alpha\boldsymbol{R};\beta\boldsymbol{R'}}^{(1)},
\end{align*}
and the inhomogeneous term in this equation is proportional to $f_{\beta\alpha}$.
As can be confirmed by the formal solution of this equation in reciprocal
space, $\eta_{\alpha\boldsymbol{R};\beta\boldsymbol{R'}}^{(1)}(t)$
will vanish unless one and only one of $\alpha$ and $\beta$ is associated
with a set of filled bands, so indeed $\eta_{\alpha\boldsymbol{R};\beta\boldsymbol{R'}}^{(1)}(t)$
will be proportional to $f_{\alpha\beta}$. But using this result
in the second term of (\ref{eq:I1work}), we see that the second term
will involve terms such as $f_{\alpha\beta}H_{\alpha\boldsymbol{R};\beta\boldsymbol{R'}}^{(0)}$,
which vanish. So we have $I^{(1)}(\boldsymbol{R},\boldsymbol{R'};t)=0$.

Thus to first order in the optical response of a topologically trivial
insulator at zero temperature the link current will vanish, and so we see from (\ref{eq:site_charge_dynamical},\ref{eq:rhoF_def},\ref{eq:jF_def})
that $\boldsymbol{j}_{F}(\boldsymbol{x},t)=0$ and $\rho_{F}(\boldsymbol{x},t)$
is independent of time. The entire optical response to first order
is described by $\boldsymbol{p}(\boldsymbol{x},t)$ and $\boldsymbol{m}(\boldsymbol{x},t)$,
even if the wavelength of light is comparable to or smaller than the
lattice spacing.

\section{Conclusion}

\label{sectionIV} We have presented a general strategy for constructing
microscopic polarization and magnetization fields, $\boldsymbol{p}(\boldsymbol{x},t)$
and $\boldsymbol{m}(\boldsymbol{x},t)$, which together with microscopic
``free'' charge and current densities, $\rho_{F}(\boldsymbol{x},t)$
and $\boldsymbol{j}_{F}(\boldsymbol{x},t)$, can be used (\ref{eq:micro})
to represent the expectation value of the microscopic electronic charge and current
density operators, $\left\langle \hat{\rho}(\boldsymbol{x},t)\right\rangle $
and $\langle\hat{\boldsymbol{j}}(\boldsymbol{x},t)\rangle$. Our goal
has been to write $\boldsymbol{p}(\boldsymbol{x},t)$ and $\boldsymbol{m}(\boldsymbol{x},t)$
as sums over contributions from different sites (\ref{eq:p_and_m}),
and we have done this by associating a set of localized orbitals with
each site. In a periodic crystal, which has been the focus of our
work, we have taken these to be maximally localized Wannier functions,
each set of such functions associated with a set of bands that is
topologically trivial; schemes exist to construct
such Wannier functions, which are primarily \textit{ab initio} based
\cite{WF1,WF2}, and can readily be used to implement the formalism
presented. The description that results is one of a lattice gauge
theory, where the free charge and current densities are described
by site charges and link currents respectively; the polarization and
magnetization fields associated with a given lattice site are then written
in terms of the single-particle density operators associated with
that site and its neighbors, and can be expected to be non-vanishing
only in the neighborhood of the site.

While our site quantities $\boldsymbol{p}_{\boldsymbol{R}}(\boldsymbol{x},t)$
and $\boldsymbol{m}_{\boldsymbol{R}}(\boldsymbol{x},t)$, together
with our free charge and current densities, have been defined to include
only valence and conduction electron contributions, the contributions from ion cores can
be identified as well. In the simple case where the ions are considered
fixed, and approximated as point particles, there is a time-independent ionic charge density 
\begin{align*}
 & \rho^{ion}(\boldsymbol{x})=\sum_{\boldsymbol{R}}\rho_{\boldsymbol{R}}^{ion}(\boldsymbol{x}),
\end{align*}
where 
\begin{align*}
 & \rho_{\boldsymbol{R}}^{ion}(\boldsymbol{x})=\sum_{N}q_{N}\delta(\boldsymbol{x}-\boldsymbol{R}-\boldsymbol{d}_{N}),
\end{align*}
and where we assume that in each unit cell there are ions with charges
$q_{N}$ located at $\boldsymbol{R}+\boldsymbol{d}_{N}$. Following
the strategy used for electrons, we can write $\rho_{\boldsymbol{R}}^{ion}(\boldsymbol{x})$
as a contribution that would arise if all the ions were at the lattice
site, and a time independent polarization, 
\begin{align*}
 & \rho_{\boldsymbol{R}}^{ion}(\boldsymbol{x})=\delta(\boldsymbol{x}-\boldsymbol{R})\sum_{N}q_{N}-\boldsymbol{\nabla}\boldsymbol{\cdot}\boldsymbol{p}_{\boldsymbol{R}}^{ion}(\boldsymbol{x}),
\end{align*}
where 
\begin{align*}
 & \boldsymbol{p}_{\boldsymbol{R}}^{ion}(\boldsymbol{x})=\int\boldsymbol{s}(\boldsymbol{x};\boldsymbol{y},\boldsymbol{R})\rho_{\boldsymbol{R}}^{ion}(\boldsymbol{y})d\boldsymbol{y}
\end{align*}
(compare (\ref{eq:site_polarization})). The contributions of the
ions can then be taken into account in our summary equations in Section \ref{Summary} by replacing the old $\rho_{F}(\boldsymbol{x},t)$ by
\begin{align*}
 & \rho_{F}^{new}(\boldsymbol{x},t)=\rho_{F}(\boldsymbol{x},t)+\sum_{N,\boldsymbol{R}}q_{N}\delta(\boldsymbol{x}-\boldsymbol{R})
\end{align*}
in (\ref{eq:summary_equations}), and by replacing the old $\boldsymbol{p}_{\boldsymbol{R}}(\boldsymbol{x},t)$
by
\begin{align*}
 & \boldsymbol{p}_{\boldsymbol{R}}^{new}(\boldsymbol{x},t)=\boldsymbol{p}_{\boldsymbol{R}}(\boldsymbol{x},t)+\boldsymbol{p}_{\boldsymbol{R}}^{ion}(\boldsymbol{x})
\end{align*}
in (\ref{eq:pm_decomp}). Note that in equilibrium $\rho_{F}^{new}(\boldsymbol{x},t)$
vanishes and all the microscopic charge is associated with the polarization
fields of valence electrons and ions. If the motion of the ions is
also considered then additional terms, including contributions to
the magnetization, appear as they did for valence and conduction electrons.

To move to macroscopic electrodynamics, spatial averages of the microscopic
polarization and magnetization can now be identified as the macroscopic
polarization and magnetization fields, and expansions of the polarization
and magnetization fields associated with a site can be used to identify
its electric and magnetic multipole moments; we plan to address these
matters in a following publication. Benefits of this approach include
the fact that the description of the interaction of the charges with
the electromagnetic field involves the electric and magnetic fields
themselves, rather than the scalar and vector potentials that describe
them, and the fact that the structure of the expressions for $\left\langle \hat{\rho}(\boldsymbol{x},t)\right\rangle $
and $\langle\hat{\boldsymbol{j}}(\boldsymbol{x},t)\rangle$ in terms
of their ``constituent sources'' $\boldsymbol{p}(\boldsymbol{x},t)$,
$\boldsymbol{m}(\boldsymbol{x},t),$ $\rho_{F}(\boldsymbol{x},t)$,
and $\boldsymbol{j}_{F}(\boldsymbol{x},t)$ are such that continuity
is guaranteed by construction, even if approximations are made in
describing those constituent sources.

We have restricted ourselves to electrons described in the independent
particle approximation, with no interactions besides those that can
be included in a mean field treatment of the electromagnetic field.
But while including interactions between the electrons will make the
description of the dynamics more difficult, much of the kinematics
associated with defining the constituent sources will remain unchanged.
Other simplifications we have employed, such as the omission of the
contribution of the electron spin to the magnetization, would be easier
to remedy. Also, although we have formulated our theory in a three
dimensional space, it is readily applicable to a two
dimensional lattice, using some care in formulating the Fourier transforms
and Wannier functions.

For a given $\left\langle \hat{\rho}(\boldsymbol{x},t)\right\rangle $
and $\langle\hat{\boldsymbol{j}}(\boldsymbol{x},t)\rangle$ there
is not a unique way to construct the constituent sources, even if
the set of localized orbitals being employed is fixed. For example,
we have restricted ourselves to line integral forms (\ref{eq:relator_definitions})
of the quantities that are responsible for relating $\boldsymbol{p}(\boldsymbol{x},t)$
and $\boldsymbol{m}(\boldsymbol{x},t)$ to $\left\langle \hat{\rho}(\boldsymbol{x},t)\right\rangle $
and $\langle\hat{\boldsymbol{j}}(\boldsymbol{x},t)\rangle$. Yet the
essential feature of those relators is only that they satisfy (\ref{eq:relator_equations}),
not that they are of line integral form. And even within the line
integral form we have used straight line paths in our examples, although
the equations we derive are more general. The straight line path does
seem the most natural, since it can be shown that it leads most naturally
to the usual multipole expansion, but other paths would be worth exploring.
The issue then is \textit{not }what are the ``correct'' constituent
sources $\boldsymbol{p}(\boldsymbol{x},t)$, $\boldsymbol{m}(\boldsymbol{x},t),$
$\rho_{F}(\boldsymbol{x},t)$, and $\boldsymbol{j}_{F}(\boldsymbol{x},t)$,
since they cannot be uniquely defined, but whether or not a particular
choice is convenient.

We have shown that our choice exhibits a number of interesting features:
First, in a limit where the lattice sites are moved further away from
each other, with the set of orbitals employed remaining fixed, our
description flows naturally into that of a set of ``isolated atoms''
on a lattice, regardless of the wavelength of light. Second, in the
long-wavelength limit of a uniform applied electric field we found
that the spatially averaged current density is the sum of a free current
part and a contribution from the time derivative of the polarization.
The first of these is responsible, for instance, for the transverse
DC conductivity in a topologically nontrivial insulator; the second
is the sole contribution to the linear response of a topologically
trivial insulator. More generally, in a topologically trivial insulator
the linear optical response is due \textit{solely }to induced microscopic
polarization and magnetization fields, again regardless of the wavelength
of light. This is as expected, since it is only to higher order that
one would physically expect that injected quasi-particles could be
driven by the electromagnetic field and lead to induced free charges
and currents. Finally, we showed that in the ground state of a topologically
trivial insulator the expressions for the polarization and the magnetization
agree with results from the ``modern theory of polarization and magnetization.''

We believe these features suggest that our choice of constituent sources
is worth developing further as a description of the ground state of
systems of interest, and of the linear and nonlinear response of matter
to radiation very generally. And we believe that the overall framework
we have established here for introducing microscopic polarization
and magnetization fields in extended systems will prove to be valuable
for studying electronic dynamics at atomic scales.

\section{Acknowledgements}

This work was supported by the Natural Sciences and Engineering Research
Council of Canada (NSERC). P.T.M. acknowledges a PGS-D scholarship
from NSERC. We thank Sylvia Swiecicki for many helpful discussions
in the early stages of this work.

\section{Appendices}

\appendix
%dummy comment inserted by tex2lyx to ensure that this paragraph is not empty

\section{Relator equations}

We give a derivation of the relations (\ref{eq:relator_equations})
between relators that follows the spirit of Healy's \cite{Healybook},
although our notation is different. We characterize the path from
$\boldsymbol{y}$ to $\boldsymbol{x}$ by a function $\boldsymbol{z}(u)$
such that $\boldsymbol{z}(u_{1})=\boldsymbol{y}$ and $\boldsymbol{z}(u_{2})=\boldsymbol{x}$.
Then from the first of (\ref{eq:relator_definitions}) we have 
\begin{align}
 & s^{i}(\boldsymbol{w};\boldsymbol{x},\boldsymbol{y})=\int_{u_{1}}^{u_{2}}du\frac{dz^{i}(u)}{du}\delta(\boldsymbol{w}-\boldsymbol{z}),\label{eq:swork}
\end{align}
and so 
\begin{align*}
-\frac{\partial s^{i}(\boldsymbol{w};\boldsymbol{x},\boldsymbol{y})}{\partial w^{i}} & =-\int_{u_{1}}^{u_{2}}du\frac{dz^{i}(u)}{du}\frac{\partial\delta(\boldsymbol{w}-\boldsymbol{z})}{\partial w^{i}}\\
 & =\int_{u_{1}}^{u_{2}}du\frac{dz^{i}(u)}{du}\frac{\partial\delta(\boldsymbol{w}-\boldsymbol{z})}{\partial z^{i}}\\
 & =\int_{u_{1}}^{u_{2}}du\frac{d}{du}\delta(\boldsymbol{w}-\boldsymbol{z})\\
 & =\delta(\boldsymbol{w}-\boldsymbol{y})-\delta(\boldsymbol{w}-\boldsymbol{x}),
\end{align*}
which is the first of (\ref{eq:relator_equations}).

Moving to the second of (\ref{eq:relator_definitions}) we have 
\begin{align}
 & \alpha^{jk}(\boldsymbol{w};\boldsymbol{x},\boldsymbol{y})=\epsilon^{jmn}\int_{u_{1}}^{u_{2}}du\frac{dz^{m}}{du}\frac{\partial z^{n}}{\partial x^{k}}\delta(\boldsymbol{w}-\boldsymbol{z}),\label{eq:alphawork}
\end{align}
so 
\begin{align*}
 & \epsilon^{ipj}\frac{\partial\alpha^{jk}(\boldsymbol{w};\boldsymbol{x},\boldsymbol{y})}{\partial w^{p}}=\epsilon^{ipj}\epsilon^{jmn}\int_{u_{1}}^{u_{2}}du\frac{dz^{m}}{du}\frac{\partial z^{n}}{\partial x^{k}}\frac{\partial\delta(\boldsymbol{w}-\boldsymbol{z})}{\partial w^{p}}.
\end{align*}
Now 
\begin{align*}
 & \epsilon^{ipj}\epsilon^{jmn}=\epsilon^{jip}\epsilon^{jmn}=\delta^{im}\delta^{pn}-\delta^{in}\delta^{pm},
\end{align*}
so 
\begin{align*}
 & \epsilon^{ipj}\frac{\partial\alpha^{jk}(\boldsymbol{w};\boldsymbol{x},\boldsymbol{y})}{\partial w^{p}}\\
 & =\left(\delta^{im}\delta^{pn}-\delta^{in}\delta^{pm}\right)\int_{u_{1}}^{u_{2}}du\frac{dz^{m}}{du}\frac{\partial z^{n}}{\partial x^{k}}\frac{\partial\delta(\boldsymbol{w}-\boldsymbol{z})}{\partial w^{p}}\\
 & =\int_{u_{1}}^{u_{2}}du\frac{dz^{i}}{du}\frac{\partial z^{p}}{\partial x^{k}}\frac{\partial\delta(\boldsymbol{w}-\boldsymbol{z})}{\partial w^{p}}\\
 & \qquad-\int_{u_{1}}^{u_{2}}du\frac{dz^{p}}{du}\frac{\partial z^{i}}{\partial x^{k}}\frac{\partial\delta(\boldsymbol{w}-\boldsymbol{z})}{\partial w^{p}}\\
 & =-\int_{u_{1}}^{u_{2}}du\frac{dz^{i}}{du}\frac{\partial z^{p}}{\partial x^{k}}\frac{\partial\delta(\boldsymbol{w}-\boldsymbol{z})}{\partial z^{p}}\\
 & \qquad+\int_{u_{1}}^{u_{2}}du\frac{dz^{p}}{du}\frac{\partial z^{i}}{\partial x^{k}}\frac{\partial\delta(\boldsymbol{w}-\boldsymbol{z})}{\partial z^{p}}\\
 & =-\int_{u_{1}}^{u_{2}}du\frac{dz^{i}}{du}\frac{\partial}{\partial x^{k}}\delta(\boldsymbol{w}-\boldsymbol{z})\\
 & \qquad+\int_{u_{1}}^{u_{2}}du\frac{\partial z^{i}}{\partial x^{k}}\frac{d}{du}\delta(\boldsymbol{w}-\boldsymbol{z}).
\end{align*}
Partially integrating the second term gives 
\begin{align*}
 & \epsilon^{ipj}\frac{\partial\alpha^{jk}(\boldsymbol{w};\boldsymbol{x},\boldsymbol{y})}{\partial w^{p}}\\
 & =-\int_{u_{1}}^{u_{2}}du\frac{dz^{i}}{du}\frac{\partial}{\partial x^{k}}\delta(\boldsymbol{w}-\boldsymbol{z})\\
 & \qquad-\int_{u_{1}}^{u_{2}}du\left(\frac{\partial}{\partial x^{k}}\frac{dz^{i}}{du}\right)\delta(\boldsymbol{w}-\boldsymbol{z})+\left[\frac{\partial z^{i}}{\partial x^{k}}\delta(\boldsymbol{w}-\boldsymbol{z})\right]_{u_{1}}^{u_{2}}\\
 & =-\frac{\partial}{\partial x^{i}}\left(\int_{u_{1}}^{u_{2}}du\frac{dz^{i}}{du}\delta(\boldsymbol{w}-\boldsymbol{z})\right)+\left[\frac{\partial z^{i}}{\partial x^{k}}\delta(\boldsymbol{w}-\boldsymbol{z})\right]_{u_{1}}^{u_{2}}.
\end{align*}
In the second term note the $z^{i}(u_{2})=x^{i}$ and $z^{i}(u_{1})=y^{i}$,
so there will only be a contribution at $u_{2}$. Then recognizing
the first term from (\ref{eq:swork}), we have 
\begin{align*}
 & \epsilon^{ipj}\frac{\partial\alpha^{jk}(\boldsymbol{w};\boldsymbol{x},\boldsymbol{y})}{\partial w^{p}}=-\frac{\partial}{\partial x^{k}}s^{i}(\boldsymbol{w};\boldsymbol{x},\boldsymbol{y})+\delta^{ik}\delta(\boldsymbol{w}-\boldsymbol{x}),
\end{align*}
which is the second of (\ref{eq:relator_equations}).

The final expression is derived in much the same way. We have 
\begin{align}
 & \beta^{jk}(\boldsymbol{w};\boldsymbol{x},\boldsymbol{y})=\epsilon^{jmn}\int_{u_{1}}^{u_{2}}du\frac{dz^{m}}{du}\frac{\partial z^{n}}{\partial y^{k}}\delta(\boldsymbol{w}-\boldsymbol{z}),\label{eq:betawork}
\end{align}
so 
\begin{align*}
 & \epsilon^{ipj}\frac{\partial\beta^{jk}(\boldsymbol{w};\boldsymbol{x},\boldsymbol{y})}{\partial w^{p}}\\
 & =\epsilon^{ipj}\epsilon^{jmn}\int_{u_{1}}^{u_{2}}du\frac{dz^{m}}{du}\frac{\partial z^{n}}{\partial y^{k}}\delta(\boldsymbol{w}-\boldsymbol{z})\\
 & =\left(\delta^{im}\delta^{pn}-\delta^{in}\delta^{pm}\right)\int_{u_{1}}^{u_{2}}du\frac{dz^{m}}{du}\frac{\partial z^{n}}{\partial y^{k}}\frac{\partial\delta(\boldsymbol{w}-\boldsymbol{z})}{\partial w^{p}},
\end{align*}
and following exactly the strategy above we have 
\begin{align*}
 & \epsilon^{ipj}\frac{\partial\beta^{jk}(\boldsymbol{w};\boldsymbol{x},\boldsymbol{y})}{\partial w^{p}}\\
 & =-\frac{\partial}{\partial y^{i}}\left(\int_{u_{1}}^{u_{2}}du\frac{dz^{i}}{du}\delta(\boldsymbol{w}-\boldsymbol{z})\right)+\left[\frac{\partial z^{i}}{\partial y^{k}}\delta(\boldsymbol{w}-\boldsymbol{z})\right]_{u_{1}}^{u_{2}}\\
 & =-\frac{\partial}{\partial y^{i}}s^{i}(\boldsymbol{w};\boldsymbol{x},\boldsymbol{y})-\delta^{ik}\delta(\boldsymbol{w}-\boldsymbol{y}),
\end{align*}
because in the last term only the contribution from $u_{1}$ will
survive. This is the third of (\ref{eq:relator_equations}).

Next we confirm that the relations (\ref{eq:symmetric}) hold for
symmetric paths $C(\boldsymbol{x},\boldsymbol{y})$, where for each
and every $\boldsymbol{x}$ and $\boldsymbol{y}$ the path $C(\boldsymbol{x},\boldsymbol{y})$
is the ``reverse'' of the path $C(\boldsymbol{y},\boldsymbol{x}).$
More precisely, if $C(\boldsymbol{x},\boldsymbol{y})$ is specified
by giving $\boldsymbol{z}(u)$ as $u$ varies from $u_{1}$ to $u_{2}$,
with $\boldsymbol{z}(u_{1})=\boldsymbol{y}$ and $\boldsymbol{z}(u_{2})=\boldsymbol{x}$,
then $C(\boldsymbol{y},\boldsymbol{x})$ is specified by the same
$\boldsymbol{z}(u)$ as $u$ varies from $u_{2}$ to $u_{1}$. We
have 
\begin{align*}
s^{i}(\boldsymbol{w};\boldsymbol{x},\boldsymbol{y}) & =\int_{C(\boldsymbol{x},\boldsymbol{y})}dz^{i}\delta(\boldsymbol{w}-\boldsymbol{z})\\
 & =\int_{u_{1}}^{u_{2}}du\frac{dz^{i}(u)}{du}\delta(\boldsymbol{w}-\boldsymbol{z}),
\end{align*}
and 
\begin{align*}
s^{i}(\boldsymbol{w};\boldsymbol{y},\boldsymbol{x}) & =\int_{C(\boldsymbol{y},\boldsymbol{x})}dz^{i}\delta(\boldsymbol{w}-\boldsymbol{z})\\
 & =\int_{u_{2}}^{u_{1}}du\frac{dz^{i}(u)}{du}\delta(\boldsymbol{w}-\boldsymbol{z})\\
 & =-\int_{u_{1}}^{u_{2}}du\frac{dz^{i}(u)}{du}\delta(\boldsymbol{w}-\boldsymbol{z})\\
 & =-s^{i}(\boldsymbol{w};\boldsymbol{x},\boldsymbol{y}),
\end{align*}
while 
\begin{align*}
\alpha^{jk}(\boldsymbol{w};\boldsymbol{x},\boldsymbol{y}) & =\epsilon^{jmn}\int_{C(\boldsymbol{x},\boldsymbol{y})}dz^{m}\frac{\partial z^{n}}{\partial x^{k}}\delta(\boldsymbol{w}-\boldsymbol{z})\\
 & =\epsilon^{jmn}\int_{u_{1}}^{u_{2}}\frac{dz^{m}(u)}{du}\frac{\partial z^{n}}{\partial x^{k}}\delta(\boldsymbol{w}-\boldsymbol{z}),
\end{align*}
and 
\begin{align*}
\beta^{jk}(\boldsymbol{w};\boldsymbol{y},\boldsymbol{x}) & =\epsilon^{jmn}\int_{C(\boldsymbol{y},\boldsymbol{x})}dz^{m}\frac{\partial z^{n}}{\partial x^{k}}\delta(\boldsymbol{w}-\boldsymbol{z})\\
 & =\epsilon^{jmn}\int_{u_{2}}^{u_{1}}du\frac{dz^{m}(u)}{du}\frac{\partial z^{n}}{\partial x^{k}}\delta(\boldsymbol{w}-\boldsymbol{z})\\
 & =-\epsilon^{jmn}\int_{u_{1}}^{u_{2}}du\frac{dz^{m}(u)}{du}\frac{\partial z^{n}}{\partial x^{k}}\delta(\boldsymbol{w}-\boldsymbol{z})\\
 & =-\alpha^{jk}(\boldsymbol{w};\boldsymbol{x},\boldsymbol{y}).
\end{align*}

Finally, we consider the special case of a straight-line path $C(\boldsymbol{x},\boldsymbol{y})$
for each $\boldsymbol{x}$ and $\boldsymbol{y}$. That is, taking
$u_{1}=0$ and $u_{2}=1$, for the path $C(\boldsymbol{x},\boldsymbol{y})$
we have 
\begin{align*}
 & \boldsymbol{z}=\boldsymbol{y}+u(\boldsymbol{x}-\boldsymbol{y}).
\end{align*}
Then we have 
\begin{align*}
\frac{dz^{i}(u)}{du} & =x^{i}-y^{i},\\
\frac{\partial z^{n}}{\partial x^{k}} & =u\delta^{nk}\\
\frac{\partial z^{n}}{\partial y^{k}} & =(1-u)\delta^{nk}
\end{align*}
and so from (\ref{eq:swork}) we have 
\begin{align}
 & s^{i}(\boldsymbol{w};\boldsymbol{x},\boldsymbol{y})=\label{eq:s-straight}\\
 & \int_{0}^{1}(x^{i}-y^{i})\delta(\boldsymbol{w}-\boldsymbol{y}-u(\boldsymbol{x}-\boldsymbol{y}))du,\nonumber 
\end{align}
while from (\ref{eq:alphawork}) we have 
\begin{align}
 & \alpha^{jk}(\boldsymbol{w};\boldsymbol{x},\boldsymbol{y})=\label{eq:alpha-straight}\\
 & \epsilon^{jmk}\int_{0}^{1}(x^{m}-y^{m})\delta(\boldsymbol{w}-\boldsymbol{y}-u(\boldsymbol{x}-\boldsymbol{y}))udu,\nonumber 
\end{align}
and from (\ref{eq:betawork}) we have 
\begin{align}
 & \beta^{jk}(\boldsymbol{w};\boldsymbol{x},\boldsymbol{y})=\label{eq:beta-straight}\\
 & \epsilon^{jmk}\int_{0}^{1}(x^{m}-y^{m})\delta(\boldsymbol{w}-\boldsymbol{y}-u(\boldsymbol{x}-\boldsymbol{y}))(1-u)du.\nonumber 
\end{align}
Since a straight-line path is symmetric, in the terminology used above,
we expect (\ref{eq:s-straight},\ref{eq:alpha-straight},\ref{eq:beta-straight})
to satisfy (\ref{eq:symmetric}), and it is easy to confirm that they
do.

\section{Orthogonalization of states}

In this Appendix we use a short-hand notation, taking 
\begin{align*}
 & (\alpha\boldsymbol{R})\rightarrow n\\
 & \boldsymbol{R}\rightarrow\boldsymbol{R}_{n}
\end{align*}
Our set of non-orthogonal states (\ref{eq:Wprime_def}) are then labelled
$\left\{ W'_{n}(\boldsymbol{x},t)\right\} $; the elements of the
overlap matrix $\mathbb{S}(t)$ characterizing them are 
\begin{align*}
S_{nm}(t) & \equiv\int W_{n}^{'*}(\boldsymbol{x},t)W'_{m}(\boldsymbol{x},t)d\boldsymbol{x}\\
 & =e^{i\Phi(\boldsymbol{R}_{n},\boldsymbol{R}_{m};t)}\int W_{n}^{*}(\boldsymbol{x})e^{i\Delta(\boldsymbol{R}_{n},\boldsymbol{x},\boldsymbol{R}_{m};t)}W_{m}(\boldsymbol{x})d\boldsymbol{x}
\end{align*}
Clearly $\mathbb{S}(t)$ is Hermitian, and although its matrix elements
are in general not gauge-invariant, they can be written as 
\begin{align*}
 & S_{nm}(t)=e^{i\Phi(\boldsymbol{R}_{n},\boldsymbol{R}_{m};t)}\hat{S}_{nm}(t),
\end{align*}
where the $\hat{S}_{nm}(t)$ are gauge-invariant, 
\begin{align*}
 & \hat{S}_{nm}(t)=\int W_{n}^{*}(\boldsymbol{x})e^{i\Delta(\boldsymbol{R}_{n},\boldsymbol{x},\boldsymbol{R}_{m};t)}W_{m}(\boldsymbol{x})d\boldsymbol{x}
\end{align*}
and since 
\begin{align*}
\hat{S}_{nm}^{*}(t) & =e^{i\Phi(\boldsymbol{R}_{n},\boldsymbol{R}_{m};t)}S_{nm}^{*}(t)\\
 & =e^{-i\Phi(\boldsymbol{R}_{m},\boldsymbol{R}_{n};t)}S_{mn}(t)\\
 & =\hat{S}_{mn}(t)
\end{align*}
the matrix $\mathbb{\hat{S}}(t)$ is also Hermitian.

We seek a set $\left\{ \bar{W}_{n}(\boldsymbol{x},t)\right\} $ spanned
by the original set that are orthogonal, 
\begin{align}
 & \bar{W}_{n}(\boldsymbol{x},t)=\sum_{p}W'_{p}(\boldsymbol{x},t)C_{pn}(t),\label{eq:Wbar_work}
\end{align}
where 
\begin{align*}
 & \int\bar{W}_{n}^{*}(\boldsymbol{x},t)\bar{W}_{m}(\boldsymbol{x},t)d\boldsymbol{x}=\sum_{p,l}C_{np}^{*}(t)S_{nl}(t)C_{lm}(t)=\delta_{nm},
\end{align*}
or in matrix notation 
\begin{align*}
 & \mathbb{C}^{\dagger}(t)\mathbb{S}(t)\mathbb{C}(t)=\mathbb{I}.
\end{align*}
There are of course many matrices $\mathbb{C}(t)$ that can be found
that satisfy this condition. However, the desired matrix $\mathbb{C}(t)$
yielding the minimization of (\ref{eq:TBminimized}) is the Hermitian
matrix satisfying 
\begin{align*}
 & \mathbb{C}(t)=\mathbb{S}^{-1/2}(t),
\end{align*}
\cite{Mayer}. That is, it is the ``Hermitian square root'' of the
inverse of the overlap matrix $\mathbb{S}(t)$.

To see the structure of $\mathbb{C}(t)$, and the nature of the resulting
$\left\{ \bar{W}_{n}(\boldsymbol{x},t)\right\} ,$ first introduce
$\mathbb{T}(t)$ as the inverse of $\mathbb{S}(t)$, 
\begin{align*}
 & \sum_{l}S_{nl}(t)T_{lm}(t)=\delta_{nm},
\end{align*}
or 
\begin{align*}
 & \sum_{l}\hat{S}_{nl}(t)e^{i\Phi(\boldsymbol{R}_{n},\boldsymbol{R}_{l};t)}T_{lm}(t)=\delta_{nm}.
\end{align*}
Introducing $\mathbb{\hat{T}}(t)$ according to 
\begin{align*}
 & \hat{T}_{lm}(t)=T_{lm}(t)e^{-i\Phi(\boldsymbol{R}_{l},\boldsymbol{R}_{m};t)}
\end{align*}
we have 
\begin{align*}
 & e^{-i\Phi(\boldsymbol{R}_{m},\boldsymbol{R}_{n};t)}\sum_{l}\hat{S}_{nl}(t)\hat{T}_{lm}(t)e^{i\Delta(\boldsymbol{R}_{n},\boldsymbol{R}_{l},\boldsymbol{R}_{m};t)}=\delta_{nm},
\end{align*}
or 
\begin{align*}
 & \sum_{l}\hat{S}_{nl}(t)\hat{T}_{lm}(t)e^{i\Delta(\boldsymbol{R}_{n},\boldsymbol{R}_{l},\boldsymbol{R}_{m};t)}=\delta_{nm},
\end{align*}
and we see that the elements of $\mathbb{\hat{T}}(t)$ must be gauge-invariant,
since everything else in the equation is. Since $\mathbb{C}(t)$ is
the square root of $\mathbb{T}(t)$ we have 
\begin{align}
 & \sum_{l}C_{nl}(t)C_{lm}(t)=T_{nm}(t)=\hat{T}_{nm}(t)e^{i\Phi(\boldsymbol{R}_{n},\boldsymbol{R}_{m};t)}.\label{eq:C_solve}
\end{align}
Now the $\mathbb{C}(t)$ we seek is Hermitian, requiring 
\begin{align}
 & C_{nl}^{*}(t)=C_{ln}(t).\label{eq:C_Hermitian}
\end{align}
Introducing $\mathbb{\hat{C}}(t)$ according to 
\begin{align}
 & \hat{C}_{nl}(t)=C_{nl}(t)e^{-i\Phi(\boldsymbol{R}_{n},\boldsymbol{R}_{l};t)},\label{eq:Cslash_def}
\end{align}
then using (\ref{eq:C_Hermitian}) we see that 
\begin{align*}
\hat{C}_{nl}^{*}(t) & =C_{nl}^{*}(t)e^{i\Phi(\boldsymbol{R}_{n},\boldsymbol{R}_{l};t)}\\
 & =C_{ln}(t)e^{-i\Phi(\boldsymbol{R}_{l},\boldsymbol{R}_{n};t)}\\
 & =\hat{C}_{ln}(t),
\end{align*}
and so the $\mathbb{\hat{C}}(t)$ we seek is Hermitian and, from (\ref{eq:C_solve}),
satisfies 
\begin{align}
 & \sum_{l}\hat{C}_{nl}(t)\hat{C}_{lm}(t)e^{i\left[\Phi\left(\boldsymbol{R}_{n},\boldsymbol{R}_{l};t\right)+\Phi(\boldsymbol{R}_{l},\boldsymbol{R}_{m};t)\right]}\nonumber \\
 & =\hat{T}_{nm}(t)e^{i\Phi(\boldsymbol{R}_{n},\boldsymbol{R}_{m};t)},
\end{align}
or 
\begin{align*}
 & \sum_{l}\hat{C}_{nl}(t)\hat{C}_{lm}(t)e^{i\Delta(\boldsymbol{R}_{n},\boldsymbol{R}_{l},\boldsymbol{R}_{m};t)}=\hat{T}_{nm}(t),
\end{align*}
a gauge-invariant equation, and so the matrix $\mathbb{\hat{C}}(t)$
we seek is gauge invariant.

In terms of our new quantities we can write (\ref{eq:Wbar_work})
as 
\begin{align*}
\bar{W}_{n}(\boldsymbol{x},t) & =\sum_{p}W_{p}(\boldsymbol{x})e^{i\Phi(\boldsymbol{x},\boldsymbol{R}_{p};t)}\hat{C}_{pn}(t)e^{i\Phi(\boldsymbol{R}_{p},\boldsymbol{R}_{n};t)}\\
 & =e^{i\Phi(\boldsymbol{x},\boldsymbol{R}_{n};t)}\sum_{p}W_{p}(\boldsymbol{x})\hat{C}_{pn}(t)e^{i\Phi(\boldsymbol{R}_{n},\boldsymbol{x};t)}\\
 & \quad\times e^{i\Phi(\boldsymbol{x},\boldsymbol{R}_{p};t)}e^{i\Phi(\boldsymbol{R}_{p},\boldsymbol{R}_{n};t)}\\
 & =e^{i\Phi(\boldsymbol{x},\boldsymbol{R}_{n};t)}\chi_{n}(\boldsymbol{x},t),
\end{align*}
where 
\begin{align}
 & \chi_{n}(\boldsymbol{x},t)\equiv\sum_{p}W_{p}(\boldsymbol{x})\hat{C}_{pn}(t)e^{i\Delta(\boldsymbol{x},\boldsymbol{R}_{p},\boldsymbol{R}_{n};t)}\label{eq:chiP_def}
\end{align}
is clearly gauge-invariant, thus establishing (\ref{eq:chi_introduce}).

We can easily work out an expansion for the $\bar{W}_{n}(\boldsymbol{x},t)$
where the overlap between nonidentical $W'_{n}(\boldsymbol{x},t)$
is small. Defining a matrix $\mathfrak{s}$, 
\begin{align*}
 & \mathfrak{s}(t)\equiv\mathbb{S}(t)-\mathbb{I},
\end{align*}
where $\mathbb{I}$ is the identity matrix, we have 
\begin{align}
\mathbb{S}^{-1/2}(t) & =(\mathbb{I}+\mathfrak{s}(t))^{-1/2}\label{eq:S_expand}\\
 & =\mathbb{I}-\frac{1}{2}\mathfrak{s}(t)+\frac{3}{8}\mathfrak{s}^{2}(t)+...\nonumber 
\end{align}
where since $\mathbb{S}$ is Hermitian $\mathfrak{s}$ will be as
well. In terms of components we have 
\begin{align*}
s_{nm}(t) & =S_{nm}(t)-\delta_{nm}\\
 & =e^{i\Phi(\boldsymbol{R}_{n},\boldsymbol{R}_{m};t)}\\
 & \quad\times\left[\left(\int W_{n}^{*}(\boldsymbol{x})e^{i\Delta(\boldsymbol{R}_{n},\boldsymbol{x},\boldsymbol{R}_{m};t)}W_{m}(\boldsymbol{x})d\boldsymbol{x}\right)-\delta_{nm}\right]\\
 & =e^{i\Phi(\boldsymbol{R}_{n},\boldsymbol{R}_{m})}\hat{s}_{nm}(t),
\end{align*}
where 
\begin{align*}
 & \hat{s}_{nm}(t)=\left(\int W_{n}^{*}(\boldsymbol{x})e^{i\Delta(\boldsymbol{R}_{n},\boldsymbol{x},\boldsymbol{R}_{m};t)}W_{m}(\boldsymbol{x})d\boldsymbol{x}\right)-\delta_{nm}
\end{align*}
is gauge invariant. Clearly the power series expansion (\ref{eq:S_expand})
yields the Hermitian square root of $\mathbb{S}$, and so we have
\begin{align*}
C_{pn}(t) & =\delta_{pn}-\frac{1}{2}s_{pn}(t)+\frac{3}{8}\sum_{u}s_{pu}(t)s_{un}(t)+...\\
 & =\delta_{pn}-\frac{1}{2}\hat{s}_{pn}(t)e^{i\Phi(\boldsymbol{R}_{p},\boldsymbol{R}_{n};t)}\\
 & +\frac{3}{8}\sum_{u}\hat{s}_{pu}(t)\hat{s}_{un}(t)e^{i\left[\Phi(\boldsymbol{R}_{p},\boldsymbol{R}_{u};t)+\Phi(\boldsymbol{R}_{u},\boldsymbol{R}_{n};t)\right]}+...
\end{align*}
Then from (\ref{eq:Cslash_def}) we have 
\begin{align*}
\hat{C}_{pn}(t) & =\delta_{pn}-\frac{1}{2}\hat{s}_{pn}(t)\\
 & +\frac{3}{8}\sum_{u}\hat{s}_{pu}(t)\hat{s}_{un}(t)e^{i\Delta(\boldsymbol{R}_{p},\boldsymbol{R}_{u},\boldsymbol{R}_{n};t)}+...
\end{align*}
which is indeed gauge-invariant. Using this in (\ref{eq:chiP_def})
yields 
\begin{align*}
\chi_{n}(\boldsymbol{x},t) & =W_{n}(\boldsymbol{x})-\frac{1}{2}\sum_{p}W_{p}(\boldsymbol{x})\hat{s}_{pn}(t)e^{i\Delta(\boldsymbol{x},\boldsymbol{R}_{p},\boldsymbol{R}_{n};t)}\\
 & +\frac{3}{8}\sum_{p,u}W_{p}(\boldsymbol{x})\hat{s}_{pu}(t)\hat{s}_{un}(t)e^{i\Delta(\boldsymbol{R}_{p},\boldsymbol{R}_{u},\boldsymbol{R}_{n};t)}\\
 & \quad\times e^{i\Delta(\boldsymbol{x},\boldsymbol{R}_{p},\boldsymbol{R}_{n};t)}+...
\end{align*}
Now defining as usual 
\begin{align*}
 & \Delta(\boldsymbol{x},\boldsymbol{R}_{p},\boldsymbol{R}_{u},\boldsymbol{R}_{n};t)\\
 & \equiv\Phi(\boldsymbol{R}_{n},\boldsymbol{x};t)+\Phi(\boldsymbol{x},\boldsymbol{R}_{p};t)+\Phi(\boldsymbol{R}_{p},\boldsymbol{R}_{u};t)+\Phi(\boldsymbol{R}_{u},\boldsymbol{R}_{n};t)\\
 & =\Delta(\boldsymbol{x},\boldsymbol{R}_{p},\boldsymbol{R}_{n};t)+\Delta(\boldsymbol{R}_{p},\boldsymbol{R}_{u},\boldsymbol{R}_{n};t),
\end{align*}
we have 
\begin{align*}
\chi_{n}(\boldsymbol{x},t) & =W_{n}(\boldsymbol{x})-\frac{1}{2}\sum_{p}W_{p}(\boldsymbol{x})\hat{s}_{pn}(t)e^{i\Delta(\boldsymbol{x},\boldsymbol{R}_{p},\boldsymbol{R}_{n};t)}\\
 & +\frac{3}{8}\sum_{p,u}W_{p}(\boldsymbol{x})\hat{s}_{pu}(t)\hat{s}_{un}(t)e^{i\Delta(\boldsymbol{x},\boldsymbol{R}_{p},\boldsymbol{R}_{u},\boldsymbol{R}_{n};t)}+...,
\end{align*}
and reverting to the original notation of the text this gives (\ref{eq:chi_expand})
as the lowest correction in the magnetic field.

\section{The treatment of an isolated atom}

Here we review the treatment of the response of an isolated atom to
an electromagnetic field, neglecting interactions between the electrons,
following the spirit of the earlier work by Healy \cite{Healybook}.
For simplicity we treat the nucleus as fixed. Introducing an electron
field operator $\psi(\boldsymbol{x},t)$ and beginning with minimal
coupling, the electronic charge and current density operators are given by 
\begin{align}
\hat{\rho}(\boldsymbol{x},t) & =e\psi^{\dagger}\left(\boldsymbol{x},t\right)\psi(\boldsymbol{x},t),\label{eq:rho_op}\\
\hat{\boldsymbol{j}}(\boldsymbol{x},t) & =\frac{1}{2}\psi^{\dagger}(\boldsymbol{x},t)\Big[\boldsymbol{J}\big(\boldsymbol{x},\mathfrak{p}_{mc}(\boldsymbol{x},t)\big)\psi(\boldsymbol{x},t)\Big]\\
 & +\frac{1}{2}\Big[\boldsymbol{J}\big(\boldsymbol{x},\mathfrak{p}_{mc}(\boldsymbol{x},t)\big)\psi(\boldsymbol{x},t)\Big]^{\dagger}\psi(\boldsymbol{x},t),\nonumber 
\end{align}
where, as in the main text, the function $\boldsymbol{J}(\boldsymbol{x},\mathfrak{p}_{mc}(\boldsymbol{x},t))$
follows from $H_{0}(\boldsymbol{x},\mathfrak{p}_{mc}(\boldsymbol{x},t))$
in the usual fashion, $\mathfrak{p}_{mc}(\boldsymbol{x},t)$ is given
by (\ref{eq:pmc_def}), and the electron field operator $\psi(\boldsymbol{x},t)$
satisfies the dynamical equation 
\begin{align*}
i\hbar\frac{\partial\psi(\boldsymbol{x},t)}{\partial t} & =\left[\psi(\boldsymbol{x},t),\mathsf{H}_{mc}(t)\right]\\
 & =\Big(H_{0}\big(\boldsymbol{x},\mathfrak{p}_{mc}(\boldsymbol{x},t)\big)+e\phi(\boldsymbol{x},t)\Big)\psi(\boldsymbol{x},t),
\end{align*}
where 
\begin{align*}
 & \mathsf{H}_{mc}(t)=\\
 & \qquad\int\psi^{\dagger}(\boldsymbol{x},t)\Big(H_{0}\big(\boldsymbol{x},\mathfrak{p}_{mc}(\boldsymbol{x},t)\big)+e\phi(\boldsymbol{x},t)\Big)\psi(\boldsymbol{x},t)d\boldsymbol{x}.
\end{align*}
Assuming the electrons involved remain in a region of space about
the nucleus, which we take to be at $\boldsymbol{R}$, we introduce
a new field operator 
\begin{align*}
 & \psi_{sp}(\boldsymbol{x},t)=e^{-i\Phi(\boldsymbol{x},\boldsymbol{R};t)}\psi(\boldsymbol{x},t),
\end{align*}
where $\Phi(\boldsymbol{x},\boldsymbol{R};t)$ is as given (\ref{eq:PHI_def})
in the main text. Now 
\begin{align*}
i\hbar\frac{\partial\psi_{sp}(\boldsymbol{x},t)}{\partial t} & =\hbar\frac{\partial\Phi(\boldsymbol{x},\boldsymbol{R};t)}{\partial t}\psi_{sp}(\boldsymbol{x},t)\\
 & +e^{-i\Phi(\boldsymbol{x},\boldsymbol{R};t)}\left(i\hbar\frac{\partial\psi(\boldsymbol{x},t)}{\partial t}\right).
\end{align*}
Since 
\begin{align*}
 & \hbar\frac{\partial\Phi(\boldsymbol{x},\boldsymbol{R};t)}{\partial t}\\
 & =\frac{e}{c}\int s^{i}(\boldsymbol{w};\boldsymbol{x},\boldsymbol{R})\frac{\partial A^{i}(\boldsymbol{w},t)}{\partial t}d\boldsymbol{w}\\
 & =-e\int s^{i}(\boldsymbol{w};\boldsymbol{x},\boldsymbol{R})E^{i}(\boldsymbol{w},t)d\boldsymbol{w}\\
 & \qquad-e\int s^{i}(\boldsymbol{w};\boldsymbol{x},\boldsymbol{R})\frac{\partial\phi(\boldsymbol{w},t)}{\partial w^{i}}d\boldsymbol{w}\\
 & =-e\Omega_{\boldsymbol{R}}^{0}(\boldsymbol{x},t)+e\int\frac{\partial s^{i}(\boldsymbol{w};\boldsymbol{x},\boldsymbol{R})}{\partial w^{i}}\phi(\boldsymbol{w},t)d\boldsymbol{w}\\
 & =-e\Omega_{\boldsymbol{R}}^{0}(\boldsymbol{x},t)-e\phi(\boldsymbol{x},t)+e\phi(\boldsymbol{R},t),
\end{align*}
where in the second to the last line we have used the definition (\ref{eq:Omega_defs})
in the last line we have used the first of (\ref{eq:relator_equations}),
we can write 
\begin{align*}
 & e^{-i\Phi(\boldsymbol{x},\boldsymbol{R};t)}\left(i\hbar\frac{\partial\psi(\boldsymbol{x},t)}{\partial t}\right)\\
 & =e^{-i\Phi(\boldsymbol{x},\boldsymbol{R};t)}\Big(H_{0}\big(\boldsymbol{x},\mathfrak{p}_{mc}(\boldsymbol{x},t)\big)+e\phi(\boldsymbol{x},t)\Big)\psi(\boldsymbol{x},t)\\
 & =\Big(H_{0}\big(\boldsymbol{x},\mathfrak{p}(\boldsymbol{x},\boldsymbol{R};t)\big)+e\phi(\boldsymbol{x},t)\Big)\psi_{sp}(\boldsymbol{x},t),
\end{align*}
(recall (\ref{eq:pref_def})) and we have 
\begin{align*}
 & i\hbar\frac{\partial\psi_{sp}(\boldsymbol{x},t)}{\partial t}=\\
 & \qquad\Big(H_{0}\big(\boldsymbol{x},\mathfrak{p}(\boldsymbol{x},\boldsymbol{R};t)\big)-e\Omega_{\boldsymbol{R}}^{0}(\boldsymbol{x},t)+e\phi(\boldsymbol{R},t)\Big)\psi_{sp}(\boldsymbol{x},t),
\end{align*}
where note the term $e\phi(\boldsymbol{R},t)$ depends only on time
and therefore will contribute only a global phase to $\psi_{sp}(\boldsymbol{x},t)$;
it will not contribute to any operator values and thus can be dropped;
we take 
\begin{align*}
 & i\hbar\frac{\partial\psi_{sp}(\boldsymbol{x},t)}{\partial t}=\Big(H_{0}\big(\boldsymbol{x},\mathfrak{p}(\boldsymbol{x},\boldsymbol{R};t)\big)-e\Omega_{\boldsymbol{R}}^{0}(\boldsymbol{x},t)\Big)\psi_{sp}(\boldsymbol{x},t),
\end{align*}
and then have 
\begin{align}
 & i\hbar\frac{\partial\psi_{sp}(\boldsymbol{x},t)}{\partial t}=\big[\psi_{sp}(\boldsymbol{x},t),\mathsf{H}_{sp}(t)\big],\label{eq:psisp_evolve}
\end{align}
where 
\begin{align}
 & \mathsf{H}_{sp}(t)=\label{eq:Htotal_sp}\\
 & \int\psi_{sp}^{\dagger}(\boldsymbol{x},t)\Big(H_{0}\big(\boldsymbol{x},\mathfrak{p}(\boldsymbol{x},\boldsymbol{R};t)\big)-e\Omega_{\boldsymbol{R}}^{0}(\boldsymbol{x},t)\Big)\psi_{sp}(\boldsymbol{x},t)d\boldsymbol{x}.\nonumber 
\end{align}

Looking at the electronic charge and current densities, we can write the first
(\ref{eq:rho_op}) as 
\begin{align}
 & \hat{\rho}(\boldsymbol{x},t)=e\psi_{sp}^{\dagger}(\boldsymbol{x},t)\psi_{sp}(\boldsymbol{x},t),\label{eq:rho_op_sp}
\end{align}
while the second becomes 
\begin{align*}
 & \hat{\boldsymbol{j}}(\boldsymbol{x},t)=\\
 & \frac{1}{2}\psi_{sp}^{\dagger}(\boldsymbol{x},t)e^{-i\Phi(\boldsymbol{x},\boldsymbol{R};t)}\left[\boldsymbol{J}\big(\boldsymbol{x},\mathfrak{p}_{mc}(\boldsymbol{x},t)\big)e^{i\Phi(\boldsymbol{x},\boldsymbol{R};t)}\psi_{sp}(\boldsymbol{x},t)\right]\\
 & +\frac{1}{2}\left[\boldsymbol{J}\big(\boldsymbol{x},\mathfrak{p}_{mc}(\boldsymbol{x},t)\big)e^{i\Phi(\boldsymbol{x},\boldsymbol{R};t)}\psi_{sp}(\boldsymbol{x},t)\right]^{\dagger}e^{i\Phi(\boldsymbol{x},\boldsymbol{R};t)}\\
 & \qquad\qquad\times\psi_{sp}(\boldsymbol{x},t),
\end{align*}
or 
\begin{align*}
 & \hat{\boldsymbol{j}}(\boldsymbol{x},t)=\\
 & \frac{1}{2}\psi_{sp}^{\dagger}(\boldsymbol{x},t)\left[\boldsymbol{J}\big(\boldsymbol{x},e^{-i\Phi(\boldsymbol{x},\boldsymbol{R};t)}\mathfrak{p}_{mc}(\boldsymbol{x},t)e^{i\Phi(\boldsymbol{x},\boldsymbol{R};t)}\big)\psi_{sp}(\boldsymbol{x},t)\right]\\
 & +\frac{1}{2}\left[\boldsymbol{J}\big(\boldsymbol{x},e^{-i\Phi(\boldsymbol{x},\boldsymbol{R};t)}\mathfrak{p}_{mc}(\boldsymbol{x},t)e^{i\Phi(\boldsymbol{x},\boldsymbol{R};t)}\big)\psi_{sp}(\boldsymbol{x},t)\right]^{\dagger}\\
 & \qquad\qquad\times\psi_{sp}(\boldsymbol{x},t),
\end{align*}
which can be written as 
\begin{align}
\hat{\boldsymbol{j}}(\boldsymbol{x},t) & =\frac{1}{2}\psi_{sp}^{\dagger}(\boldsymbol{x},t)\left[\boldsymbol{J}\big(\boldsymbol{x},\mathfrak{p}(\boldsymbol{x},\boldsymbol{R};t)\big)\psi_{sp}(\boldsymbol{x},t)\right]\label{eq:j_op_sp}\\
 & +\frac{1}{2}\left[\boldsymbol{J}\big(\boldsymbol{x},\mathfrak{p}(\boldsymbol{x},\boldsymbol{R};t)\big)\psi_{sp}(\boldsymbol{x},t)\right]^{\dagger}\psi_{sp}(\boldsymbol{x},t).\nonumber 
\end{align}

For an isolated atom we can define polarization and magnetization
$operators$ as 
\begin{align}
\boldsymbol{\hat{p}}(\boldsymbol{x},t) & \equiv\int\boldsymbol{s}(\boldsymbol{x};\boldsymbol{w},\boldsymbol{R})\hat{\rho}(\boldsymbol{w},t)d\boldsymbol{w},\label{eq:p_and_m-1}\\
\hat{m}^{j}(\boldsymbol{x},t) & \equiv\frac{1}{c}\int\alpha^{jk}(\boldsymbol{x};\boldsymbol{w},\boldsymbol{R})\hat{j}^{k}(\boldsymbol{w},t)d\boldsymbol{w}.\nonumber 
\end{align}
Then from the properties of the relators we find immediately that
\begin{align}
\hat{\rho}(\boldsymbol{x},t) & =-\boldsymbol{\nabla}\boldsymbol{\cdot}\boldsymbol{\hat{p}}(\boldsymbol{x},t)+Q\delta(\boldsymbol{x}-\boldsymbol{R}),\label{eq:rhoj_first}\\
\boldsymbol{\hat{j}}(\boldsymbol{x},t) & =\frac{\partial\boldsymbol{\hat{p}}(\boldsymbol{x},t)}{\partial t}+c\boldsymbol{\nabla\times\hat{m}}(\boldsymbol{x},t)\nonumber \\
 & -\int\boldsymbol{s}(\boldsymbol{x};\boldsymbol{y},\boldsymbol{R})\hat{K}(\boldsymbol{y},t)d\boldsymbol{y},\nonumber 
\end{align}
where 
\begin{align*}
Q & =\int\hat{\rho}(\boldsymbol{x},t)d\boldsymbol{x},\\
\hat{K}(\boldsymbol{x},t) & =\frac{\partial\hat{\rho}(\boldsymbol{x},t)}{\partial t}+\boldsymbol{\nabla}\boldsymbol{\cdot}\boldsymbol{\hat{j}}(\boldsymbol{x},t).
\end{align*}
In arriving at (\ref{eq:rhoj_first}) we have only used (\ref{eq:relator_equations})
and the fact that the charge-current operators are only nonzero in
a confined region of space near $\boldsymbol{R}$. We write the quantity
$Q$ without an operator hat because it is a conserved quantity and
can be taken as a number; it is the total electron charge. Note however
that local charge conservation (or the $ansatz$ that it holds if
we make various approximations in our equations) leads to $\hat{K}(\boldsymbol{x},t)=0$.
Hence we can write (\ref{eq:rhoj_first}) as 
\begin{align}
\hat{\rho}(\boldsymbol{x},t) & =-\boldsymbol{\nabla}\boldsymbol{\cdot}\boldsymbol{\hat{p}}(\boldsymbol{x},t)+Q\delta(\boldsymbol{x}-\boldsymbol{R}),\label{eq:rhoj_second}\\
\boldsymbol{\hat{j}}(\boldsymbol{x},t) & =\frac{\partial\boldsymbol{\hat{p}}(\boldsymbol{x},t)}{\partial t}+c\boldsymbol{\nabla\times\hat{m}}(\boldsymbol{x},t),\nonumber 
\end{align}
the standard form. There is a great advantage of determining $\hat{\rho}(\boldsymbol{x},t)$
and $\boldsymbol{\hat{j}}(\boldsymbol{x},t)$ (or their expectation
values) by first determining $\boldsymbol{\hat{p}}(\boldsymbol{x},t)$
and $\boldsymbol{\hat{m}}(\boldsymbol{x},t)$ (or their expectation
values). For if the former are found from the latter, regardless of
how many approximations are involved in determining the latter we
will still automatically have charge conservation.

For comparison with the next section of the appendices we here define
\begin{align*}
\boldsymbol{p}_{\boldsymbol{R}}(\boldsymbol{x},t) & \equiv\int\boldsymbol{s}(\boldsymbol{x};\boldsymbol{y},\boldsymbol{R})\left\langle \hat{\rho}(\boldsymbol{y},t)\right\rangle d\boldsymbol{y},\\
m_{\boldsymbol{R}}^{j}(\boldsymbol{x},t) & \equiv\frac{1}{c}\int\alpha^{jk}(\boldsymbol{x};\boldsymbol{y},\boldsymbol{R})\left\langle \hat{j}^{k}(\boldsymbol{y},t)\right\rangle d\boldsymbol{y},
\end{align*}
and then from (\ref{eq:rhoj_second}) we can write 
\begin{align}
\left\langle \hat{\rho}(\boldsymbol{x},t)\right\rangle  & =-\boldsymbol{\nabla}\boldsymbol{\cdot}\boldsymbol{p}_{\boldsymbol{R}}(\boldsymbol{x},t)+Q\delta(\boldsymbol{x}-\boldsymbol{R}),\label{eq:rhoj_second-1}\\
\left\langle \boldsymbol{\hat{j}}(\boldsymbol{x},t)\right\rangle  & =\frac{\partial\boldsymbol{p}_{\boldsymbol{R}}(\boldsymbol{x},t)}{\partial t}+c\boldsymbol{\nabla\times m}_{\boldsymbol{R}}(\boldsymbol{x},t).\nonumber 
\end{align}
Introducing a set of basis function $\left\{ W_{\alpha\boldsymbol{R}}(\boldsymbol{x})\right\} $,
where $\alpha$ varies but $\boldsymbol{R}$ is fixed, 
\begin{align*}
 & \psi_{sp}(\boldsymbol{x},t)=\sum_{\alpha}c_{\alpha}(t)W_{\alpha\boldsymbol{R}}(\boldsymbol{x}),
\end{align*}
where 
\begin{align*}
\left\{ c_{\alpha}(t),c_{\beta}(t)\right\}  & =0,\\
\Big\{ c_{\alpha}(t),c_{\beta}^{\dagger}(t)\Big\}  & =\delta_{a\beta},
\end{align*}
we can write 
\begin{align}
\rho_{\boldsymbol{R}}(\boldsymbol{x},t) & =\sum_{\alpha,\beta}\rho_{\beta\alpha}(\boldsymbol{x},\boldsymbol{R})\eta_{\alpha\beta}(t),\label{eq:rho_and_j_decomp-1-1}\\
\boldsymbol{j}_{\boldsymbol{R}}(\boldsymbol{x},t) & =\sum_{\alpha,\beta}\boldsymbol{j}_{\beta\alpha}(\boldsymbol{x},\boldsymbol{R};t)\eta_{\alpha\beta}(t),\nonumber 
\end{align}
with $\rho_{\beta\alpha}(\boldsymbol{x},\boldsymbol{R})$ and $\boldsymbol{j}_{\beta\alpha}(\boldsymbol{x},\boldsymbol{R};t)$
given by (\ref{eq:simplified_rhoj}), and 
\begin{align*}
 & \eta_{\alpha\beta}(t)\equiv\left\langle c_{\beta}^{\dagger}(t)c_{\alpha}(t)\right\rangle .
\end{align*}
From the evolution equation (\ref{eq:psisp_evolve}) we find that
the dynamical of $\eta_{\alpha\beta}(t)$ are given by 
\begin{align}
 & i\hbar\frac{\partial\eta_{\alpha\beta}(t)}{\partial t}=\sum_{\lambda}\Big(\bar{H}_{\alpha\lambda}(\boldsymbol{R};t)\eta_{\lambda\beta}(t)-\eta_{\alpha\lambda}(t)\bar{H}_{\lambda\beta}(\boldsymbol{R};t)\Big),\label{eq:atom_dynamical-1}
\end{align}
where 
\begin{align}
\bar{H}_{\alpha\lambda}(\boldsymbol{R};t) & =\frac{1}{2}\int W_{\alpha\boldsymbol{R}}^{*}(\boldsymbol{x})H_{0}\big(\boldsymbol{x},\mathfrak{p}(\boldsymbol{x},\boldsymbol{R};t)\big)W_{\lambda\boldsymbol{R}}(\boldsymbol{x})d\boldsymbol{x}\nonumber \\
 & +\frac{1}{2}\int\Big(H_{0}\big(\boldsymbol{x},\mathfrak{p}(\boldsymbol{x},\boldsymbol{R};t)\big)W_{\alpha\boldsymbol{R}}(\boldsymbol{x})\Big)^{*}W_{\lambda\boldsymbol{R}}(\boldsymbol{x})d\boldsymbol{x}\nonumber \\
 & -e\int W_{\alpha\boldsymbol{R}}^{*}(\boldsymbol{x})\Omega_{\boldsymbol{R}}^{0}(\boldsymbol{x},t)W_{\lambda\boldsymbol{R}}(\boldsymbol{x})d\boldsymbol{x}.\label{eq:HbarRt}
\end{align}

\section{The ``isolated atom limit'' of a crystal}

The isolated atom limit of our equations in Section II is identified
by the assumption that $W_{\alpha\boldsymbol{R}}(\boldsymbol{x})$
and $W_{\beta\boldsymbol{R'}}(\boldsymbol{x})$ have no common support
if $\boldsymbol{R}\neq\boldsymbol{R'}$. In this limit it follows
from the definition (\ref{eq:Wprime_def}) that the set of functions
$\left\{ W'_{\alpha\boldsymbol{R}}(\boldsymbol{x},t)\right\} $ are
mutually orthogonal, so $\bar{W}_{\alpha\boldsymbol{R}}(\boldsymbol{x},t)\rightarrow W'_{\alpha\boldsymbol{R}}(\boldsymbol{x},t)$
and, from (\ref{eq:Wprime_def},\ref{eq:chi_introduce}), we see $\chi_{\alpha\boldsymbol{R}}(\boldsymbol{x},t)\rightarrow W_{\alpha\boldsymbol{R}}(\boldsymbol{x})$.
Then from (\ref{eq:Hmatrix_def}) we have 
\begin{align}
 & \bar{H}_{\alpha\boldsymbol{R};\lambda\boldsymbol{R''}}(t)\rightarrow\delta_{\boldsymbol{RR''}}\bar{H}_{\alpha\lambda}(\boldsymbol{R};t),\label{eq:Hbar_reduce}
\end{align}
with $\bar{H}_{\alpha\lambda}(\boldsymbol{R};t)$ is given by (\ref{eq:HbarRt}).
Assuming no initial correlation between the electronic motion in the
individual atoms, from (\ref{eq:sigma_up},\ref{eq:toGI}) we have
\begin{align}
 & \eta_{\alpha\boldsymbol{R;}\beta\boldsymbol{R'}}(t)\rightarrow\delta_{\boldsymbol{RR'}}\eta_{\alpha\beta}(\boldsymbol{R};t)\label{eq:sigma_reduce}
\end{align}
at least initially, and this condition will then be maintained as
the dynamics evolve according to (\ref{eq:density_operator_dynamical_result}),
which reduces to 
\begin{align}
 & i\hbar\frac{\partial\eta_{\alpha\beta}(\boldsymbol{R};t)}{\partial t}=\label{eq:atom_dynamical}\\
 & \qquad\sum_{\lambda}\Big(\bar{H}_{\alpha\lambda}(\boldsymbol{R};t)\eta_{\lambda\beta}(\boldsymbol{R};t)-\eta_{\alpha\lambda}(\boldsymbol{R};t)\bar{H}_{\lambda\beta}(\boldsymbol{R};t)\Big).\nonumber 
\end{align}
for each $\boldsymbol{R}.$ Comparing with (\ref{eq:atom_dynamical-1})
we see that this is indeed the dynamics expected for a collection
of isolated atoms. Finally, (\ref{eq:GR_def}) becomes 
\begin{align*}
 & G_{\boldsymbol{R}}(\boldsymbol{x},\boldsymbol{y};t)=i\sum_{\alpha,\beta}\eta_{\alpha\beta}(\boldsymbol{R};t)W_{\beta\boldsymbol{R}}^{*}(\boldsymbol{y})W_{\alpha\boldsymbol{R}}(\boldsymbol{x}),
\end{align*}
and for use in (\ref{eq:rho_and_j_decomp}) we can take $\rho_{\beta\boldsymbol{R'};\alpha\boldsymbol{R''}}(\boldsymbol{x},\boldsymbol{R};t)\rightarrow\rho_{\beta\alpha}(\boldsymbol{x},\boldsymbol{R})$
and $\boldsymbol{j}_{\beta\boldsymbol{R'};\alpha\boldsymbol{R''}}(\boldsymbol{x},\boldsymbol{R};t)\rightarrow\boldsymbol{j}_{\beta\alpha}(\boldsymbol{x},\boldsymbol{R};t)$
, where from (\ref{eq:rho_decomp_term}) and (\ref{eq:j_decomp_term})
we obtain 
\begin{align}
\rho_{\beta\alpha}(\boldsymbol{x},\boldsymbol{R}) & =eW_{\beta\boldsymbol{R}}^{*}(\boldsymbol{x})W_{\alpha\boldsymbol{R}}(\boldsymbol{x}),\label{eq:simplified_rhoj}\\
\boldsymbol{j}_{\beta\alpha}(\boldsymbol{x},\boldsymbol{R};t) & =\frac{1}{2}W_{\beta\boldsymbol{R}}^{*}(\boldsymbol{x})\Big(\boldsymbol{J}\big(\boldsymbol{x},\mathfrak{p}(\boldsymbol{x},\boldsymbol{R};t)\big)W_{\alpha\boldsymbol{R}}(\boldsymbol{x})\Big)\nonumber \\
 & +\frac{1}{2}\Big(\boldsymbol{J}^{*}\big(\boldsymbol{x},\mathfrak{p}(\boldsymbol{x},\boldsymbol{R};t)\big)W_{\beta\boldsymbol{R}}^{*}(\boldsymbol{x})\Big)W_{\alpha\boldsymbol{R}}(\boldsymbol{x}),\nonumber 
\end{align}
so (\ref{eq:rho_and_j_decomp}) become 
\begin{align}
\rho_{\boldsymbol{R}}(\boldsymbol{x},t) & \rightarrow\sum_{\alpha,\beta}\rho_{\beta\alpha}(\boldsymbol{x},\boldsymbol{R})\eta_{\alpha\beta}(\boldsymbol{R};t)\label{eq:rho_and_j_decomp-1}\\
 & =-ie\left[G_{\boldsymbol{R}}(\boldsymbol{x},\boldsymbol{y};t)\right]_{\boldsymbol{y}\rightarrow\boldsymbol{x}},\nonumber \\
\boldsymbol{j}_{\boldsymbol{R}}(\boldsymbol{x},t) & \rightarrow\sum_{\alpha,\beta}\boldsymbol{j}_{\beta\alpha}(\boldsymbol{x},\boldsymbol{R};t)\eta_{\alpha\beta}(\boldsymbol{R};t)\nonumber \\
 & =-ie\left[\mathcal{J}_{\boldsymbol{R}}(\boldsymbol{x},\boldsymbol{y};t)G_{\boldsymbol{R}}(\boldsymbol{x},\boldsymbol{y};t)\right]_{\boldsymbol{y\rightarrow}x}.\nonumber 
\end{align}

From (\ref{eq:Hbar_reduce},\ref{eq:sigma_reduce}) it follows from
(\ref{eq:I_identify}) that $I(\boldsymbol{R},\boldsymbol{R'};t)=0$,
and from (\ref{eq:site_charge_dynamical}) each site charge (see (\ref{eq:QR_def}))
is independent of time, 
\begin{align*}
 & Q_{\boldsymbol{R}}=\int\rho_{\boldsymbol{R}}(\boldsymbol{x},t)d\boldsymbol{x}=e\sum_{\alpha}\eta_{\alpha\alpha}(\boldsymbol{R};t),
\end{align*}
and from (\ref{eq:rhoF_def},\ref{eq:jF_def}) we have 
\begin{align}
\rho_{F}(\boldsymbol{x}) & =\sum_{\boldsymbol{R}}Q_{\boldsymbol{R}}\delta(\boldsymbol{x}-\boldsymbol{R}),\label{eq:free_limit}\\
\boldsymbol{j}_{F}(\boldsymbol{x},t) & =0,\nonumber 
\end{align}
where the first is independent of time. Now since the Wannier functions
associated with different sites are assumed to have no common support,
from charge conservation (\ref{eq:KRterm}) we must have $K_{\boldsymbol{R}}(\boldsymbol{x},t)=0$
for all $\boldsymbol{R}$, since at any given $\boldsymbol{x}$ at
most one $K_{\boldsymbol{R}}(\boldsymbol{x},t)$ can contribute to
the sum (\ref{eq:KRterm}); together with the second of (\ref{eq:free_limit}),
this guarantees that $\boldsymbol{\tilde{j}}(\boldsymbol{x},t)=0$
(see (\ref{eq:jbar_def})). With that, and the use of (\ref{eq:rho_and_j_decomp-1}),
our general expressions (\ref{eq:pRgeneral},\ref{eq:mRgeneral})
reduce to 
\begin{align}
 & \boldsymbol{p}_{\boldsymbol{R}}(\boldsymbol{x},t)=\label{eq:pRgeneral-1}\\
 & \quad\sum_{\alpha,\beta,\boldsymbol{R'},\boldsymbol{R}''}\left[\int\boldsymbol{s}(\boldsymbol{x};\boldsymbol{y},\boldsymbol{R})\rho_{\beta\alpha}(\boldsymbol{y},\boldsymbol{R})d\boldsymbol{y}\right]\eta_{\alpha\beta}(\boldsymbol{R};t)\nonumber 
\end{align}
and 
\begin{align}
 & m_{\boldsymbol{R}}^{j}(\boldsymbol{x},t)=\label{eq:mRgeneral-1}\\
 & \quad\frac{1}{c}\sum_{\alpha,\beta,\boldsymbol{R'},\boldsymbol{R''}}\left[\int\alpha^{jk}(\boldsymbol{x};\boldsymbol{y},\boldsymbol{R})j_{\beta\alpha}^{k}(\boldsymbol{y},\boldsymbol{R};t)d\boldsymbol{y}\right]\eta_{\alpha\beta}(\boldsymbol{R};t).\nonumber 
\end{align}
With the aid of the first of (\ref{eq:first_pm_def}) and (\ref{eq:second_m_def}),
\begin{align*}
\boldsymbol{p}(\boldsymbol{x},t) & =\sum_{\boldsymbol{R}}\boldsymbol{p}_{\boldsymbol{R}}(\boldsymbol{x},t),\\
\boldsymbol{m}(\boldsymbol{x},t) & =\sum_{\boldsymbol{R}}\boldsymbol{m}_{\boldsymbol{R}}(\boldsymbol{x},t),
\end{align*}
so from the first of (\ref{eq:expectation_write1}) and (\ref{eq:full_j}),
together with (\ref{eq:free_limit}), we can write 
\begin{align}
\left\langle \hat{\rho}(\boldsymbol{x},t)\right\rangle  & =\sum_{\boldsymbol{R}}\Big(-\boldsymbol{\nabla}\boldsymbol{\cdot}\boldsymbol{p}_{\boldsymbol{R}}(\boldsymbol{x},t)+Q_{\boldsymbol{R}}\delta(\boldsymbol{x}-\boldsymbol{R})\Big),\nonumber\\
\left\langle \boldsymbol{\hat{j}}(\boldsymbol{x},t)\right\rangle  & =\sum_{\boldsymbol{R}}\left(\frac{\partial\boldsymbol{p}_{\boldsymbol{R}}(\boldsymbol{x},t)}{\partial t}+\boldsymbol{\nabla\times m}_{\boldsymbol{R}}(\boldsymbol{x},t)\right)\label{eq:isolated_limit}, 
\end{align}
with of course the dynamics of $\boldsymbol{p}_{\boldsymbol{R}}(\boldsymbol{x},t)$
and $\boldsymbol{m}_{\boldsymbol{R}}(\boldsymbol{x},t)$ given by
the use of the solution of (\ref{eq:atom_dynamical}) in (\ref{eq:pRgeneral-1},\ref{eq:mRgeneral-1}).
Comparing (\ref{eq:rhoj_second-1}) with (\ref{eq:isolated_limit})
we see that the latter is indeed what we would expect for the charge-current
density of a collection of isolated atoms.

Naturally, in a calculation with a finite set of Wannier functions
one cannot expect $K_{\boldsymbol{R}}(\boldsymbol{x},t)=0,$ or even
the sum to be zero. But this situation arose as well for an isolated
atom in Appendix C. There, and indeed as was done in the general derivation
in Section \ref{sectionII}, the approach is to envision a calculation with an infinite
set of basis functions, and construct equations for the charge and
current densities in terms of the polarization and magnetization fields.
By their very structure the equations guarantee charge conservation,
even if the set of basis functions is truncated and the resulting
expressions for the polarization and magnetization fields are only
approximate.

\bibliographystyle{apsrev4-1}
\bibliography{PMformalism_Updated}

\end{document}